\documentclass{aa}  

\usepackage{graphicx}
\usepackage[varg]{txfonts}
\usepackage{lscape}
\usepackage{natbib}
\usepackage{rotating}
\usepackage{multirow}
\usepackage{hhline}
\usepackage{tabularx}
\usepackage{booktabs}
\usepackage{caption}
\usepackage{pict2e,slashbox}
\usepackage{multicol}
\usepackage{supertabular}
\usepackage{setspace}
\usepackage{threeparttable}
\usepackage{enumerate}
\usepackage{textcomp}
\usepackage[utf8]{inputenc}
\usepackage[]{textpos}   % showboxes
\usepackage{color}

\def\deg{\ifmmode^\circ\else$^\circ$\fi}
\def\alphaTF{\ifmmode{\alpha_{\mathrm{\,{\small TF}}}}\else{$\alpha_{\mathrm{\,{\small TF}}}$}\fi}

\def\Msun{\ifmmode{\mathrm M_\odot}\else{M$_\odot$}\fi}
\newcommand{\rbreak}{\ensuremath{R_\mathrm{brk III}}}
\newcommand{\mubreak}{\ensuremath{\mu_\mathrm{brk III}}}
\newcommand{\hi}{\ensuremath{h_{\mathrm{i}}}}
\newcommand{\mui}{\ensuremath{\mu_{0,\mathrm{i}}}}
\newcommand{\ho}{\ensuremath{h_{\mathrm{o}}}}
\newcommand{\muo}{\ensuremath{\mu_{0,\mathrm{o}}}}
\newcommand{\risoph}{\ensuremath{R_\mathrm{25}}}

\begin{document}

%%\title{Antitruncated stellar discs in S0 galaxies\\formed in major mergers}
%\title{Formation of S0 galaxies with antitruncated stellar discs \\through major mergers}

\title{Formation of S0 galaxies through mergers}
\subtitle{Antitruncated stellar discs resulting from major mergers}

\author{Alejandro Borlaff\inst{1}, M.~Carmen Eliche-Moral\inst{1}, Cristina Rodr\'iguez-P\'erez\inst{1}, Miguel Querejeta\inst{2}, Trinidad Tapia\inst{3}, \\Pablo G.~P\'erez-Gonz\'alez\inst{1},  Jaime Zamorano\inst{1}, Jes\'us Gallego\inst{1}, and John Beckman\inst{4,5,6}} 

\institute{Departamento de Astrof\'{\i}sica y CC.~de la Atm\'osfera, Universidad Complutense de Madrid, E-28040 Madrid, Spain
\\\email{asborlaff@ucm.es}
\and
Max-Planck-Institut f\"{u}r Astronomie, K\"{o}nigstuhl, 17, 69117 Heidelberg, Germany
\and
Instituto de Astronom\'{i}a, Universidad Nacional Aut\'onoma de M\'exico, Apdo.~106, Ensenada BC 22800, Mexico 
\and
Instituto de Astrof\'{i}sica de Canarias, C/ V\'{i}a L\'actea, E-38200 La Laguna, Tenerife, Spain
\and
Facultad de F\'{i}sica, Universidad de La Laguna, Avda. Astrof\'{i}sico Fco. S\'{a}nchez s/n, 38200, La Laguna, Tenerife, Spain
\and
Consejo Superior de Investigaciones Cient\'{i}ficas, Spain
}

   \date{Received 28 May 2014; accepted 11 July 2014}

\abstract{%Context
Lenticular galaxies (S0s) are more likely to host antitruncated (Type III) stellar discs than galaxies of later Hubble types. Major mergers are popularly considered too violent to make these breaks.
}{%Aims
We have investigated whether major mergers can result into S0-like remnants with realistic antitruncated stellar discs or not.
}{%Methods
We have analysed 67 relaxed S0 and E/S0 remnants resulting from dissipative N-body simulations of major mergers from the GalMer database. We have simulated realistic $R$-band surface brightness profiles of the remnants to identify those with antitruncated stellar discs. Their inner and outer discs and the breaks have been quantitatively characterized to compare with real data. 
}{% Results
Nearly $70$\% of our S0-like remnants are antitruncated, meaning that major mergers that result in S0s have a high probability of producing Type III stellar discs. Our remnants lie on top of the extrapolations of the observational trends (towards brighter magnitudes and higher break radii) in several photometric diagrams, because of the higher luminosities and sizes of the simulations compared to observational samples. In scale-free photometric diagrams, simulations and observations overlap and the remnants reproduce the observational trends, so the physical mechanism after antitruncations is highly scalable. We have found novel photometric scaling relations between the characteristic parameters of the antitruncations in real S0s, which are also reproduced by our simulations. We show that the trends in all the photometric planes can be derived from three basic scaling relations that real and simulated Type III S0s fulfill: $\hi \propto \rbreak$, $\ho \propto \rbreak$, and $\mubreak \propto \rbreak$, where $\hi$ and $\ho$ are the scalelengths of the inner and outer discs, and $\mubreak$ and $\rbreak$ are the surface brightness and radius of the breaks. Bars and antitruncations in real S0s are structurally unrelated phenomena according to the studied photometric planes.
}{% Conclusions
Major mergers provide a feasible mechanism to form realistic antitruncated S0 galaxies.
}

\keywords{Galaxies: formation -- galaxies: fundamental parameters -- galaxies: evolution -- galaxies: elliptical and lenticular, cD --  galaxies: interactions --  galaxies: structure}

\titlerunning{Antitruncated stellar discs in S0 galaxies formed through major mergers}
\authorrunning{Borlaff et al.}

\maketitle

\section{Introduction}
\label{Sec:introduction}

Lenticular galaxies (S0s) have traditionally been considered as transition types between elliptical and spiral galaxies since the galaxy classification proposed by \citet{1926ApJ....64..321H}, because their discs show no significant spiral structure nor any signs of recent star formation. Lenticulars are common in the inner parts of galaxy clusters \citep{1980ApJ...236..351D}, so they may provide valuable information on the processes driving galaxy evolution in these regions. The fraction of S0s in the clusters at $z\sim 0.5$ is 2--3 times smaller than in low-redshift clusters, with a proportional increase of the spiral fraction \citep{1997ApJ...490..577D}. This is evidence that S0 galaxies must have evolved from later morphological types as they have fallen into the clusters. However, the mechanisms leading to the formation of S0s cannot be exclusive of the intracluster medium, because $\sim 50$\% of this galaxy population resides in groups and the field \citep[see][]{1982ApJ...257..423H, 2006ApJS..167....1B,2007ApJ...655..790C, 2009ApJ...692..298W}. In addition, accounting for the diversity of properties of this galaxy population, this result shows the existence of multiple evolutionary pathways to build them up \citep{2010MNRAS.405.1089L,2010MNRAS.407.1231R,2010ApJ...708..841W,2013MNRAS.432..430B,2013MNRAS.432.1010C}, triggering a heated debate over the past years about the mechanisms that have led their formation and evolution. 

Lenticular galaxies have discs that usually do not follow the typical exponential surface brightness profile \citep{2009IAUS..254..173S,2012ApJS..198....2K,2012A&AT...27..313I}. \citet[E08 hereafter]{2008AJ....135...20E} classified the discs of galaxies into three classes according to the shape of their profiles. Type-I discs are well modelled with an exponential profile for all radii. Type-II galaxies present a brightness deficit at the outer parts of the disc with respect to the extrapolated trend of the inner regions (down-bending profile), becoming steeper after certain radius (this feature is known as truncation). Finally, the surface brightness profile of Type III discs becomes less steeper than the extrapolation of the exponential trend of the inner parts after the break radius (antitruncation). Consequently, antitruncated discs present an excess of brightness from the break radius outwards compared to the inner exponential profile (up-bending profile). 

\citet{2008AJ....135...20E} and \citet[G11 henceforth]{2011AJ....142..145G} analysed the frequency of each disc profile class as a function of the Hubble type in samples of nearby barred and unbarred galaxies, respectively. They found that the fraction of Type III profiles is significantly higher in early-type than in late-type galaxies, peaking in the S0s: while $30$--50\% of S0s are antitruncated, only $\sim 20$\% of Sc-Sd's have Type III breaks \citep[see also][]{2012A&AT...27..313I}. On the contrary, Type-II profiles are more frequent in late-type galaxies \citep[$\sim 25$\% in S0s against $\sim 80$\% in Sc-Sd's, see][]{2002MNRAS.334..646K}. This indicates that the mechanisms for antitruncations are much more frequent in S0s than in spirals. Two questions thus arise from these results: 1) what these mechanisms are, and 2) whether these are important processes for the evolution of spirals into S0s or not. 

\citet{2012ApJ...744L..11E} also derived the relative frequency of profile types in S0 galaxies for the Virgo cluster and the field. They observed a complete lack of Type-II profiles in cluster S0s, whereas Type III discs were equally common in S0s in both environments. In the case that the processes responsible of the formation of a disc antitruncation are similar in both environments, the result above would imply that this process occurred with similar relative frequency in clusters and in the field. This would exclude all processes that take place preferentially in dense or in sparse environments as candidates to form Type III S0s (such as strangulation or gas stripping). However, if there were different ways to produce a Type III break in S0s depending on the environment, the results by \citeauthor{2012ApJ...744L..11E} would imply the existence of a cosmic agreement between these different mechanisms to build up a similar relative fraction of antitruncated S0s in both high and low density environments. \citet{2012MNRAS.419..669M} reported that the structure of the outer discs of spirals is not significantly affected by the galaxy environment either. These results seem more compatible with a scenario in which there is just one mechanism after the formation of antitruncations in any galaxy type, which takes place with similar likelihood in both high- and low-density environments, but that has occurred more frequently in S0s than in spirals. Anyway, more data are required to robustly establish the dependence of the shape of the discs on environment in both S0s and spiral galaxies \citep[see][]{2014arXiv1404.0559L}.

The processes that can produce antitruncations have been poorly studied. One candidate would be bars, which might produce antitruncations in the discs by means of a secular redistribution of stars and gas, or inducing different star formation thresholds as a function of radial location in the disc \citep[][]{2006ApJ...636..712E,2011MSAIS..18..113B}. \citet{2009IAUS..254..173S} argued against this mechanism, because there is a higher percentage of Type III profiles in unbarred S0 galaxies than in barred ones (50\% vs. 30\%). This is also supported by E08 and G11 results: antitruncated stellar discs appear in $\sim 38${\%} of unbarred S0-Sb galaxies, but only in $\sim 24$\% of barred ones. No study has demonstrated the feasibility of forming antitruncations through bars, successive stellar formation phases taking place at different radii, disc tides, or radial migration either. 

There is observational evidence suggesting that interactions may be responsible for some antitruncations \citep{2005ApJ...626L..81E,2014arXiv1404.0559L}. In fact, minor mergers are the most invoked mechanism since \citet[Y07 hereafter]{2007ApJ...670..269Y} proved with numerical simulations that merger experiments with mass ratios above 7:1 and large gas fractions can result in stable antitruncated discs \citep[see also][]{2001MNRAS.324..685L}. In fact, a scenario in which S0s have grown through a higher number of minor mergers than spirals seems compatible with: 1) the higher percentage of tidal tails and merger relics observed in early types than in late ones \citep{2010AJ....140..962M,2011A&A...532A..74B, 2011MNRAS.417..863D}; 2) the higher bulge-to-disc ratios exhibited by S0s than by later types on average \citep[because minor mergers are known to induce bulge growth, see][]{2001A&A...367..428A,2006ApJ...639..644E}; and 3) the higher fraction of Type III discs found in S0s than in spirals (see references above). However, a set of minor mergers would progressively exhaust the gas in the remnant galaxy, and according to the simulations by Y07, the lack of gas prevents antitruncations from forming. Therefore, a formation scenario of a Type III S0 through a set of successive minor mergers would be feasible only if gas infall is extraordinarily efficient.

No study has analysed yet whether major mergers can form Type III S0s or not. N-body simulations have shown that, for typical gas contents in the progenitors, major mergers present mild probabilities of producing well-defined discs in the remnants \citep[][]{2003ApJ...597..893N,2005A&A...437...69B}. Even if a major merger resulted in an S0-like remnant, its disc would form independently of the central bulge (the bulge would result from the merger, whereas the disc would grow around it later through the re-accretion of expelled gas and stellar material). The bulge is thus expected to be structurally decoupled from the underlying disc in S0s that result from major mergers, but this is in strong disagreement with real data because local S0s exhibit a bulge-disc structural coupling as strong as spirals \citep{2010MNRAS.405.1089L}. If major mergers explain with difficulty the global structure of discs in S0s, they cannot be good candidates for explaining the formation of antitruncations either. Arguments like these have pushed major mergers into the background of the formation scenarios of S0s.

However, hierarchical models of galaxy formation predict that E-S0 galaxies have undergone at least one major merger in the last $\sim 9$\,Gyr regardless of the environment, a scenario that is also supported by several observational studies \citep[see][and references therein]{2010A&A...519A..55E,2010arXiv1003.0686E,2011MNRAS.412..684B,2011MNRAS.412L...6B,2013MNRAS.428..999P,2014arXiv1403.4932C}. In fact, many authors find observational evidence of a major-merger origin of many S0s at $z<1$ \citep[see][]{2009A&A...496...51P,2009A&A...501..437Y,2009A&A...507.1313H,2009A&A...496..381H,2012MPLA...2730034H,2014A&A...565A..31T}. So, the question of whether major mergers can produce Type III S0s remains unsettled. In order to shed some light on this question, we have investigated whether major mergers can generate S0 remnants with realistic antitruncated disc profiles using N-body simulations. In a forthcoming paper, we will analyse the conditions and physical mechanisms after the formation of these features. 

The outline of this paper is as follows. The methodology followed is described in detail in Sect.\,\ref{Sec:methodology}. The results are presented in Sect.\,\ref{Sec:results}, where we compare the properties of the antitruncations of our S0-like remnants with those exhibited by real data and perform a modelling of the trends in all the photometric parameters on the basis of three basic scaling relations. The discussion of the results and final conclusions can be found in Sects.\,\ref{Sec:discussion} and \ref{Sec:conclusions}. We comment on the limitations of the models in Appendix\,\ref{Sec:limitations}. We assume a concordant cosmology \citep[$\Omega_M = 0.3$, $\Omega_\Lambda = 0.7$, $H_0 = 70$ km s$^{-1}$ Mpc$^{-1}$, see][]{2007ApJS..170..377S}. All magnitudes are in the Vega system.

\section{Methodology} 
\label{Sec:methodology}

In order to investigate if major mergers can result in realistic S0s with antitruncated stellar discs, we have analysed the remnants available in the GalMer database, which is a library of hydrodynamic N-body simulations of galaxy mergers. It is briefly described in Sect.\,\ref{Sec:galmer}. We have selected the major merger experiments that are dynamically relaxed and which present typical properties of S0 galaxies at the end of the simulation. The selection is described in detail in Eliche-Moral et al.\,(in preparation, Paper I hereafter), but we summarize it in Sect.\,\ref{Sec:S0identification}. The simulation of realistic surface brightness profiles from the mass density maps of the selected remnants is presented in Sect.\,\ref{Sec:profiles}. Finally, Sect.\,\ref{Sec:fits} shows the procedure carried out to identify and characterize the antitruncations in the remnants. 

\subsection{The GalMer database} 
\label{Sec:galmer}

The GalMer database\footnote{The GalMer database is available at: http://galmer.obspm.fr/} is a public library of hydrodynamic N-body galaxy simulations of galaxy mergers developed under the Horizon Project \citep{2010A&A...518A..61C}. Here, we will provide a brief summary of the most relevant aspects of these simulations, but we refer the reader to \citeauthor{2010A&A...518A..61C} for detailed information. 

GalMer contains $\sim$1000 dissipative simulations of binary galaxy encounters, considering progenitors with different morphologies (E0, S0, Sa, Sb, Sd) and sizes (g: giant, i: intermediate, d: dwarf), which interact according to different orbital configurations. The database presents a web form which allows the selection of the simulations on the basis of the types, sizes of the progenitors, and the orbital parameters: inclination of the galaxies with respect to the orbital plane, initial distance between progenitors, pericentre, motion energy, and spin-orbit coupling type (prograde or retrograde encounters). The mass ratios of the encounters range from 1:1 to 20:1 depending on the couple of progenitors under consideration, although the majority of the experiments contained in the database are major encounters, and exclusively between giant progenitors. As we are interested in studying whether major mergers can give rise to antitruncations, we have centred our analysis on the available major merger experiments, which sum up 876 models in total. We emphasize that these major merger models consider only giant progenitors (there are no intermediate -- intermediate or dwarf -- dwarf major encounters in GalMer up to date). 

Each progenitor galaxy is modelled as an spherical non-rotating dark-matter halo, a stellar disc (except for E0 progenitors), a gaseous disc (except for E0 and S0 progenitors), and a central non-rotating bulge (except for Sd progenitors). The bulge-to-disc ratios in the S0, Sa, and Sb progenitors are 2.0, 0.7, and 0.4 respectively (see Paper I). Spherical components are modelled as Plummer spheres, with characteristic mass and radius $M_\mathrm{B}$ and $r_\mathrm{B}$ for the bulge, and $M_\mathrm{H}$ and $r_\mathrm{H}$ for the dark matter halo (see Table\,\ref{tab:morph_params}). The stellar and gaseous discs follow the Miyamoto-Nagai density profile \citep{1975PASJ...27..533M}, with masses $M_\mathrm{\star}$ and $M_\mathrm{g}$ respectively, vertical and radial scalelengths given by $h_\mathrm{\star}$ and $a_\mathrm{\star}$ in the stellar one and by $h_\mathrm{g}$ and $a_\mathrm{g}$ in the gaseous one (we list them in Table\,\ref{tab:morph_params}). The total number of particles for the giant -- giant major 
merger simulations is 240,000, with 120,000 particles per galaxy distributed among its components depending on the morphology. Total stellar masses in the progenitors range $\sim 0.5$ -- $1.5\times 10^{11}\Msun$. The characteristic parameters and number of particles of each component in the five types of giant progenitors are shown in Table\,\ref{tab:morph_params}. Mass ratios range from 1:1 to 3:1 for the considered simulations, as indicated in Table\,\ref{tab:massratios}. There are no major merger simulations including gS0 progenitors in the database.

Each merger experiment within the database has been evolved for 3 -- 3.5\,Gyr using a Tree-SPH code, in which gravitational forces are calculated using a hierarchical tree method and the gas evolution is simulated by means of smoothed particle hydrodynamics. The softening length is fixed to $\epsilon = 280$\,pc in the giant -- giant encounters. Pressure gradients, gas viscous forces, radiative forces, and star formation are considered. Gas particles are modelled as hybrid particles, following the method described in \citet{1994ApJ...437..611M}. These particles are characterized by two mass values at each time $t$ during the whole period of the simulation: the total gravitational mass of the particle, $M_{\mathrm{i}}$, and the gas mass remaining in it at time $t$, $M_{\mathrm{i,gas}}$. When the gas fraction falls below 5\% of the initial $M_{\mathrm{i,gas}}$ content, the hybrid particle turns into a stellar particle and the remaining gas is dispersed among its neighbours. Each hybrid particle presents its own star formation rate (SFR) and stellar mass loss history. The simulations assume a certain Initial Mass Function (IMF) and a prescription for star formation to account for the mass returned to the ISM by the young stars. GalMer simulations assume isothermal gas with a temperature $ T_\mathrm{gas} = 10^{4}$\,K. 

\begin{figure*}[t!]
\center
\includegraphics[width = \textwidth]{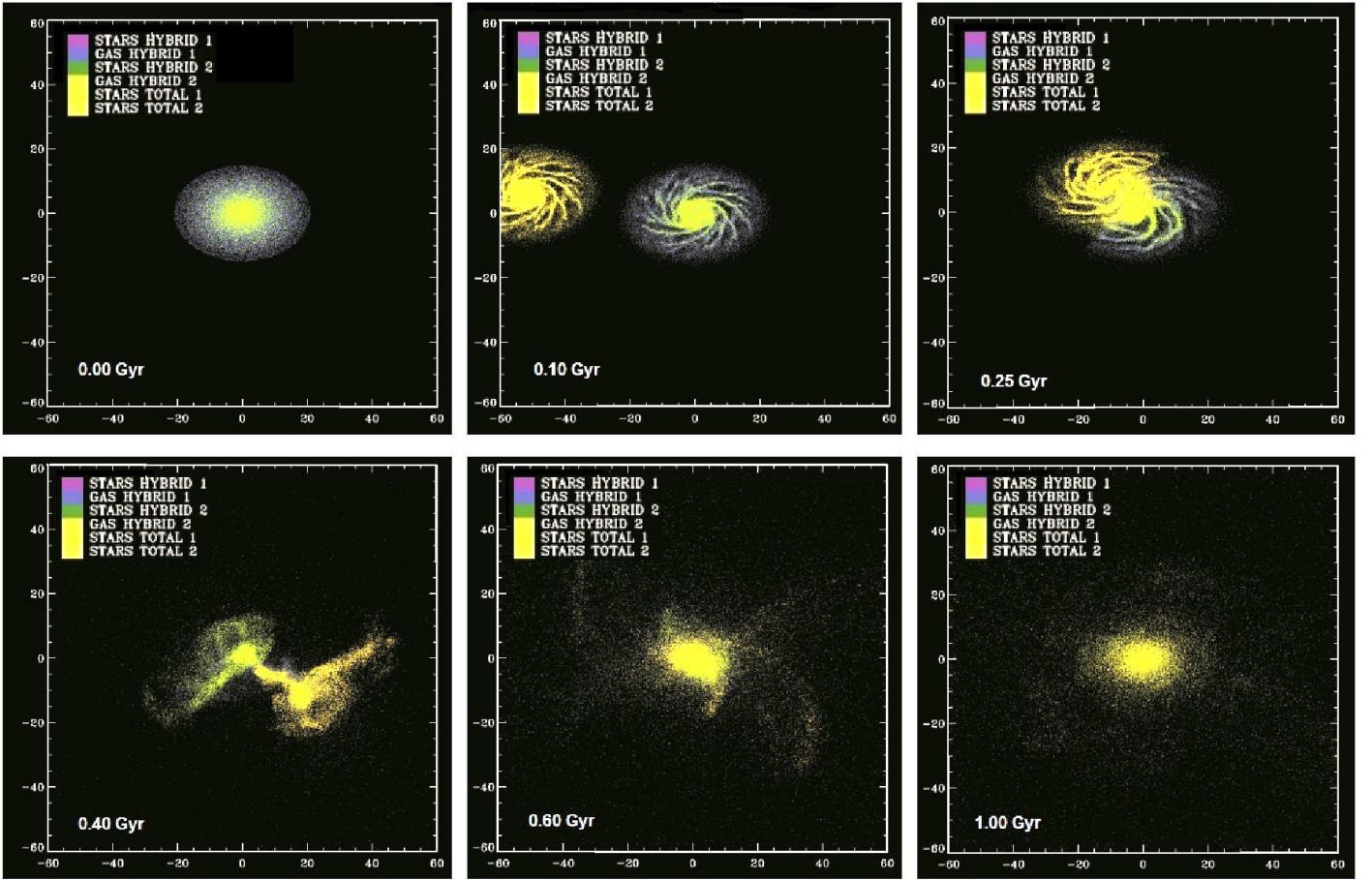}  
\caption{Time evolution of the baryonic material in a merger model in which the two progenitors are gSd galaxies (experiment gSdgSdo23).  Time is given in the bottom left corner of each frame. At each snapshot, the view is centred on progenitor 1 with $\theta = 45^\mathrm{o}$ and $\phi = 90^\mathrm{o}$, assuming the original coordinates system of the GalMer simulation. Particles belonging to different galaxy components in the two merging galaxies are coded with different colours according to the legend in the panels. The label ''Star total'' refers to the collisionless stellar content of each progenitor (i.e. the ''old stars''). [\emph{A colour version of this figure is available in the online edition.}]
}
\label{fig:sim_frames}
\end{figure*}

Each simulation will be referred from now on with the code g[\emph{type1}]g[\emph{type2}]o[\emph{orbit}], where [\emph{type1}] and [\emph{type2}] are the morphological types of the progenitor galaxies, and [\emph{orbit}] refers to the numerical identifier of the orbit used in the GalMer database, which is unique for each set of orbital parameters. This nomenclature is simpler than the original one used by \citet{2010A&A...518A..61C}, which specifies the orbital parameters in the name of the experiment (g[type1]g[type2][orbit id][dir/ret][inclination]). A Table with the equivalence of both nomenclatures will be provided in Paper I. The simulations are classified within the database considering the progenitor of the earlier Hubble type as the primary galaxy in each experiment, in order to avoid duplicated models in the database. GalMer provides the simulations in binary FITS tables, one per time $t$ from the starting of the simulation to the final time, in time intervals of 50\,Myr. The position, velocity, mass, and relevant properties at time $t$ for each particle in the simulation are stored in each FITS file. We show the time evolution of one of these experiments in Fig.\,\ref{fig:sim_frames}. 

We remark that, in the present study, we have analysed the major merger simulations available in the database, which only consist of giant -- giant encounters. The stellar mass of the remnants ranges between $\sim 1$ -- $3\times 10^{11}\Msun$. We will take this into account when comparing with observations in Sect.\,\ref{Sec:results}. 

\begin{table}
%\begin{minipage}[t]{0.5\textwidth}
\caption{Masses, radial and vertical scalelengths, and number of particles of the progenitor galaxies in the major merger experiments available in GalMer}
\label{tab:morph_params}
{\footnotesize
\begin{center}
\begin{tabular}{llrrrr}
\toprule
\multicolumn{2}{c}{Characteristic parameters} & gE0  & gSa & gSb & gSd\\
\midrule
(a)& $M_\mathrm{B}$ [$2.3\times 10^{9} M_\odot$] & 70 & 10 & 5 & 0\\
& $M_\mathrm{H}$ [$2.3\times 10^{9} M_\odot$] & 30 & 50 & 75 & 75\\
& $r_\mathrm{B}$ [kpc] & 4  & 2 & 1 & -\\
& $r_\mathrm{H}$ [kpc] & 7  & 10 & 12 & 15\\\vspace{-0.2cm}\\\hline \vspace{-0.2cm}\\
(b)& $M_\mathrm{\star}$ [$2.3\times 10^{9} M_\odot$] & 0  & 40 & 20 & 25\\
& $M_\mathrm{g}/M_\mathrm{\star}$ & 0 & 0.1 & 0.2 & 0.3\\
& $a_\mathrm{\star}$ [kpc] & -  & 4 & 5 & 6\\
& $h_\mathrm{\star}$ [kpc] & -  & 0.5 &0.5 & 0.5\\
& $a_\mathrm{g}$ [kpc] & - & 5 & 6 & 7\\
& $h_\mathrm{g}$ [kpc] & -  & 0.2 & 0.2 & 0.2\\\vspace{-0.2cm}\\\hline \vspace{-0.2cm}\\
(c) & $N_\mathrm{g}$ & -  & 20K & 40K & 60K\\   %\vspace{-0.2cm}\\\hline \vspace{-0.2cm}\\
     & $N_\mathrm{stellar}$ & 80K & 60K & 40K & 20K\\   %\vspace{-0.2cm}\\\hline \vspace{-0.2cm}\\
     & $N_\mathrm{DM}$ & 40K & 40K & 40K & 40K\\
\bottomrule
\end{tabular}
\begin{minipage}[t]{0.45\textwidth}{\vspace{0.2cm}
\emph{Rows}: (a) Characteristic parameters of the Plummer spheres used to model the bulge and the halo of the progenitor galaxies, for the different morphological types under consideration. (b) Characteristic parameters of the Miyamoto-Nagai density profiles used for the gaseous and stellar discs for the different morphological types of the progenitors. (c) Number of hybrid (initially gaseous), collisionless stellar, and dark matter particles used to simulate each progenitor galaxy, as a function of its morphological type.}
\end{minipage}
\end{center}
} \vspace{-0.2cm}
\end{table}

\begin{table}
\caption{Mass ratios of the major merger simulations available in the GalMer database, as a function of the progenitor types}
\label{tab:massratios}
{\normalsize
\begin{center}
\begin{tabular}{cccccc}
\toprule
\backslashbox[8mm]{Type 1}{Type 2}    &     & gE0 &  gSa   & gSb & gSd\\
\midrule
gE0    &      &      1:1          & 1.5:1  & 3:1 & 3:1\\
gSa    &     &        --             & 1:1   & 2:1   & 2:1\\
gSb    &     &        --           & --   & 1:1      & 1:1\\
gSd    &     &        --          & --   &  --        & 1:1\\
\bottomrule
\end{tabular}
\tablefoot{The GalMer database classifies the binary merger simulations considering the progenitor of earlier morphological type as the primary galaxy (\emph{type1}). This study was carried out with the 876 major merger simulations available in the database up to February 2014, which only involved giant progenitors and did not contain any experiments with a gS0 progenitor.
}
\end{center}
}\vspace{-0.8cm}
\end{table}

\subsection{Selection of S0-like relaxed remnants} 
\label{Sec:S0identification}

We have selected the major merger experiments from GalMer which result in S0-like relaxed remnants at the end of the simulation. This selection is described in detail in Paper I. Here, we briefly describe the procedure and the final sample of merger experiments used in the present study.

From the initial sample of 876 simulations of major mergers available in GalMer, we first removed those that do not produce a unique remnant body at the end of the simulation (i.e. those that are flyby interactions or encounters that take longer to merge than the considered time interval). In most cases, the galaxies are merged into a single body soon after the second pericentre passage. Then we identified the experiments that result in a final remnant with a noticeable disc component and apparently relaxed, inspecting the mass density maps of the stellar material. These maps were analysed visually with the previewer of the GalMer database, considering different views and distances to the galaxies. Using the classifications performed by three independent co-authors, we obtained an initial subsample of 215 candidates to be relaxed S0-like remnants, which represents $\sim 25$\% of all the major mergers stored in the database.

We performed a set of quantitative tests on each remnant on the basis of strict dynamical criteria to discard the remnants that were not dynamically relaxed from this subsample, obtaining a subsample of 173 final relaxed remnants with a high probability of having a disc component that might be detectable in broad-band images. 

\begin{figure}[t!]
\center
\includegraphics[width = 0.5\textwidth]{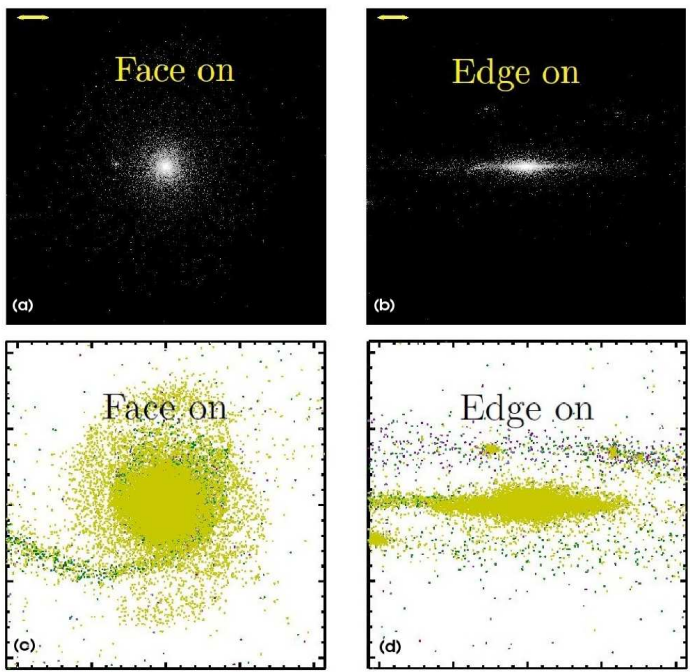}  
\caption{Relevance of accounting for observational effects in the identification and characterization of morphological features in the remnants that result from N-body simulations. \emph{Top panels}: Simulated R-band images of the final stellar remnant of model gSdgSdo9, for face-on and edge-on views. The simulations assume $D = 30$\,Mpc, $\mu_{R,\mathrm{lim}}= 27.5$\,mag\,arcsec$^{-2}$, and a FWHM$=$0.7\arcsec. The segment represents a length of 10\,kpc. \emph{Bottom panels}: Maps of projected locations of the stellar particles in the same remnant, for the same views. The colour code of the particles is the same as in Fig.\,\ref{fig:sim_frames}. The field of view in all panels is 100\,kpc${\times}$100\,kpc.  [\emph{A colour version of this figure is available in the online edition.}] }
\label{fig:remnant_fe_on}
\end{figure}

In order to test whether the discs of our S0-like remnants present antitruncations coherent with those observed in real S0 galaxies, we first need to assess that the analysed remnants look like real E/S0 and S0 galaxies in simulated broad-band photometric images that mimic real observations. Realistic photometric images of these stellar remnants were simulated in several broad bands ($B$, $V$, $R$, $I$, and $K$), in both face-on and edge-on views, in order to visually classify them according to their morphologies (E, E/S0, S0, or spiral). This step is extremely relevant for our study, because previous theoretical studies usually identify morphological features in simulated remnants only through density maps or in plots of the projected locations of the particles, which do not account for the limited spatial resolution and depth, the cosmological dimming, and the noise inherent in real astronomical data. These maps do not consider the effects of distance (which typically fade out the external regions of galaxies) or the different mass-to-light ratios ($M/L$) across the galaxy, so it cannot be ensured that a disc component observable in these maps would be detectable by an observer accounting for all these effects. An example of the loss of disc components in our remnants when observational effects are considered is shown in Fig.\,\ref{fig:remnant_fe_on}. We compare identical views of the remnant of model gSdgSdo9 in realistic simulated deep $R$-band images and in simple plots of projected positions of the particles. The outer disc regions and the tidal tails that are clearly observed in the plots of projected locations are practically lost in the $R$-band images. The satellite located at $\sim 15$\,kpc from the remnant centre becomes difficult to detect in the image, even though the limiting magnitude assumed 
is quite deep ($\mu_R= 27.5$\,mag\,arcsec$^{-2}$). 

We have converted the stellar mass contained in each particle into light flux in a given band by assuming the $M/L$ value in the band corresponding to a stellar population with the average age and metallicity of the stars contained in the particle, according to the stellar population synthesis models by \citet{2003MNRAS.344.1000B}. For collisionless stellar particles (old stellar particles, henceforth), we considered evolutionary models that assumed a SFH characteristic of the morphological type of the progenitor the particle originally belonged to. The parametrization of the SFH for each progenitor type (E, Sa, Sb, or Sd) was taken from \citet[]{2010A&A...519A..55E}, and it is based on observational studies. The age assumed for old stellar particles (which basically make up the whole structure of the remnants, and in particular, the discs) has been set to 10\,Gyr, according to the results of recent studies on the old stellar content of the discs in nearby S0 galaxies \citep{2012MNRAS.427..790S,2013MSAIS..25...93S}. The effects of this assumption are analysed in detail in Sect.\,\ref{Sec:results}. For hybrid particles with some stellar content at the end of the simulation, we instead considered simple stellar population (SSP) models for describing the SFH occurring in them during the whole simulation. This is justified by the fact that the star formation in most mergers basically takes place in two short starbursts: one occurring soon after the first pericentre passage, and the other after the full-merger moment \citep{2008A&A...492...31D}. Therefore, a SSP model is a reasonable approximation for the SFH experienced by hybrid particles. The IMF was set to Chabrier type and the stellar evolution to the prescription of Padova 1994 models in all cases. For the simulated images, we have also considered observing conditions (such as limiting magnitude, photon noise, spatial resolution, and seeing) analogous to those of the data in the observational studies by E08 and G11, which we are going to consider as the observational reference samples throughout the paper. We assumed a luminosity distance of $D=30$\,Mpc to convert between spatial into sky projected angular ones, because this is approximately the maximum distance exhibited by the S0 galaxies contained in the reference observational sample by E08 and G11. This ensures that any break found in our simulated remnants would be detectable by these observers too. Dust extinction has not been considered in the simulated images (Sect.\,\ref{Sec:dust}).

Five co-authors performed independent morphological classifications attending to the photometric images of each remnant. The final morphological type of each remnant corresponds to the median value of the five classifications. A complete agreement between all classifiers was obtained in 85\% of the remnants, ensuring the robustness of the final classification. In the sample of 173 relaxed major merger remnants with possible detectable discs, we finally identified 106 E's, 25 E/S0s, and 42 S0s, which correspond to the following percentages: 61.3\% of E's, 14.4\% of E/S0s, and 24.3\% of S0s. The elliptical galaxies contained in this subsample are particular, in the sense that, even though they have no significant or recognizable disc in realistic broad-band photometric images, they host a disc component which can be identified in their density maps. 

We confirmed the morphological classification (elliptical or disc galaxy) by identifying the two galaxy components (bulge$+$disc) in realistic simulations of the radial surface brightness profiles of the remnants in the $V$, $R$, and $K$ bands (for more information, see Sect.\,\ref{Sec:profiles}). We also studied the global kinematics, final SFR levels, and gas contents of the remnants, to assess that they are typical of the assigned morphological types. We found that all the relaxed remnants in our sample with S0-like morphology (25 E/S0s and 42 S0s) exhibited morphological, structural, and kinematic properties coherent with those observed in real E/S0 and S0 systems, as well as gas contents and SFRs typical of these types. For more details, see Paper I.

\subsection{Simulation of realistic surface brightness profiles}
\label{Sec:profiles}

In order to identify breaks in the discs of our final sample of 67 E/S0 and S0 relaxed remnants, we have simulated realistic surface brightness profiles from projected radial mass density profiles, reproducing the photometric band and observing conditions of the real data used in E08 and G11 and following a procedure similar to the one commented in Sect.\,\ref{Sec:S0identification} to simulate realistic photometric images. This ensures that the features identified as Type III breaks are comparable to those observed in real data. 

To convert mass into light, we adopted for each particle the $M/L$ ratio in the $R$ band corresponding to a stellar population of the same age and metallicity of the particle, as described in Sect.\,\ref{Sec:S0identification}, where specific details are provided. We remark that we have not included dust extinction. We have obtained azimuthally-averaged 1D surface brightness profiles of the stellar material in the remnants, considering face-on views of them centred on their mass centroids. By assumption, the face-on view of each remnant corresponds to the direction of the total angular momentum of its baryonic content. Therefore, any direction perpendicular to this will provide an edge-on view of it. We have chosen the direction that is in the XY plane of our original coordinates system.

The $R$-band images in E08 and G11 studies reached depths of $\mu_R\sim 27$ -- 28\,mag\,arcsec$^{-2}$ for a limiting signal-to-noise ratio of $S/N =5$. We have thus assumed a limiting surface brightness of $\mu_R \sim 27.5$\,mag\,arcsec$^{-2}$ for $S/N=5$ in our simulations and added Poissonian noise to the data considering the limiting signal-to-noise ratio. The projected spatial resolution has been fixed to the average seeing of their data too (FWHM$\sim 0.7$\arcsec). To ensure that our profiles can be compared to those obtained by E08 and G11, we have assumed a distance to the remnants of $D=30$\,Mpc, which is the maximum distance within this observational sample (see also Sect.\,\ref{Sec:S0identification}). We have assumed the concordant $\Lambda$CDM cosmological model to convert from physical lengths to projected angular values and to correct for cosmological dimming. 

We have also reproduced the radial average performed by E08 and G11 to the surface brightness profiles of their data to improve the $S/N$ ratio in the outskirts of the discs. Their spatial radial bins in the profiles fulfilled the non-linear relation: $R_{{i+1}} = 1.03 {\times }R_{{i}}$ (i.e. they are logarithmically equispaced). Therefore, the spatial resolution in the profiles decreases as the radial position increases in the discs. We note that neither these non-equispaced radial bins nor the seeing considered affect our analysis, since we are dealing with large radial extensions in the galaxy. We trace our galaxy discs until 40 - 70 kpc in radius, so the largest radial bin considered (located at the end of the disc) is narrower than $\sim 2$\,kpc in any case for our simulations. Additionally, the minimum spatial resolution (0.7\arcsec\ at the initial radial bin, at the galaxy centre) is equivalent to $\sim 100$\,pc for the distance considered to our remnants ($D=30$\,Mpc). This scale is even lower than the softening length used in the experiments, which really sets the minimum spatial resolution of the simulations. Therefore, the discs in our remnants are adequately sampled, with spatial resolutions between $\sim 0.3$ -- 2\,kpc as we move towards the disc outskirts.

\begin{figure}[t!]
\center
\includegraphics[width = 0.5\textwidth]{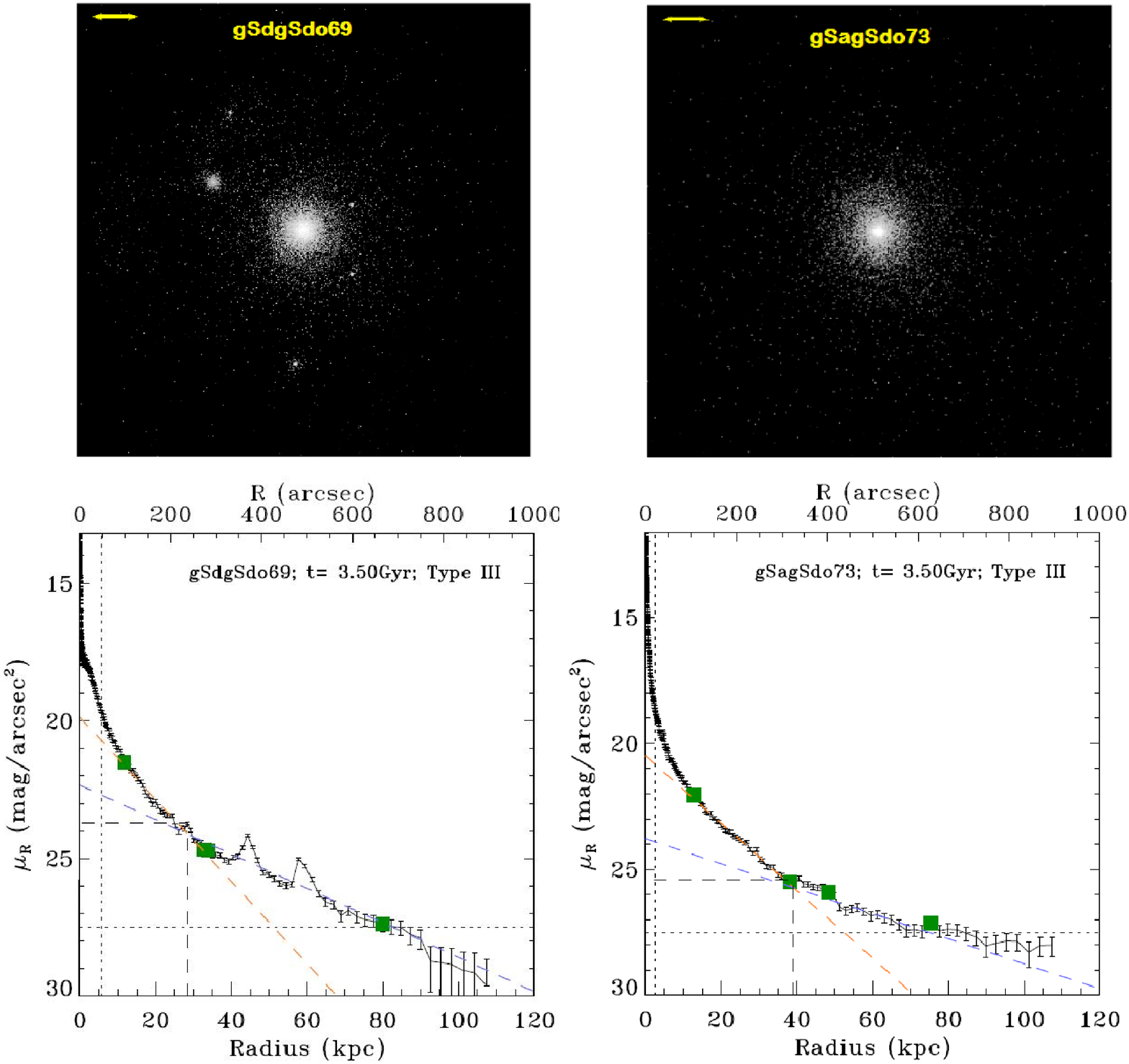}  
\caption{Comparison between the $R$-band images and surface brightness profiles of a remnant that develops several tidal satellites (model gSdgSdo69, on the left) and one that does not (gSagSdo73, on the right). \emph{Top panels}: $R$-band simulated image for the face-on view of each remnant, where tidal satellites can be easily identified. \emph{Bottom panels}: $R$-band surface brightness profiles simulated in each case, showing peaks at the radial positions where the remnants present a satellite.  [\emph{A colour version of this figure is available in the online edition.}]
}%\addtocounter{figure}{-1}
\label{fig:satellites_comp}
\end{figure}

Some examples of the resulting $R$-band radial surface brightness profiles are plotted in Fig.\,\ref{fig:ejemplos}. All profiles exhibit a clear bulge$+$disc structure, with evidence of additional galaxy components in some of them, such as lenses (see, e.g. model gSagSao5). The profiles are quite regular and smooth in general, and correspond to one of the three profile types defined by E08 (Type-I, Type-II, or Type III). Some of the mergers develop a significant population of tidal satellites around the final remnant, which produce noticeable peaks in the profile at their radial locations. In Fig.~\ref{fig:satellites_comp}, we compare the surface brightness profiles of two Type III remnants. The remnant of model gSdgSdo69 presents several tidal satellites, which produce the peaks in the surface brightness profiles at $r\sim 40$ and $r\sim 60$\,kpc. The remnant of model gSagSdo73 does not have any tidal satellite, so its profile presents a smooth decay along the whole radial range under consideration. We have ignored the existence of these peaks in the analysis of the structure of the remnant discs.

In Paper I, we show that our S0-like remnants are bluer than nearby quiescent S0s by $\sim 0.6$\,mag as a result of the recent bursts of star formation induced by the mergers. This effect makes the average $M/L$ ratios used in our models to be $\sim 0.5$ times smaller than the typical ones in quiescent early-type galaxies. Therefore, our remnants are inherently twice as bright as the brightest galaxies in the reference samples by E08 and G11 in general, despite the fact that they may have similar stellar masses, just because of these recent merger-induced starbursts. Moreover, other assumptions adopted to estimate the $M/L$ ratios used in the models also affect the total luminosity of the remnants, such as the selected ages of the old stellar particles, the SFHs assumed for the old stellar particles of each progenitors, the IMF, or the effects of dust extinction (see Appendix\,\ref{Sec:limitations}). 

Additionally, our remnants are also $\sim 2$ times larger than the S0s in E08 and G11 samples. The optical size of the remnants in our simulations is $\risoph \sim 20$\,kpc on average, whereas $\risoph \sim 10$\,kpc typically in the Type III S0s of the observational samples  (see Table\,\ref{tab:antitrunc}). Therefore, our S0-like remnants are twice as bright and larger than the most massive S0s present in E08 and G11 samples, despite of having similar stellar masses (Sect.\,\ref{Sec:IMF_SFH}). 

This does not mean that our remnants have unrealistic luminosities or sizes, because present-day massive S0s seem to have evolved passively during the last $\sim 6$\,Gyr (see references in Sect.\,\ref{Sec:introduction}), whereas our remnants have experienced passive evolution for $\sim 1$ -- 2 \,Gyr at most (Paper I). Consequently, the stars formed in the merger-induced starbursts are quite young. Considering the evolution of the SFHs typically assumed as representative of E-S0s, these galaxies can fade by up to $\sim 1$\,mag and $\sim 2$\,mag in the $K$ and $B$ bands respectively during the $\sim 2$\,Gyr after its buildup \citep{2013MNRAS.428..999P}. Therefore, the offset between the data and the simulation in the total luminosities might be reduced just allowing the remnants to relax for a longer time period. In fact, there are S0 galaxies in the NIRS0S sample \citep{2010MNRAS.405.1089L,2011MNRAS.418.1452L} with masses and scalelengths similar to our remnants (Querejeta et al., in preparation, Paper III of this series), but the reference observational samples used in this study (E08; G11) lack these systems due to volume limitations. Anyway,  we will show that the physical mechanism after the formation of antitruncations in real S0s and in our major merger simulations is highly scalable, because both simulations and observations overlap in scale-free photometric planes, showing very similar trends. This scalability ensures that the results obtained in our simulations can be extrapolated to other mass ranges and validate their comparison with real data. 

All these differences in the size and luminosity of our remnants compared with the data in E08 and G11 are accounted in the discussion of the results in Sect.\,\ref{Sec:results}.

\vspace{0.5cm}

\begin{figure*}[!t]
\includegraphics[width=\textwidth]{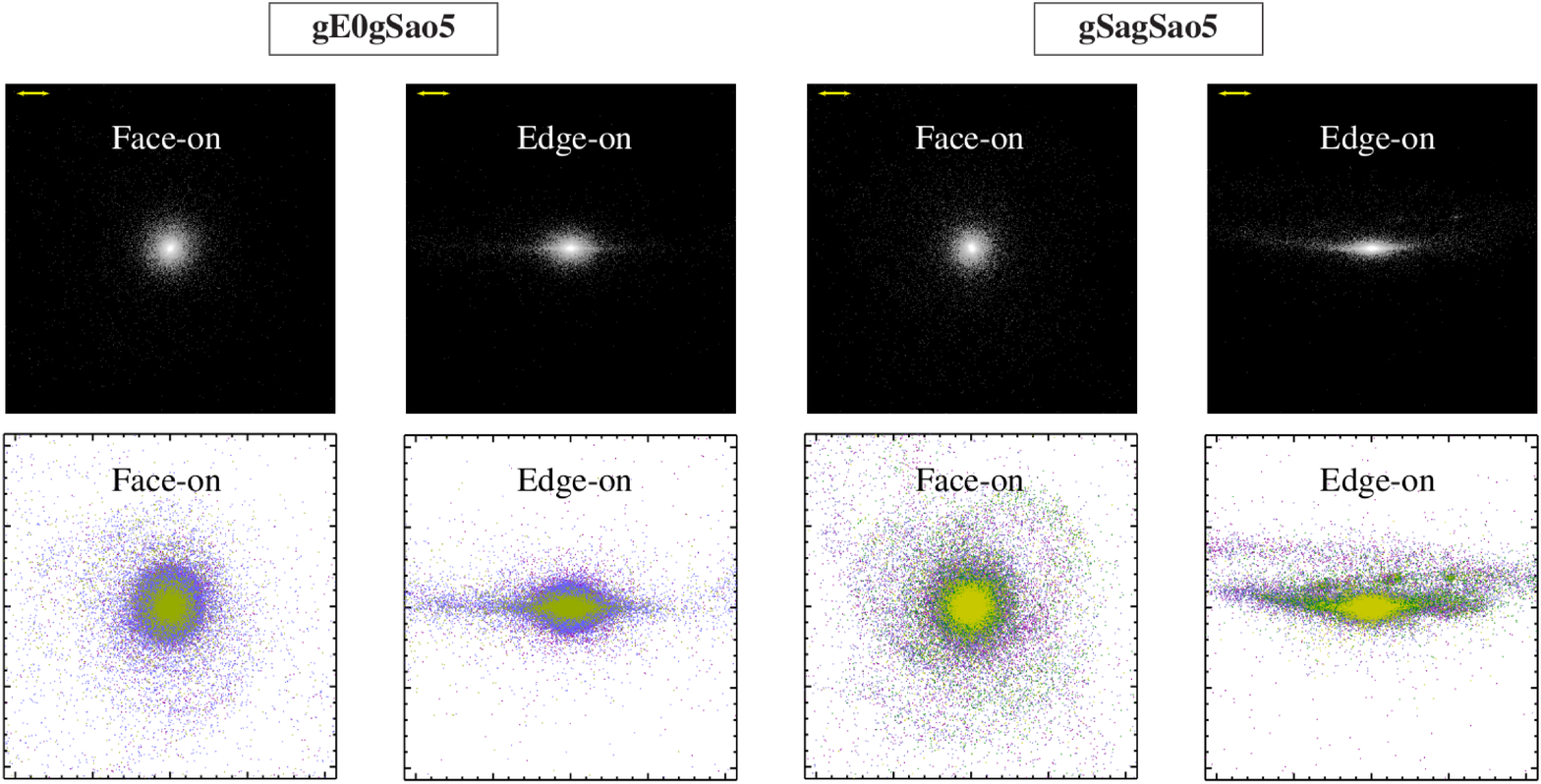}
\includegraphics[width=0.45\textwidth, bb = 0 12 290 430, clip]{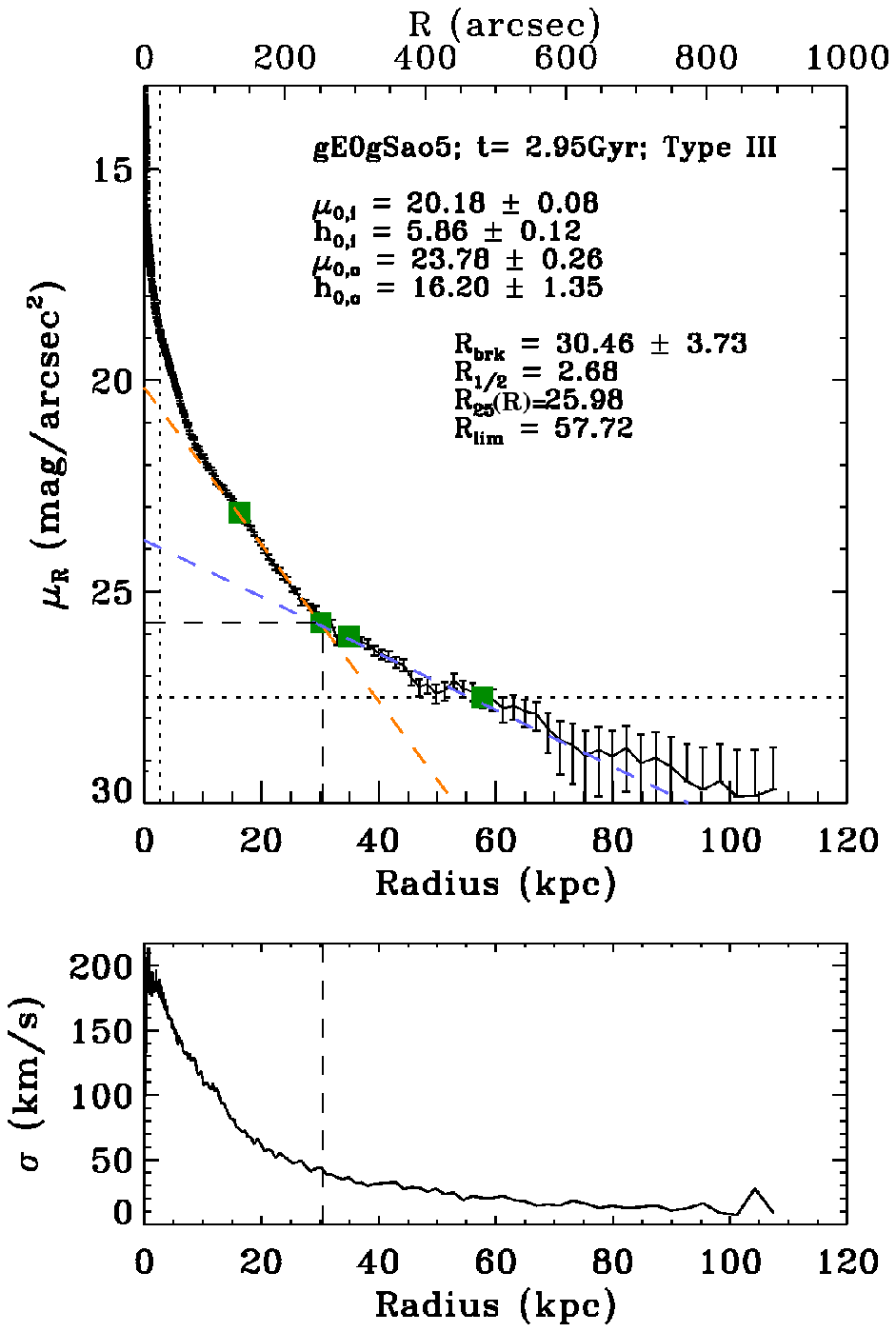}
\includegraphics[width=0.45\textwidth, bb = 0 12 290 430, clip]{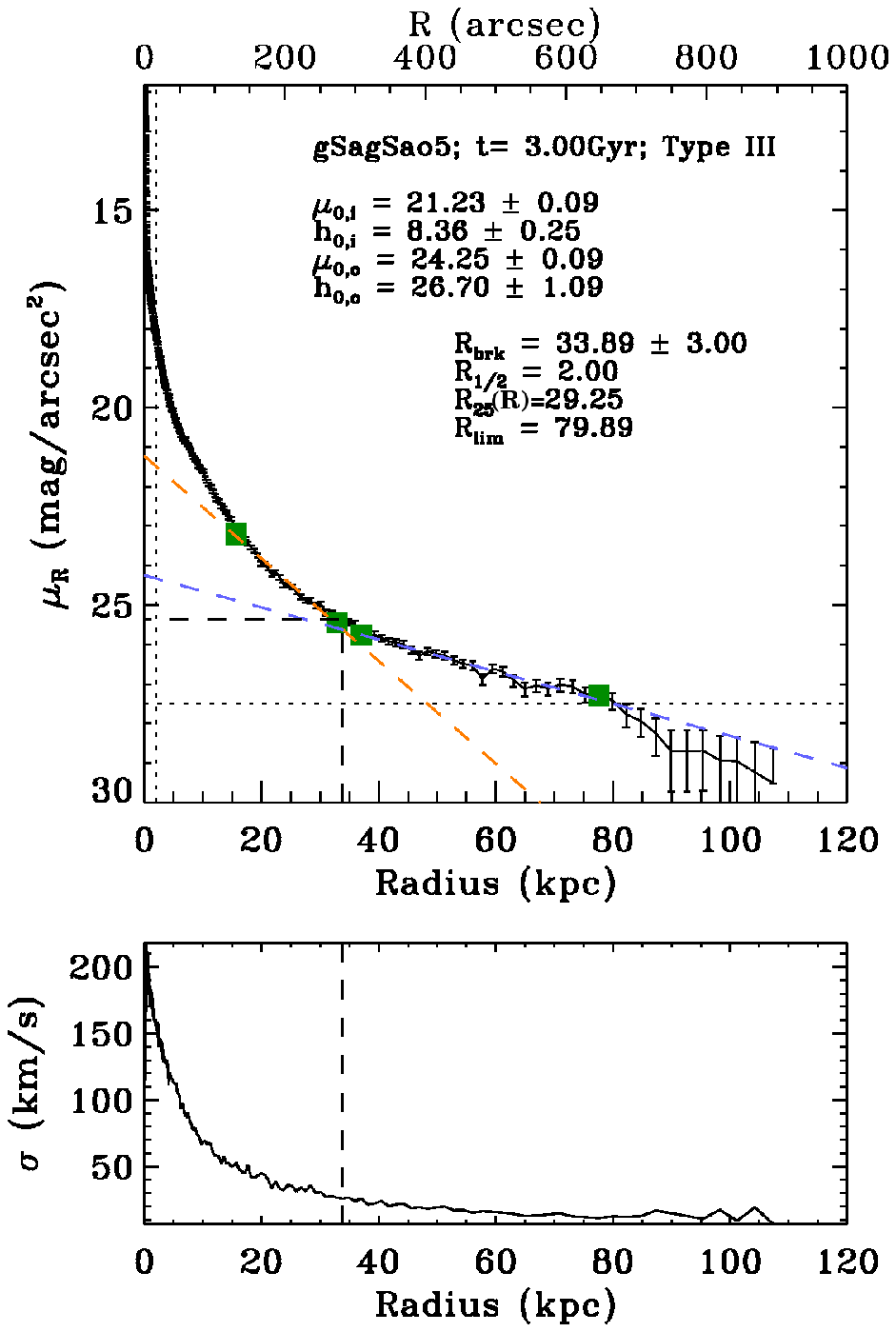}
\caption{Simulated $R$-band images and radial surface brightness profiles of some Type III S0-like remnants. The simulations assume $D = 30$\,Mpc, $\mu_{R,lim}= 27.5$\,mag\,arcsec$^{-2}$, and a spatial resolution of 0.7\arcsec. \textbf{{Top panels}}: Simulated $R$-band photometric images of the final remnants for face-on and edge-on views (see Sect.\,\ref{Sec:S0identification}). The yellow segment represents a physical length of 10\,kpc. The field of view is 100\,kpc${\times}$100\,kpc in all panels. We use logarithmic grey scale. \textbf{{Intermediate panels}}: Maps of projected locations of the stellar particles in the same remnant, for the same projections and fields of view as in the top panels. The colour code of the particles is the same as in Fig.\,\ref{fig:sim_frames}. \textbf{{Bottom panels}}: $R$-band surface brightness profile ($\mu_R(r)$) and radial profile of the dispersion of velocities in the remnant ($\sigma$). \emph{Black dots}: Simulated data. \emph{Dotted horizontal line}: Limiting surface brightness. \emph{Green squares}: Minimum and maximum radial limits considered for the piecewise fits. [\emph{The legend continues on the next page}].
}
\label{fig:ejemplos}

\end{figure*}

\begin{figure*}[!]
\includegraphics[width=\textwidth]{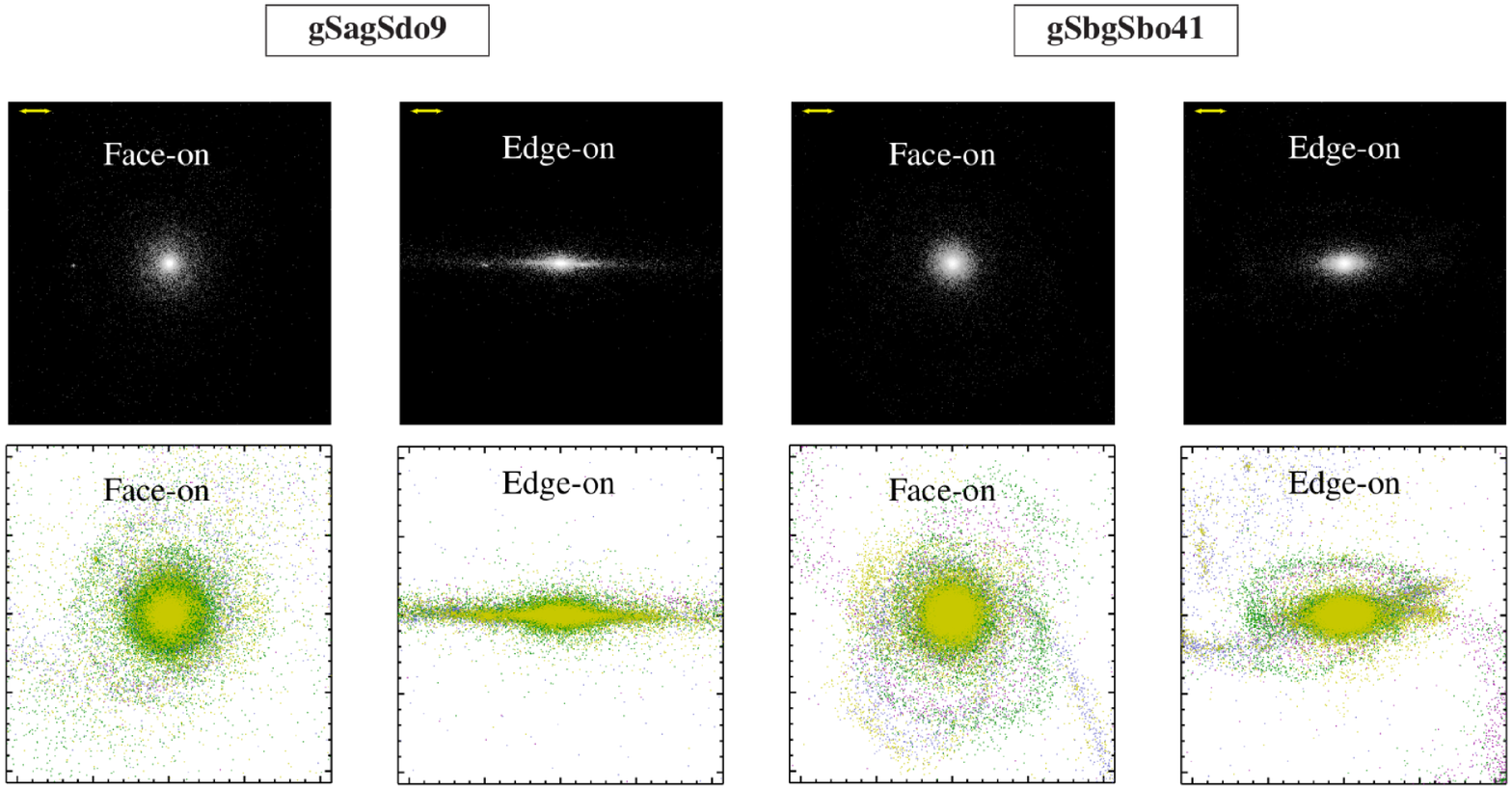}
\includegraphics[width=0.45\textwidth, bb = 0 12 290 430, clip]{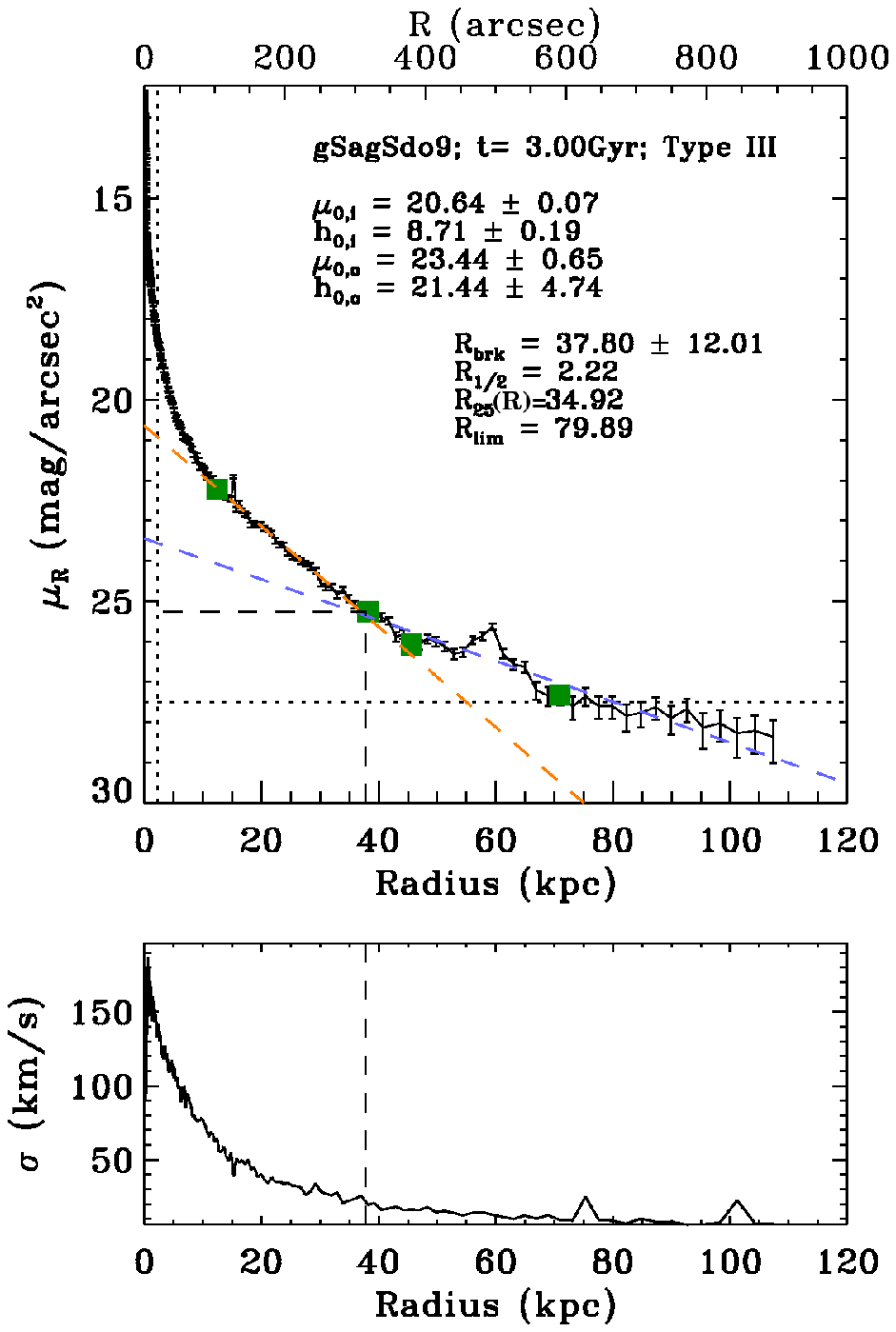}
\includegraphics[width=0.45\textwidth, bb = 0 12 290 430, clip]{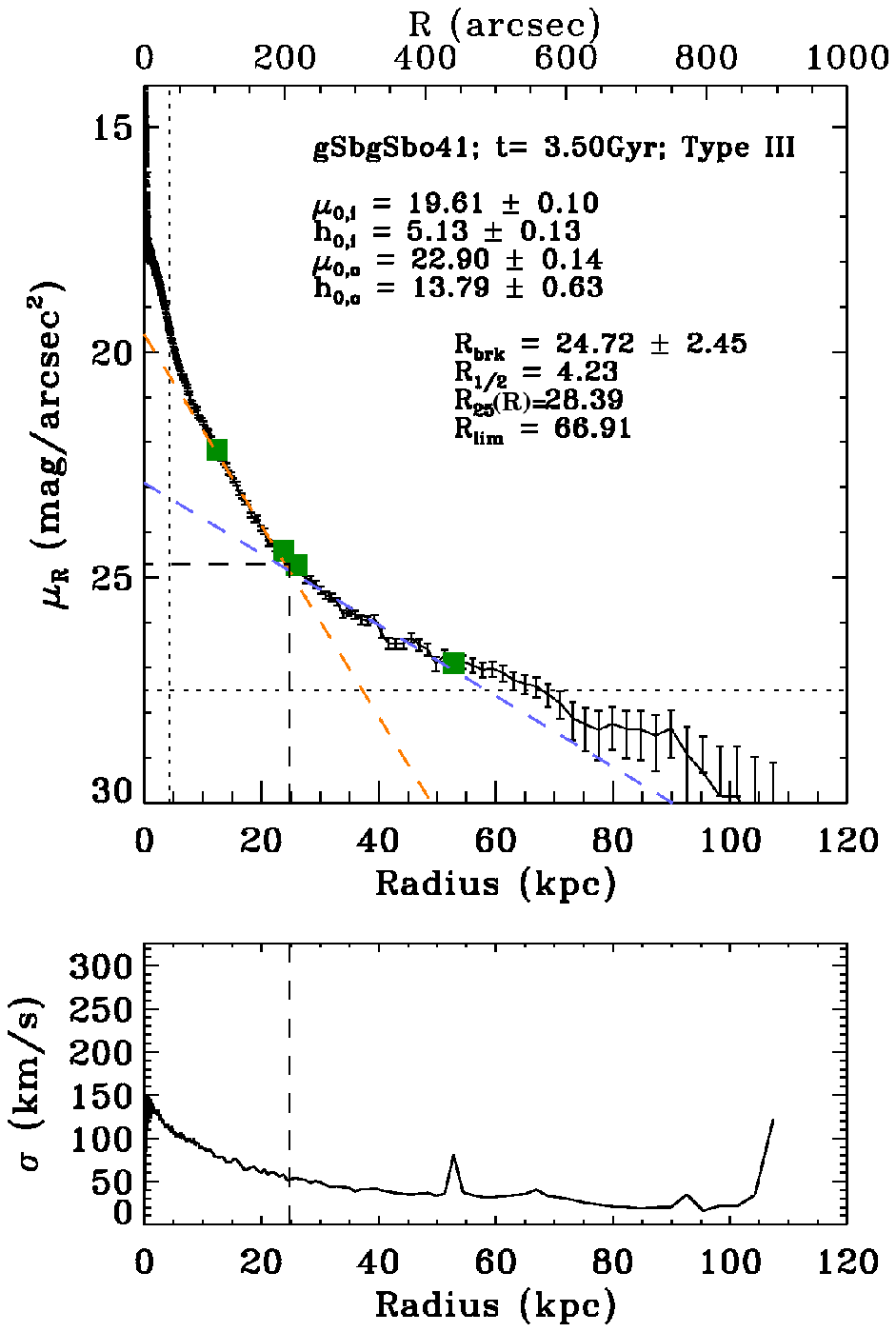}
\addtocounter{figure}{-1}
\vspace{0.2cm}
\caption{(\textbf{Continued figure}). [\emph{Continuation of the legend on the previous page}]. \emph{Red and blue dashed lines}: Linear fits performed to data within the selected radial limits defining the inner and outer discs, respectively. \emph{Black dashed vertical and horizontal lines}: Values of the radius and surface brightness of the antitruncation (\rbreak, \mubreak). The photometric parameters of the breaks and the inner and outer discs resulting from the piecewise fits are shown in each panel (the surface brightness values are provided in mag\,arcsec$^{-2}$ and the scalelengths are in kpc). $R_\mathrm{1/2}$ corresponds to the half-light (or effective) radius of the whole galaxy and $R_\mathrm{lim}$ is the radius at which the limiting magnitude of the profile is achieved. \risoph $(R)$ refers to the radius of the isophote with $\mu_{R}= 25$\,mag\,arcsec$^{-2}$. \risoph $(R)$ is not equivalent to \risoph\ used throughout the manuscript (defined in the $B$ band instead). }
\end{figure*}

\begin{figure*}[!]
\includegraphics[width=\textwidth]{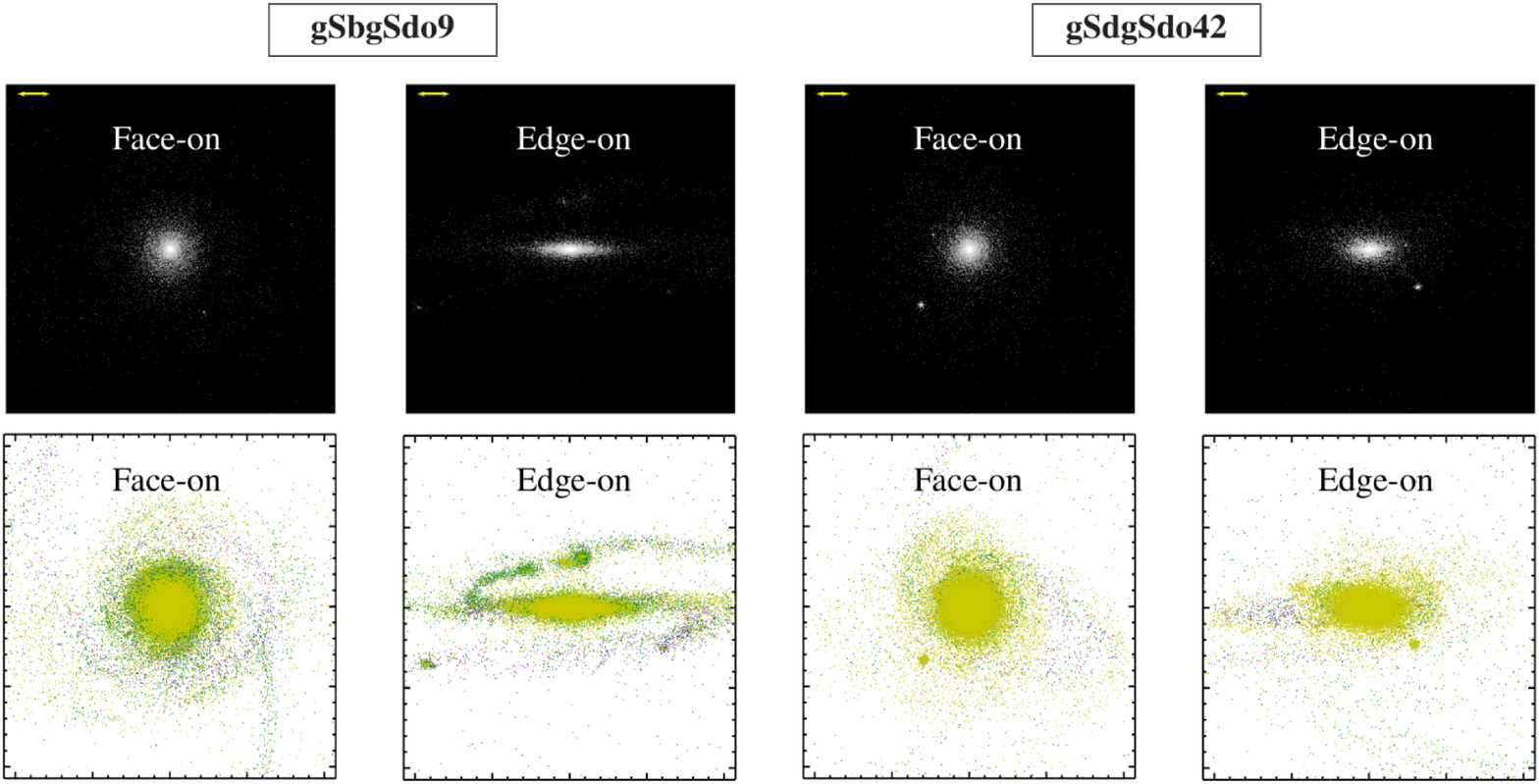}
\includegraphics[width=0.45\textwidth, bb = 0 12 290 430, clip]{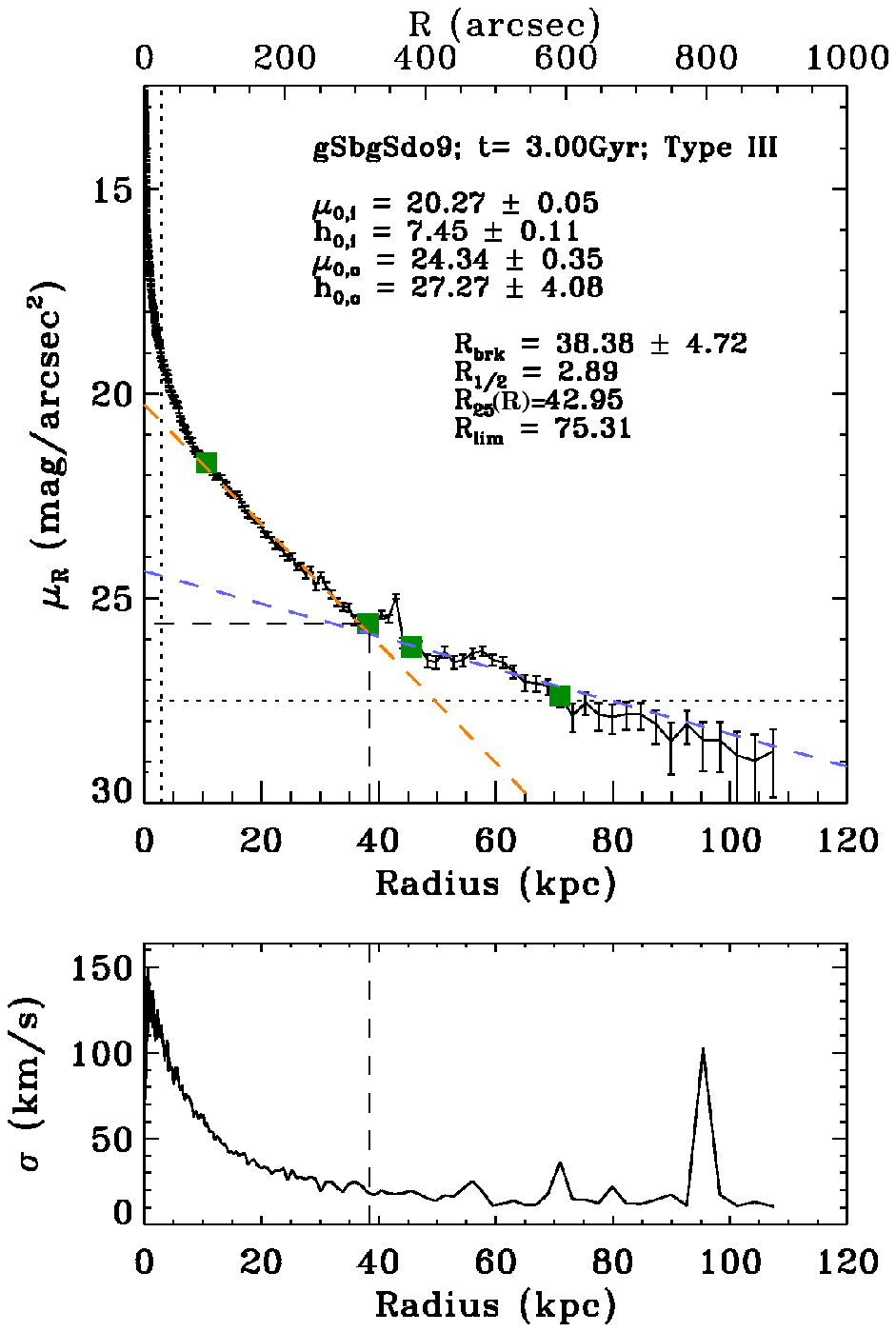}
\includegraphics[width=0.45\textwidth, bb = 0 12 290 430, clip]{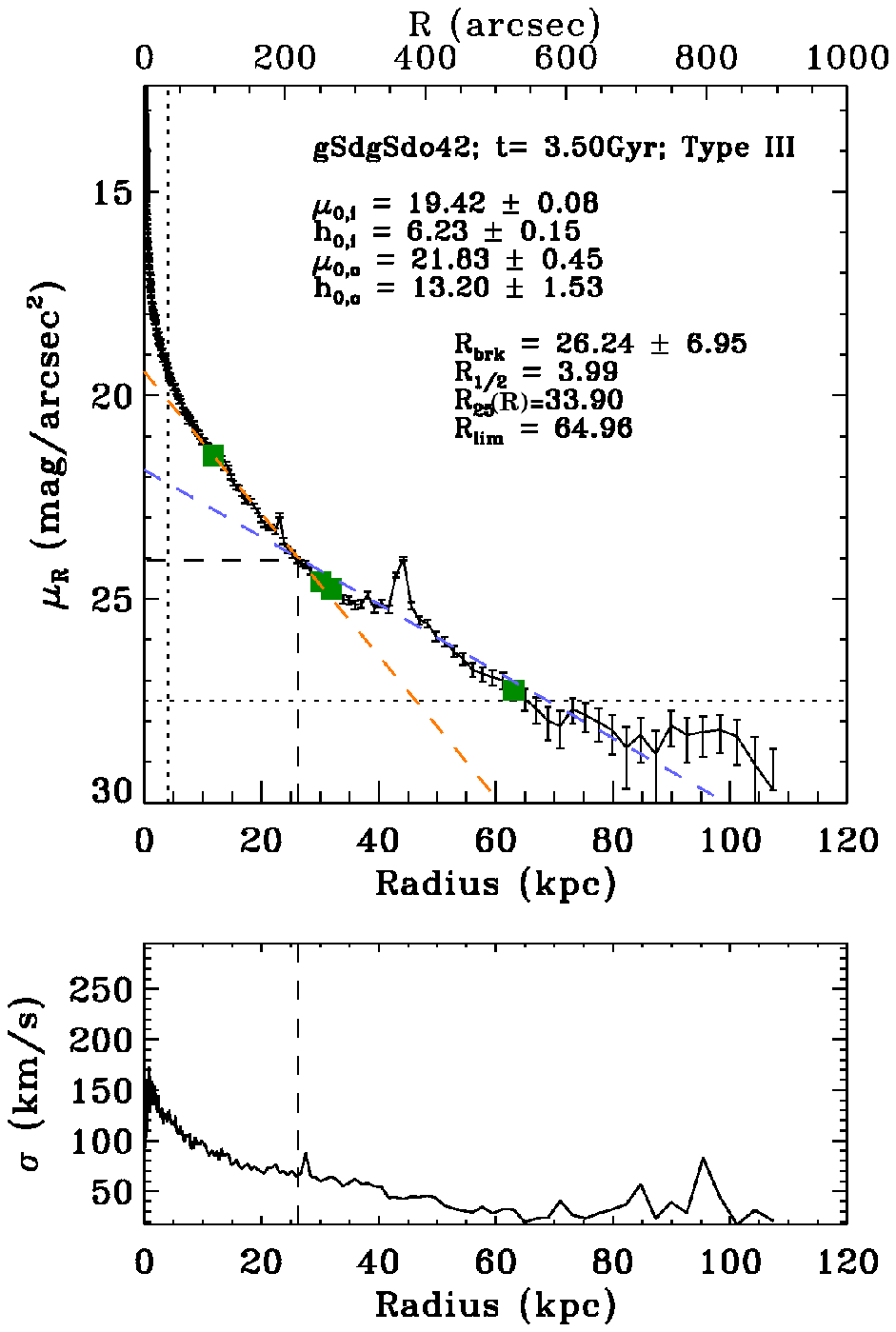}
\addtocounter{figure}{-1}
\caption{(\textbf{Continued figure}).  [\emph{A colour version of this figure is available in the online edition.}]}
\end{figure*}

\subsection{Identification and characterization of the antitruncated remnants}
\label{Sec:fits}

We have visually classified the structure of the discs in our sample of 67 S0-like remnants according to the classification scheme by E08 (Type-I, Type-II, or Type III). We finally obtained a sample of 47 major merger remnants with S0-like morphology (i.e. E/S0 or S0) and stable Type III discs (i.e. antitruncated) from the initial sample of 67 relaxed S0-like remnants. These features are kinematically stable (Borlaff et al., in preparation). The major merger models that result in Type III S0-like remnants are listed in Table\,\ref{tab:antitrunc}. Some examples of surface brightness profiles are plotted in Fig.\,\ref{fig:ejemplos}. This means that $\sim 70$\% of the major merger simulations that result in a relaxed S0-like remnant have antitruncated stellar discs. We note that this percentage cannot be compared with the fraction of S0s with antitruncated stellar discs estimated by observers, because our simulations do not consider any cosmological context.

We have characterized the detected antitruncations quantitatively following the procedure by G11 \citep[see also][]{2012MNRAS.419..669M}. These authors fitted Freeman profiles to the inner and outer discs defining the break to derive their photometric parameters, so they perform ''piecewise fits''. The Freeman exponential profile corresponds to a  S\'{e}rsic profile with a characteristic index $n=1$ \citep[][]{1970ApJ...160..811F}, and turns into a linear relation in $r$ in magnitude units, 

\begin{equation} \label{eq:Freemanmu}
\mu(r)=\mu_\mathrm{0} + \frac{2.5}{\ln(10)}\cdot\frac{r}{h_\mathrm{0}},
\end{equation}

\noindent where the surface brightness ${\mu}(r)$ at a certain radius $r$ depends on two parameters: $\mu_0$ (the central surface brightness provided in mag\,arcsec$^{-2}$) and $h_\mathrm{0}$ (the e-folding scalelength for the disc expressed in arcsec). The piecewise fits provided the central surface brightness for the inner and outer discs in mag\,arcsec$^{-2}$ (\mui\ and \muo), as well as their scalelengths in arcsec (\hi\ and \ho, respectively). The break radius, \rbreak, was defined in G11 as the point where the fitted profiles cross, and the break surface brightness, \mubreak, as the surface brightness of the disc at $r=\rbreak$.

\citet{2008AJ....135...20E} fitted the whole profile of the discs using ''broken-exponential'' functions, instead of performing piecewise fits \citep[see also][]{2014arXiv1404.0559L}. They concluded  that the characteristic photometric parameters of the inner and outer discs obtained by both procedures agreed pretty well (the scalelengths differed by only 1--5\% on average). As the G11 sample contains many more Type III S0s than that of E08, we decided to reproduce the procedure followed by G11 (piecewise fits), even though we have compared our results with both observational samples. 

\citet{2011AJ....142..145G} also defined the minimum and maximum radii delimiting the inner and outer discs around each break visually, and used the data within these limits to perform the linear fits to each disc region. We have mimicked this procedure, visually selecting the radial upper and lower limits for the inner and outer discs of our antitruncated remnants, and then performing least-squares linear fits to the data corresponding to each part of the disc. We have selected the radial limits tracing disc regions before and after the break with a relatively constant slope in the $R$-band surface brightness profile of each remnant.

We have carefully overridden the bulge region to define the lower radius of the inner disc region, paying attention to the $R$-band photometric images of the remnant (mostly, in edge-on view), to the shape of the bulge profile in the centre of the galaxy, and to the radial profile of the velocity dispersion of the remnant ($\sigma (r)$, see Fig.\,\ref{fig:ejemplos}).  

As commented above, some major mergers generate many tidal satellites around the main remnant body which remain strongly bound at the end of the simulation (this is frequent in gas-rich encounters, see models gSagSdo9 and gSdgSdo42 in Fig.\,\ref{fig:ejemplos}). These peaks obviously bias the fits slightly towards brighter \mui\ or \muo\ values, but the change is negligible and is contained within the typical observational errors ($<0.2$\,mag\,arcsec$^{-2}$). Therefore, we have ignored the existence of these peaks in the fits. Moreover, we have checked that tidal satellites do not affect either to the break radius or the break surface brightness of the remnants significantly.

Disc galaxies usually present secondary components embedded within the global bulge$+$disc structure, such as bars, rings, and/or lenses \citep{2001A&A...367..405P}. In particular, these structures are specially common in S0s \citep{2010MNRAS.405.1089L, 2011MNRAS.418.1452L,2011MNRAS.414.3645S}. Many of our S0-like remnants host inner bars and discs, as well as lenses and systems of lenses (see Paper I). Sometimes, lenses in real galaxies are associated with a spherical component, and others, they are related with disc components, as part of a system of nested discs \citep{2005ApJ...626L..81E}. \citet{2014arXiv1404.0559L} have found a structural relation between Type III breaks in galaxies and the existence of lenses or rings in them. As we are interested in breaks that really trace the transition between nested discs, we have analysed if any of our breaks may be related with spheroidal components that result from the merger, in order to remove it from our sample.

We have first determined the Type III breaks in our S0-like remnants that correspond to lens -- disc transitions and to genuinely disc -- disc ones. While rings and bars can be easily identified in simulated broad-band images through visual inspection, the identification of lenses may be a subtle process. We have identified the lens components in our remnants as done in observations: lenses are regions external to the bulge with constant surface brightness, that stand out from the exponentially decaying profile of the external disc in optical images \citep{2005MNRAS.362.1319L,2009IAUS..254..173S}. This identification has been corroborated through photometric decompositions, ensuring that the surface brightness distribution in the lens component could be fitted with a S\'{e}rsic profile with $n \sim 0.2 - 0.5$ \citep{2009ApJ...692L..34L,2010MNRAS.405.1089L}. We have found 8 out of our 47 Type III S0-like remnants in which the inner discs correspond to a lens component. The antitruncations in the rest of remnants trace pure disc-to-disc transitions.

%%%%%%%%%%%%%%%%%%%%%%%%%%%%%%%%%%%%%%%%%%%%%%%%%%%%%%%%%%%%%%%%
%%%%%%%%%%%%%%%%%%%%%%%%%%%%%%%%%%%%%%%%%%%%%%%%%%%%%%%%%%%%%%%%
%%%%%%%%%%%%%%%%%%%%%%%%%%%%%%%%%%%%%%%%%%%%%%%%%%%%%%%%%%%%%%%%

\clearpage
%\newpage
\onecolumn
 \begin{landscape}

{\normalsize
\begin{center}

\begin{minipage}[t]{21cm}
\vspace{0.5cm}
\end{minipage}
  \begin{longtable}{cl l r@{\,$\pm$\,}l r@{\,$\pm$\,}l r@{\,$\pm$\,}l r@{\,$\pm$\,}l r@{\,$\pm$\,}l r@{\,$\pm$\,}l c r@{\,$\pm$\,}l c}
\caption{Characteristic photometric parameters of the antitruncations and the inner and outer discs in our S0-like remnants resulting from major mergers}
\label{tab:antitrunc}
\\ \hline
\\\vspace{-0.5cm}\\
\multirow{2}{*}{No.} & \multirow{2}{*}{Model} & \multicolumn{1}{c}{\multirow{2}{*}{Morph}} & \multicolumn{2}{c}{ \rbreak}  &  \multicolumn{2}{c}{\mubreak}  & \multicolumn{2}{c}{\hi}  & \multicolumn{2}{c}{\mui} & \multicolumn{2}{c}{\ho} & \multicolumn{2}{c}{\muo} & ${M}_{\mathrm{B}}$ & \multicolumn{2}{c}{${V}_{\mathrm{rot}}$} & $\risoph$\\
 &   &   &  \multicolumn{2}{c}{[kpc]} &  \multicolumn{2}{c}{[mag\,arcsec$^{-2}$]} &  \multicolumn{2}{c}{[kpc]} &  \multicolumn{2}{c}{[mag\,arcsec$^{-2}$]} &  \multicolumn{2}{c}{[kpc]} &  \multicolumn{2}{c}{[mag\,arcsec$^{-2}$]} & [mag] &  \multicolumn{2}{c}{[km\,s$^{-1}$]} & [kpc] \\ 
(1) & \multicolumn{1}{l}{(2)} & \multicolumn{1}{c}{(3)} & \multicolumn{2}{c}{(4)} & \multicolumn{2}{c}{(5)} & \multicolumn{2}{c}{(6)} & \multicolumn{2}{c}{(7)} & \multicolumn{2}{c}{(8)} & \multicolumn{2}{c}{(9)} & (10) & \multicolumn{2}{c}{(11)}  & (12)\vspace{0.1cm}\\\hline
\\\vspace{-0.5cm}\\

\endfirsthead
%Adding extra space at the start of the second caption
\\
\\
\\
\caption{Characteristic photometric parameters of the antitruncations and the inner and outer discs in our S0-like remnants resulting from major mergers\\\emph{(Continued)}}
\\ \hline
\\\vspace{-0.5cm}\\
\multirow{2}{*}{No.} & \multirow{2}{*}{Model} & \multicolumn{1}{c}{\multirow{2}{*}{Morph}} & \multicolumn{2}{c}{ \rbreak}  &  \multicolumn{2}{c}{\mubreak}  & \multicolumn{2}{c}{\hi}  & \multicolumn{2}{c}{\mui} & \multicolumn{2}{c}{\ho} & \multicolumn{2}{c}{\muo} & ${M}_{\mathrm{B}}$ & \multicolumn{2}{c}{${V}_{\mathrm{rot}}$} & $\risoph$\\
 &   &   &  \multicolumn{2}{c}{[kpc]} &  \multicolumn{2}{c}{[mag\,arcsec$^{-2}$]} &  \multicolumn{2}{c}{[kpc]} &  \multicolumn{2}{c}{[mag\,arcsec$^{-2}$]} &  \multicolumn{2}{c}{[kpc]} &  \multicolumn{2}{c}{[mag\,arcsec$^{-2}$]} & [mag] &  \multicolumn{2}{c}{[km\,s$^{-1}$]} & [kpc] \\ 
(1) & \multicolumn{1}{l}{(2)} & \multicolumn{1}{c}{(3)} & \multicolumn{2}{c}{(4)} & \multicolumn{2}{c}{(5)} & \multicolumn{2}{c}{(6)} & \multicolumn{2}{c}{(7)} & \multicolumn{2}{c}{(8)} & \multicolumn{2}{c}{(9)} & (10) & \multicolumn{2}{c}{(11)}  & (12)\vspace{0.1cm}\\\hline
\\\vspace{-0.5cm}\\
\endhead
1 & gE0gSao5 & S0 & 30.5 & 3.7 & 25.74 & 0.09 & 5.86 & 0.12 & 20.18 & 0.08 & 16.2 & 1.4 & 23.78 & 0.26 & -22.32 & 142 & 5 & 18.17\\
2 & gE0gSbo44 & S0 & 39 & 11 & 26.51 & 0.13 & 8.32 & 0.31 & 21.21 & 0.13 & 15.4 & 1.6 & 23.56 & 0.32 & -22.17 & 93.9 & 5.7 & 18.83\\
3 & gSagSao1 & S0 & 49.8 & 7.2 & 26.95 & 0.16 & 10.72 & 0.21 & 21.69 & 0.06 & 29.5 & 3.8 & 24.90 & 0.26 & -22.69 & 149.9 & 3.3 & 20.89\\
4 & gSagSao5 & S0 & 34.3 & 3.3 & 25.36 & 0.07 & 8.34 & 0.28 & 21.17 & 0.10 & 26.2 & 1.1 & 24.21 & 0.10 & -22.74 & 187.7 & 3.6 & 21.07\\
5 & gSagSao9 & S0 & 49.7 & 4.3 & 26.43 & 0.11 & 12.55 & 0.44 & 22.28 & 0.10 & 52.0 & 3.0 & 25.53 & 0.08 & -22.72 & 197.5 & 4.4 & 19.30\\
6 & gSagSbo2 & S0 & 22.0 & 1.9 & 24.16 & 0.06 & 5.32 & 0.14 & 19.90 & 0.09 & 13.94 & 0.37 & 22.68 & 0.08 & -22.67 & 143.5 & 3.0 & 21.64\\
7 & gSagSbo5 & S0 & 29.7 & 2.1 & 25.00 & 0.07 & 7.01 & 0.16 & 20.55 & 0.08 & 23.24 & 0.82 & 23.76 & 0.08 & -22.65 & 173.2 & 3.6 & 20.98\\
8 & gSagSbo9 & S0 & 45.9 & 5.5 & 26.76 & 0.14 & 8.79 & 0.33 & 21.18 & 0.13 & 36.8 & 4.8 & 25.50 & 0.21 & -22.65 & 183.3 & 3.2 & 24.63\\
9 & gSagSbo21 & S0 & 40.1 & 4.0 & 26.06 & 0.09 & 8.76 & 0.24 & 21.21 & 0.10 & 28.6 & 1.9 & 24.66 & 0.15 & -22.67 & 138.1 & 3.4 & 23.60\\
10 & gSagSbo22 & S0 & 46.6 & 5.7 & 26.46 & 0.11 & 10.36 & 0.24 & 21.70 & 0.07 & 30.8 & 3.1 & 24.94 & 0.21 & -22.67 & 125.1 & 5.1 & 21.92\\
11 & gSagSbo24 & E/S0 & 37.5 & 5.1 & 25.76 & 0.08 & 7.77 & 0.13 & 20.72 & 0.06 & 17.2 & 1.3 & 23.60 & 0.23 & -22.65 & 37 & 10 & 24.44\\
12 & gSagSbo42 & E/S0 & 37.3 & 3.8 & 25.66 & 0.08 & 8.67 & 0.22 & 21.04 & 0.08 & 20.62 & 0.83 & 23.75 & 0.11 & -22.63 & 149.3 & 4.1 & 23.04\\
13 & gSagSbo43 & E/S0 & 30.2 & 2.5 & 25.05 & 0.07 & 6.41 & 0.15 & 20.14 & 0.08 & 17.67 & 0.65 & 23.40 & 0.11 & -22.63 & 159.7 & 2.9 & 22.29\\
14 & gSagSbo71 & E/S0 & 42.9 & 7.1 & 25.77 & 0.08 & 15.16 & 0.65 & 22.83 & 0.09 & 31.6 & 1.3 & 24.42 & 0.09 & -22.61 & 181.1 & 3.7 & 19.30\\
15 & gSagSdo2 & S0 & 31.6 & 8.4 & 24.65 & 0.08 & 6.60 & 0.13 & 19.95 & 0.06 & 15.3 & 2.4 & 22.92 & 0.58 & -23.07 & 149.7 & 2.8 & 26.79\\
16 & gSagSdo9 & S0 & 38 & 12 & 25.26 & 0.09 & 8.71 & 0.19 & 20.64 & 0.07 & 21.4 & 4.7 & 23.44 & 0.65 & -23.04 & 180.1 & 3.4 & 28.00\\
17 & gSagSdo18 & E/S0 & 24.8 & 2.9 & 23.81 & 0.07 & 6.30 & 0.21 & 19.77 & 0.10 & 16.35 & 0.66 & 22.39 & 0.13 & -23.03 & 28 & 10 & 31.00\\
18 & gSagSdo41 & E/S0 & 30.8 & 8.2 & 24.68 & 0.08 & 8.08 & 0.34 & 20.56 & 0.13 & 15.5 & 1.2 & 22.54 & 0.27 & -23.01 & 178.9 & 4.7 & 24.73\\
19 & gSagSdo43 & E/S0 & 25.7 & 4.5 & 24.01 & 0.07 & 6.68 & 0.20 & 19.82 & 0.09 & 18.0 & 1.6 & 22.45 & 0.29 & -23.02 & 149.5 & 2.3 & 30.06\\
20 & gSagSdo70 & E/S0 & 35.6 & 7.3 & 24.83 & 0.08 & 8.82 & 0.35 & 20.58 & 0.10 & 18.4 & 1.4 & 22.86 & 0.23 & -23.01 & 184.3 & 3.4 & 28.38\\
21 & gSagSdo73 & E/S0 & 39.1 & 5.6 & 25.42 & 0.11 & 8.12 & 0.12 & 20.50 & 0.05 & 21.9 & 2.3 & 23.78 & 0.33 & -23.02 & 35.2 & 7.8 & 30.06\\
22 & gSbgSbo17 & E/S0 & 19.0 & 1.0 & 23.58 & 0.05 & 4.15 & 0.06 & 18.95 & 0.05 & 11.61 & 0.26 & 22.13 & 0.08 & -22.54 & 40.7 & 5.6 & 22.01\\
23 & gSbgSbo19 & E/S0 & 21.0 & 1.3 & 23.98 & 0.06 & 4.59 & 0.09 & 19.23 & 0.06 & 11.96 & 0.25 & 22.29 & 0.07 & -22.53 & 14.8 & 2.4 & 20.98\\
24 & gSbgSbo22 & S0 & 32.2 & 2.4 & 25.69 & 0.09 & 6.67 & 0.13 & 20.49 & 0.07 & 20.5 & 1.1 & 24.03 & 0.12 & -22.56 & 14.4 & 5.7 & 21.64\\
25 & gSbgSbo41 & E/S0 & 24.7 & 2.5 & 24.7 & 0.07 & 5.13 & 0.13 & 19.61 & 0.10 & 13.79 & 0.63 & 22.90 & 0.14 & -22.54 & 139.3 & 5.8 & 21.54\\
26 & gSbgSbo42 & E/S0 & 30.2 & 2.7 & 25.24 & 0.07 & 6.28 & 0.12 & 20.25 & 0.07 & 17.74 & 0.95 & 23.62 & 0.15 & -22.54 & 129.8 & 4.5 & 22.57\\
27 & gSbgSbo69 & E/S0 & 38.4 & 4.3 & 25.82 & 0.09 & 8.26 & 0.15 & 20.96 & 0.06 & 24.8 & 2.5 & 24.33 & 0.20 & -22.52 & 145.9 & 5.9 & 26.04\\
28 & gSbgSbo70 & E/S0 & 38.4 & 5.7 & 26.39 & 0.12 & 5.22 & 0.37 & 18.91 & 0.41 & 38.4 & 7.3 & 25.80 & 0.26 & -22.50 & 139.7 & 7.2 & 25.01\\
29 & gSbgSdo9 & S0 & 38.4 & 4.7 & 25.62 & 0.11 & 7.45 & 0.11 & 20.27 & 0.05 & 27.3 & 4.1 & 24.34 & 0.35 & -22.97 & 165.8 & 3.7 & 43.07\\
30 & gSbgSdo14 & S0 & 32 & 10 & 24.62 & 0.08 & 8.03 & 0.21 & 20.44 & 0.08 & 16.7 & 2.6 & 22.65 & 0.51 & -22.97 & 85.0 & 4.0 & 29.22\\

\hline\\
\pagebreak
31 & gSbgSdo17 & S0 & 22.0 & 3.2 & 23.75 & 0.08 & 5.15 & 0.29 & 19.26 & 0.19 & 14.37 & 0.49 & 22.24 & 0.11 & -22.96 & 39.57 & 0.40 & 27.07\\
32 & gSbgSdo18 & S0 & 17.1 & 2.0 & 22.95 & 0.07 & 4.22 & 0.12 & 18.86 & 0.09 & 11.70 & 0.56 & 21.67 & 0.17 & -22.96 & 53.1 & 3.2 & 25.66\\
33 & gSbgSdo19 & S0 & 25.5 & 2.2 & 23.99 & 0.08 & 5.62 & 0.10 & 19.38 & 0.05 & 12.47 & 0.39 & 22.08 & 0.12 & -22.95 & 29 & 10 & 30.34\\
34 & gSbgSdo41 & S0 & 28.2 & 3.0 & 24.52 & 0.09 & 6.60 & 0.12 & 19.98 & 0.06 & 14.66 & 0.60 & 22.53 & 0.15 & -22.96 & 125.5 & 2.4 & 28.47\\
35 & gSbgSdo69 & S0 & 30.8 & 3.5 & 24.66 & 0.08 & 6.83 & 0.13 & 19.98 & 0.06 & 15.06 & 0.74 & 22.65 & 0.17 & -22.96 & 138.0 & 3.1 & 29.31\\
36 & gSbgSdo70 & S0 & 32.8 & 5.0 & 24.78 & 0.08 & 7.72 & 0.25 & 20.35 & 0.09 & 17.37 & 0.97 & 22.92 & 0.19 & -22.95 & 129.5 & 3.9 & 28.75\\
37 & gSdgSdo2 & S0 & 43.0 & 4.1 & 25.88 & 0.13 & 7.81 & 0.16 & 20.19 & 0.07 & 25.1 & 2.1 & 24.31 & 0.21 & -23.33 & 26.9 & 8.1 & 31.00\\
38 & gSdgSdo5 & S0 & 24.2 & 3.0 & 23.55 & 0.07 & 6.86 & 0.21 & 19.82 & 0.08 & 14.09 & 0.37 & 21.78 & 0.09 & -23.29 & 148.8 & 2.2 & 34.09\\
39 & gSdgSdo9 & S0 & 32.5 & 2.8 & 24.38 & 0.08 & 6.90 & 0.18 & 19.79 & 0.08 & 23.3 & 1.3 & 23.39 & 0.13 & -23.30 & 138.3 & 3.8 & 28.28\\
40 & gSdgSdo16 & E/S0 & 26.4 & 4.7 & 23.30 & 0.06 & 5.94 & 0.13 & 19.12 & 0.06 & 11.17 & 0.62 & 21.38 & 0.26 & -23.31 & 27 & 10 & 34.09\\
41 & gSdgSdo17 & S0 & 19.4 & 1.7 & 22.86 & 0.06 & 4.75 & 0.12 & 18.67 & 0.08 & 12.27 & 0.34 & 21.39 & 0.10 & -23.28 & 28.5 & 9.8 & 32.21\\
42 & gSdgSdo21 & E/S0 & 39.9 & 12 & 25.16 & 0.09 & 9.06 & 0.30 & 20.46 & 0.10 & 20.5 & 3.5 & 23.13 & 0.51 & -23.25 & 48.5 & 7.5 & 32.31\\
43 & gSdgSdo42 & S0 & 26.2 & 7.0 & 24.04 & 0.08 & 6.23 & 0.15 & 19.42 & 0.08 & 13.2 & 1.5 & 21.83 & 0.45 & -23.27 & 144.1 & 2.1 & 29.31\\
44 & gSdgSdo45 & E/S0 & 28.3 & 2.0 & 24.21 & 0.08 & 6.17 & 0.13 & 19.51 & 0.07 & 17.56 & 0.55 & 22.74 & 0.10 & -23.28 & 28.8 & 5.3 & 33.90\\
45 & gSdgSdo69 & S0 & 28.5 & 5.5 & 23.73 & 0.05 & 7.28 & 0.22 & 19.87 & 0.10 & 17.4 & 1.4 & 22.34 & 0.29 & -23.26 & 137.7 & 2.1 & 30.90\\
46 & gSdgSdo71 & S0 & 29.5 & 5.9 & 24.21 & 0.08 & 7.66 & 0.18 & 19.96 & 0.06 & 12.29 & 0.52 & 21.54 & 0.17 & -23.27 & 98.2 & 4.8 & 35.58\\
47 & gSdgSdo74 & S0 & 39.4 & 4.8 & 25.39 & 0.11 & 7.79 & 0.08 & 19.98 & 0.03 & 20.3 & 2.0 & 23.37 & 0.28 & -23.27 & 41.7 & 5.2 & 32.68\\[0.05cm]\hline\\
 \end{longtable} 
\end{center}
}
%\footnotetext[ ]
\centering
\vspace{-0.5cm}
\begin{minipage}[t]{21cm}
\emph{Columns}: (1) Number ID. (2) Model code: g[\emph{type1}]g[\emph{type2}]o[\emph{No. orbit}], see Sect\,\ref{Sec:galmer}. (3) Visual morphological type derived from realistic broad-band simulated images (see details in Paper I). (4) Antitruncation radius in the $R$ band, \rbreak. (5) Surface brightness at the break radius in the $R$ band, \mubreak. (6) Scalelength of the inner disc in the $R$ band, \hi. (7) Central R-band surface brightness of the inner disc, \mui. (8) Scalelength of the outer disc in the $R$ band, \ho. (9) Central R-band surface brightness of the outer disc, \muo. (10) Total absolute magnitude in the Johnson-Cousin $B$ band (Vega system). (11) Maximum rotational velocity of the disc, $V_\mathrm{rot}$ (Tapia et al., in preparation, Paper VI of this series). (12) Optical radius of the remnant, corresponding to the radius of the ${\mu}_B= 25$\,mag\,arcsec$^{-2}$ isophote.
\end{minipage}

\end{landscape}

\clearpage
\twocolumn

%%%%%%%%%%%%%%%%%%%%%%%%%%%%%%%%%%%%%%%%%%%%%%%%%%%%%%%%%%%%%%%%%
%%%%%%%%%%%%%%%%%%%%%%%%%%%%%%%%%%%%%%%%%%%%%%%%%%%%%%%%%%%%%%%%%
%%%%%%%%%%%%%%%%%%%%%%%%%%%%%%%%%%%%%%%%%%%%%%%%%%%%%%%%%%%%%%%%%

We have then analysed the 3D structure of these lenses in face-on and edge-on simulated images of the remnants in optical broad bands (see Fig.\,\ref{fig:ejemplos}). We find that these lenses are so flat that we can deal with them as inner discs. Therefore, all the antitruncations detected in our S0-like remnants really trace inner-to-outer disc transitions (i.e. they are related with flat components). 

The higher luminosities of the S0-like remnants due to the recent starburts and the assumptions adopted to estimate the $M/L$ ratio imply that the remnants present surface brightness values brighter than nearby quiescent S0s by $\sim 1$\,mag\,arcsec$^{-2}$ in $R$ (see Sect.\,\ref{Sec:rbreak} and comments in Sect.\,\ref{Sec:profiles}). We have checked whether we would detect the breaks if the profiles were weaker by $\sim 1$\,mag\,arcsec$^{-2}$, finding that the breaks of only two models would be lost. This means that the youth of the recent stellar populations in our models does not affect significantly the detection of the breaks.

Some examples of fits performed to the inner and outer discs of each Type III S0-like remnant are plotted in Fig.\,\ref{fig:ejemplos}. We have listed the resulting values of \muo, \mui, \ho, \hi, \rbreak, and \mubreak\ for our 47 antitruncated S0-like remnants in Table\,\ref{tab:antitrunc}.

\section{Results}
\label{Sec:results}

In the next sections, we compare the photometric properties of the breaks and the inner and outer discs found in our remnants  with those observed in real Type III S0 galaxies. In Sect.\,\ref{Sec:generalresults}, we provide a global description of the properties of the antitruncations detected in our S0-like remnants of major mergers. Section\,\ref{Sec:photplanes} is devoted to the analysis of the trends followed by real and simulated Type III S0s in several photometric planes involving the parameters of the breaks (Sect.\,\ref{Sec:rbreak}) and of their inner and outer discs (Sect.\,\ref{Sec:innerouterdiscs}). We study these trends in scale-free analogs of these planes in Sect.\,\ref{Sec:scalefreeplanes}, showing that, while real and simulated data do not follow any clear correlation in some diagrams (Sect.\,\ref{Sec:scalefree}), others exhibit strong scaling relations in both real data and simulations (Sect.\,\ref{Sec:scaling}). The influence of bars in these trends is analysed in Sect.\,\ref{Sec:bars}. Section\,\ref{Sec:KS} shows the results of the Kolmogorov-Smirnov test in two dimensions to the distributions of real and simulated data in the planes where they overlap. Finally, we demonstrate that the trends in all the photometric planes can be derived just on the basis of three simple scaling relations fulfilled by both real and simulated Type III S0s (Sect.\,\ref{Sec:basic_scaling}). 

As commented above, the samples by E08 and G11 are used as observational references. As we have simulated their observational conditions to derive the surface brightness profiles of our remnants, we can ensure a fair comparison of models with data. 

\subsection{Global description of the antitruncations resulting in the simulations}
\label{Sec:generalresults}

As commented in Sect.\,\ref{Sec:fits}, we have found that 47 out of 67 relaxed S0-like remnants ($\sim 70$\%) from major mergers develop clear antitruncated stellar discs. These antitruncations are visually detectable in realistic $R$-band surface brightness profiles (see Fig.\,\ref{fig:ejemplos}). Table\,\ref{tab:antitrunc} shows that the Type III S0-like remnants resulting from our major merger simulations have scalelengths ranging from $\sim 5$\,kpc up to $\sim 15$\,kpc for the inner discs and between $\sim 12$\,kpc and $\sim 52$\,kpc for the outer discs. Typical central surface brightness values of the inner and outer discs are $\sim 19$ -- 23 mag\,arcsec$^{-2}$ and $\sim 21$ -- 25 mag\,arcsec$^{-2}$, respectively. These values imply $\mubreak \sim$ 22 -- 26 mag\,arcsec$^{-2}$ and $\rbreak \sim 20$ -- 50\,kpc in our remnants.

This proves that, once a major merger results in a relaxed E/S0 and S0 galaxy, it has a high likelihood to have an antitruncated stellar discs. Therefore, major mergers are a feasible mechanism to explain the formation of Type III S0 galaxies in less than 3.5\,Gyr.

\begin{figure}[!th]
\center
\includegraphics[angle=270, width = 0.48\textwidth, bb = 60 50 540 750, clip]{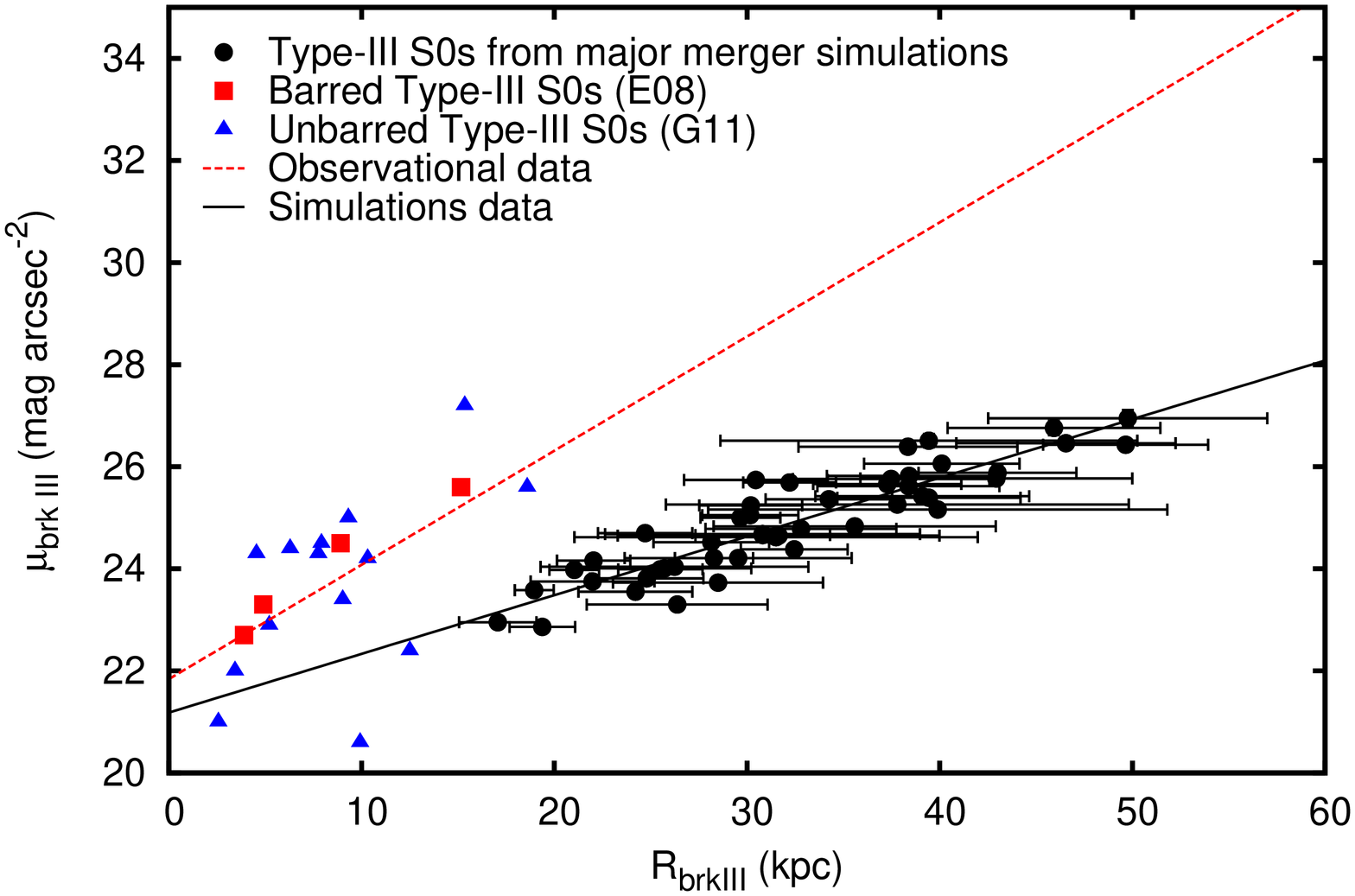}  
\includegraphics[angle=270, width = 0.48\textwidth, bb = 60 50 540 750, clip]{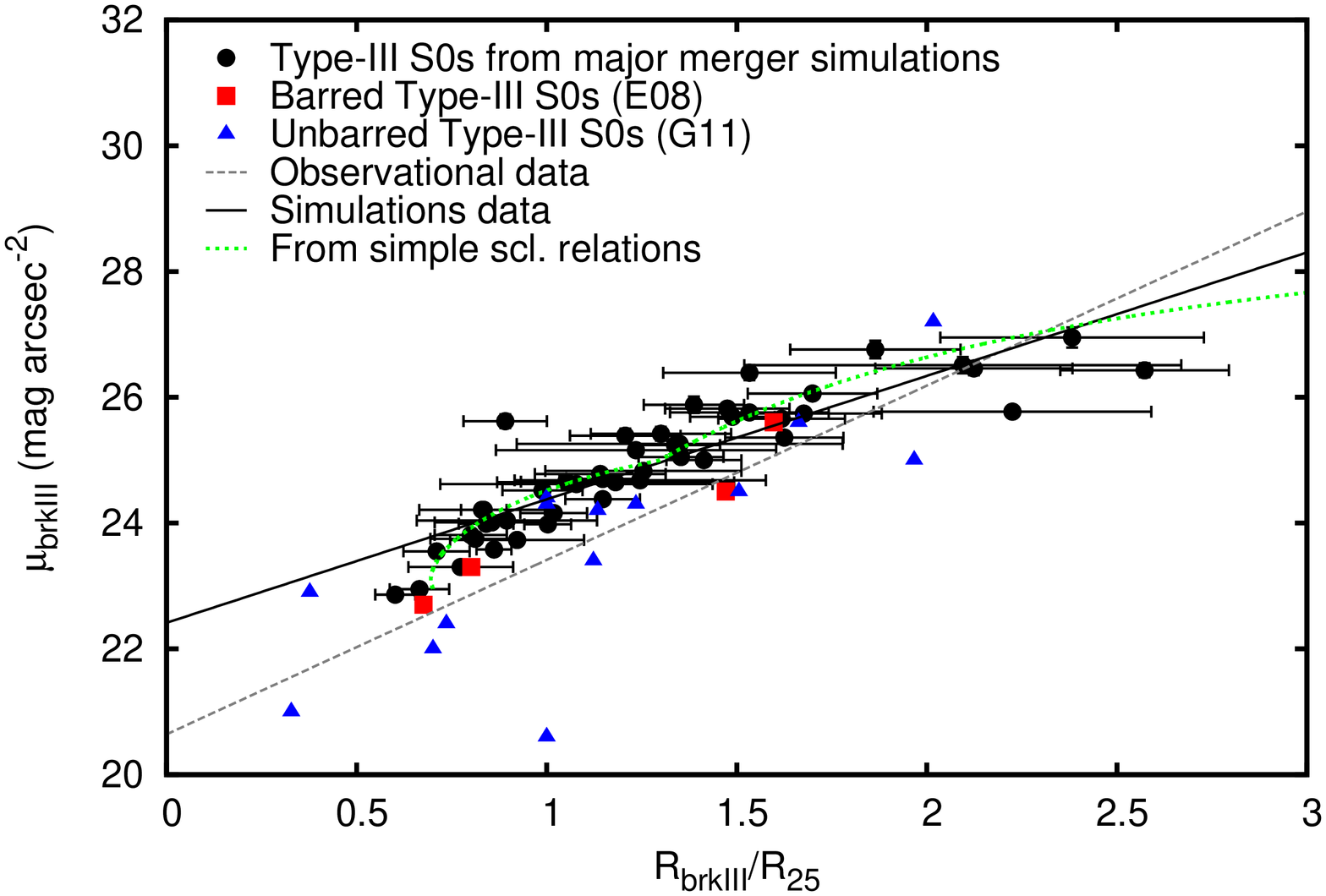}  
\includegraphics[angle=270, width = 0.48\textwidth, bb = 60 50 540 750, clip]{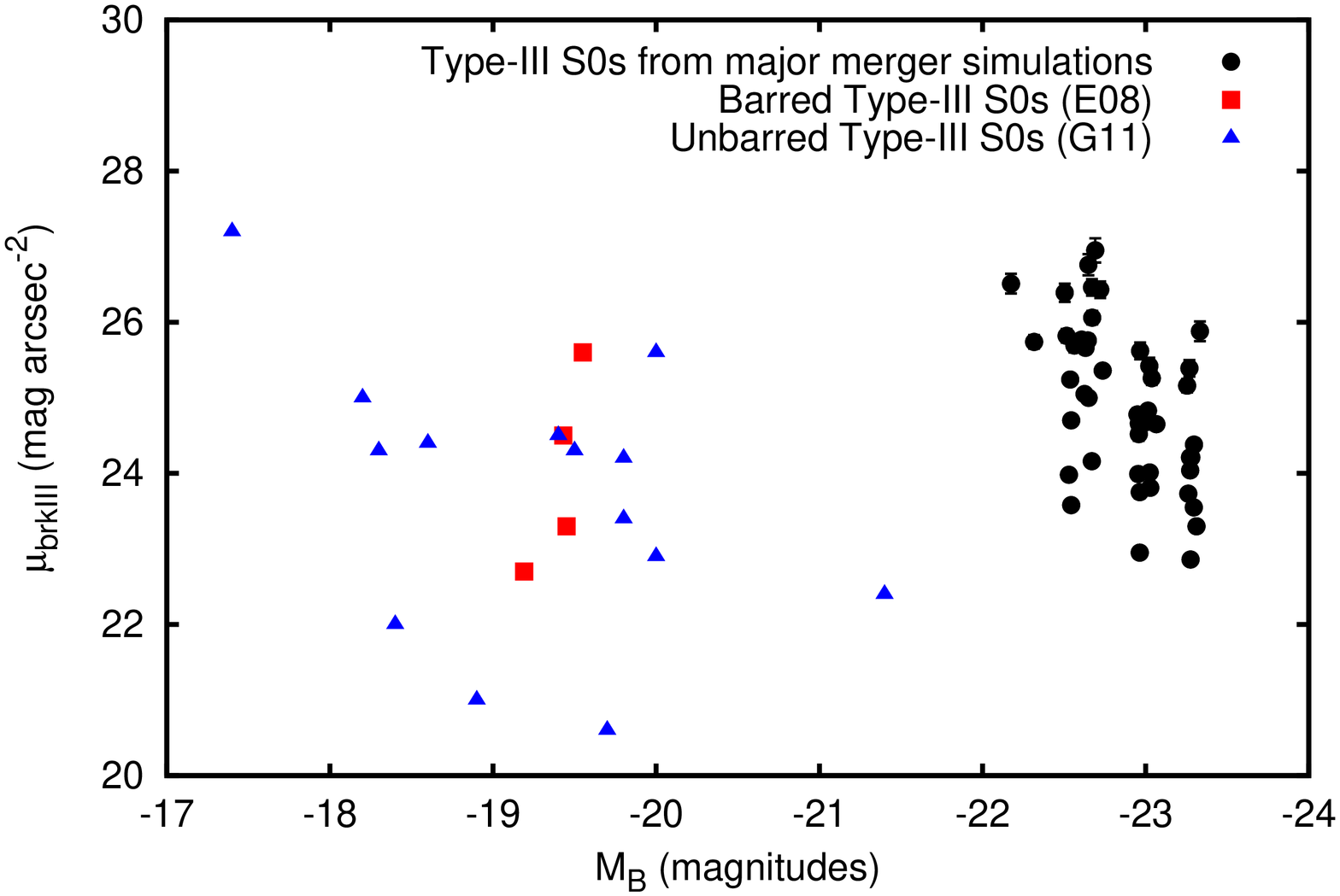}  
\caption{Distributions of our Type III S0-like remnants in the diagrams of $\mubreak$ vs. $\rbreak$,  $\rbreak/\risoph$, and $M_B$, compared to real antitruncated S0 galaxies (E08; G11). The lines correspond to linear fits performed to the major merger remnants (\emph{solid line}) and to the observational sample (\emph{dashed line}). The fit performed to observations in the plane $\mubreak$ -- $\rbreak$ is one of the three simple scaling relations from which the relations observed in the rest of photometric planes can be derived (see Sect.\,\ref{Sec:basic_scaling}), so it has been printed in red. \emph{Dotted line}: Expected relations assuming the simple scaling relations $\hi \propto \rbreak$, $\ho \propto \rbreak$, and $\mubreak \propto \rbreak$ obtained for real data. Consult the legend in the figure.  [\emph{A colour version of this figure is available in the online edition.}]} 
\label{fig:mubreak_Rbrk}
\end{figure}

\subsection{Photometric planes of antitruncations}
\label{Sec:photplanes}

\subsubsection{Planes involving the break parameters}
\label{Sec:rbreak}

In Fig.\,\ref{fig:mubreak_Rbrk}, we represent the distribution of our S0-like remnants in several photometric planes involving the \mubreak\ parameter, compared to observational data of Type III S0s (E08; G11). The top panel of the figure shows that our S0-like remnants exhibit break radii $\sim\,2$--3 times larger than those observed in nearby S0s, but similar values of the break surface brightness ($22 < \mubreak < 26$\,mag\,arcsec$^{-2}$). However, if \mubreak\ is plotted against the break radius normalized to the total optical extent of the galaxy, \risoph, the data and the simulations overlap, showing similar linear trends, slightly offset by $\sim 1$\,mag\,arcsec$^{-2}$ (see the intermediate panel of the figure).

We have performed least-squares linear fits to the observational and simulated samples in the \mubreak\ -- \rbreak\ and \mubreak\ -- $\rbreak / \risoph$ diagrams. They are overplotted in the two first panels of Fig.\,\ref{fig:mubreak_Rbrk} (dashed lines for the fits to the observational data, and solid lines for the fits to the major merger simulations). The results are listed in Table~\ref{tab:fits_results}. The Pearson coefficients indicate clear linear correlations in all cases ($\rho \gtrsim 0.8$), except for the observational \mubreak\ -- \rbreak\ relation, which is weaker. 

The fact that the major merger simulations do not lie on top of the extrapolation of the observational trend in the top panel of Fig.\,\ref{fig:mubreak_Rbrk} might imply that, 1) if our remnants had similar sizes and luminosities as observational data (and thus similar average surface brightnesses), they would have their breaks located at unrealistically high radial locations. However, 2) if they are larger than real galaxies in the reference observational sample, it could be that the breaks are located at higher radii just because of a question of scaling, as we will see. We note that, in this last case, the distribution of our simulations in the top panel of the figure implies that they are $\sim 1$\,mag\,arcsec$^{-2}$ brighter than they should for their \rbreak\ values, compared to real data. The former possibility would directly reject major mergers as feasible mechanisms to explain the formation of realistic antitruncations, whereas the latter would not, because the offset in $\mu$ might be due to the assumptions adopted to estimate the $M/L$ ratios in the models or to the blue colours of the remnants compared to quiescent galaxies with similar masses, and thus could be easily overriden (see Sect.\,\ref{Sec:profiles} and Appendix\,\ref{Sec:limitations}). 

\begin{figure}[th!]
\center
\includegraphics[angle=270, width = 0.49\textwidth, bb = 60 50 540 750, clip]{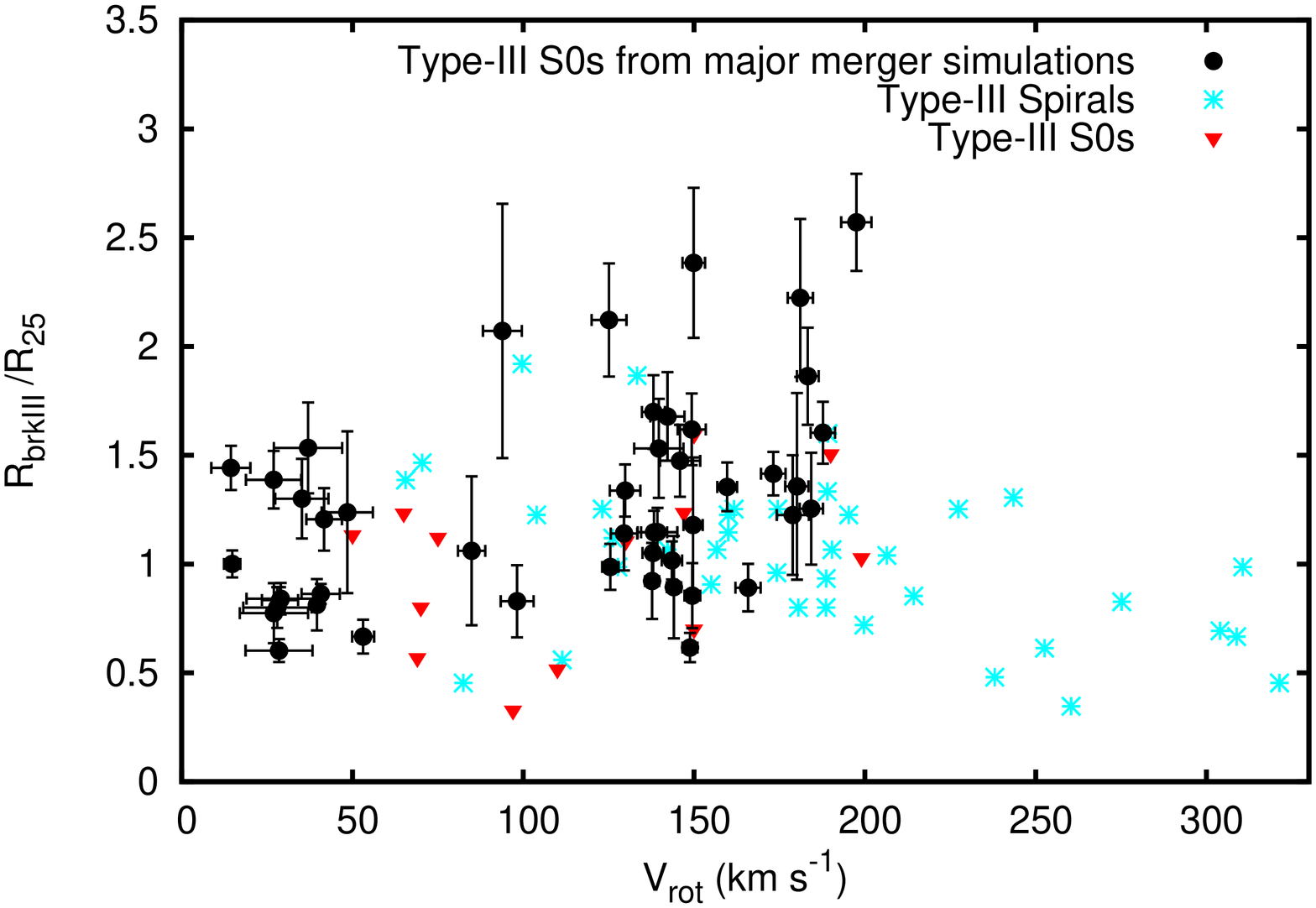}  
\caption{Distributions of our antitruncated S0-like remnants in the $\rbreak/\risoph$ -- ${V}_{\mathrm{rot}}$ diagram, compared to those of real Type III spirals and S0s from  \citet{2006A&A...454..759P}, E08, and G11.  [\emph{A colour version of this figure is available in the online edition.}]}
\label{fig:Rbrk_R25_vrot}
\end{figure}

The intermediate panel of Fig.\,\ref{fig:mubreak_Rbrk} directly discards the first possibility. It indicates that the relative radial location of the antitruncations with respect to the total galaxy size (as provided by $\risoph$) scales linearly with \mubreak\ in both real data and simulations. By normalizing \rbreak\ by the typical size of the galaxy, we remove the effects of the different scale of real S0s and of our remnants. Both data sets overlap and show similar linear trends in the \mubreak\ -- $\rbreak/\risoph$ plane. If our remnants presented breaks at unrealistic radial locations in the discs compared to their sizes, they would show a different slope from the observations in this diagram. 

This supports the idea that our simulations present an excess of brightness by $\sim 1$\,mag\,arcsec$^{-2}$ (according to their \rbreak\ values) with respect to observational data in the \mubreak\ -- \rbreak\ diagram (top panel of Fig.\,\ref{fig:mubreak_Rbrk}), but that the radial locations of the breaks are realistic accounting for the size of the S0-like remnants. If our simulations are displaced by $\sim 1$\,mag\,arcsec$^{-2}$ towards fainter magnitudes, the low-\rbreak\ end of the simulated S0s distribution overlaps with the high-\rbreak\ end of the observational distribution in the \mubreak\ -- \rbreak. This dimming would also affect the location of the simulations in the \mubreak\ -- $\rbreak/\risoph$ plane (intermediate panel of Fig.\,\ref{fig:mubreak_Rbrk}). Besides making the surface brightness values of the simulations fainter, it would probably decrease \risoph.Consequently, the simulations would move towards fainter \mubreak\ and higher $\rbreak/\risoph$ ratios in the diagram, still being consistent with observational data. 

The bottom panel of Fig.\,\ref{fig:mubreak_Rbrk} compares the distribution of observations and simulations in the \mubreak\ vs.\,$B$-band absolute magnitude ($M_B$) diagram. There is no clear trend in this plane either for the real data or for the major merger simulations. Our remnants are brighter than the brightest galaxies in the reference observational sample by $\sim 1$\,mag, as already commented. We remark that the offsets in total magnitude and in surface brightness between real data and simulations are basically a result of the assumptions adopted to estimate the $M/L$ values used in the models and to the presence of young stellar populations formed in the merger-induced starbursts induced (Sect.\,\ref{Sec:profiles} and Appendix\,\ref{Sec:limitations}).

In Fig.\,\ref{fig:Rbrk_R25_vrot} we compare the distribution of our antitruncated S0-like remnants in the plane $R_\mathrm{brk}/\risoph$ vs.\,rotational velocity ($V_\mathrm{rot}$) with the one exhibited by Type III galaxies in the samples by \citet{2006A&A...454..759P}, E08, and G11. We have distinguished between Type III S0s and spirals. The ${V}_{\mathrm{rot}}$ values for the real Type III S0s have been taken from the Leda database\footnote{HyperLeda database for physics of galaxies: http://leda.univ-lyon1.fr} \citep{2003A&A...412...45P}, and correspond to ${V}_{\mathrm{max}}$ values obtained from absorption kinematic data by several authors \citep{1978ApJ...222...84P,1989A&A...221..236B,1995ApJ...448L..13B,1996AJ....112..438J,1997A&AS..124...61B,1997AJ....113..950F, 1998A&AS..131..287S,2000A&AS..145..263S,2002A&A...384..371S}. The major merger models and the real galaxies disperse widely in the diagram, without showing any clear trend. Our S0-like remnants overlap with the distribution of real Type III S0s, which accumulate towards low-to-intermediate values of $V_\mathrm{rot}$. This means that major mergers are in agreement with the distribution of real data in this plane too.

In conclusion, the global properties of the breaks in the S0-like remnants coming from major mergers are consistent with those of real antitruncated S0 galaxies. This supports major mergers as a feasible mechanism for producing S0 galaxies with realistic antitruncated stellar discs. 

\subsubsection{Planes involving the parameters of the inner and outer discs}
\label{Sec:innerouterdiscs}

\begin{figure}[th!]
\center
\includegraphics[angle=270, width = 0.49\textwidth, bb = 60 50 540 750, clip]{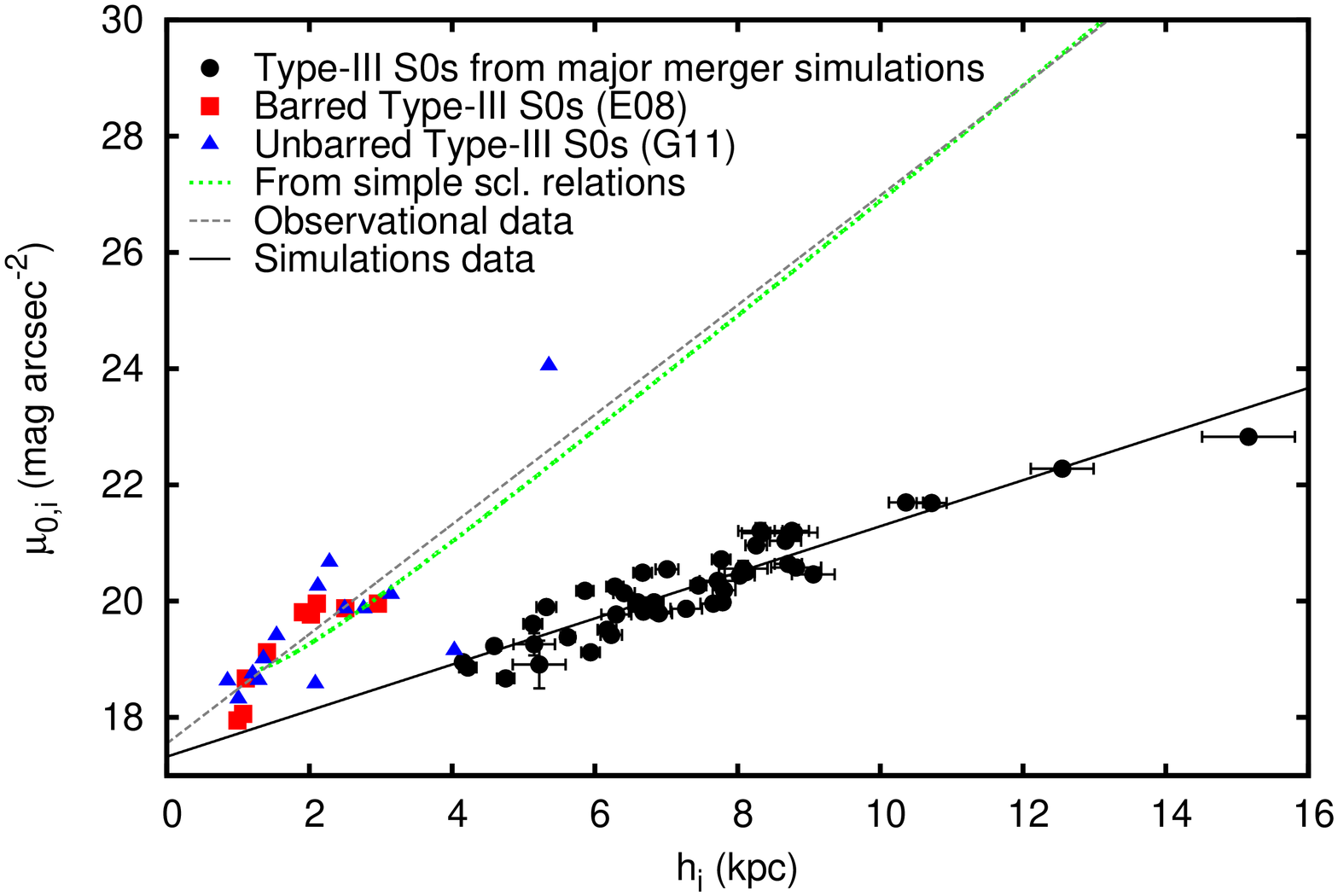}  
\includegraphics[angle=270, width = 0.49\textwidth, bb = 60 50 540 750, clip]{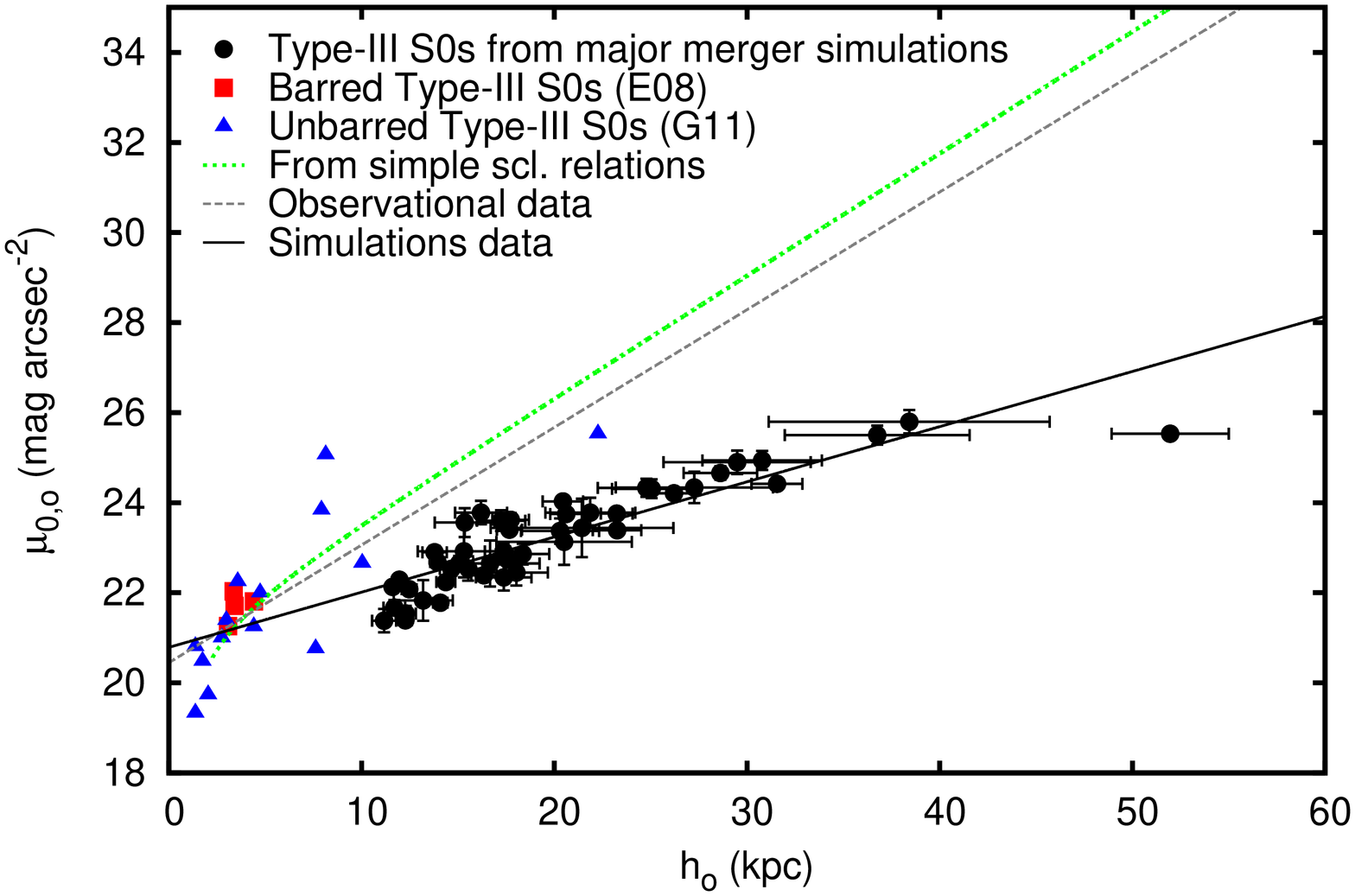}  
\caption{Distributions of our Type III S0-like remnants in the diagrams of $\mui$ vs. $\hi$ and $\muo$ vs. $\ho$ compared to the antitruncated S0s in the reference observational sample (E08; G11). \emph{Solid lines}: Linear fits to our simulations sample. \emph{Dashed lines}: Linear fits to the observational data by E08 and G11. \emph{Dotted line}: Expected relations derived from the simple observational scaling relations $\hi \propto \rbreak$, $\ho \propto \rbreak$, and $\mubreak \propto \rbreak$ (see Sect.\,\ref{Sec:basic_scaling}). Consult the legend in the figure.  [\emph{A colour version of this figure is available in the online edition.}]
}
\label{fig:trends_mui_hi_muo_ho}
\end{figure}

Figure \,\ref{fig:trends_mui_hi_muo_ho} shows the different trends of simulated and observational data in the $\mui$ -- \hi\ and $\muo$ -- \ho\ diagrams. Linear fits to the observational and the simulated sample are overplotted (see Table\,\ref{tab:fits_results}). We find that both samples follow linear scale relations in these two planes. Our remnants present higher \hi\ and \ho\ than the Type III S0s in the observational sample, again because of the different size of the remnants and the real galaxies. We will show that, once these scalelengths are normalized to the total galaxy size, our simulations also overlap with observations in the scale-free analogs of these photometric planes (see Sect.\,\ref{Sec:scalefreeplanes}). The simulations are displaced from the extrapolation of the observational trends by $\sim 1$\,mag\,arcsec$^{-2}$ in \mui\ and \muo, as also happened with \mubreak\ (see Fig.\,\ref{fig:mubreak_Rbrk}). However, as commented before, this offset is a consequence of the assumptions adopted for modelling the $M/L$ ratios and of the bluer colours of the remnants compared to real galaxies, because of their recent starbursts (see Sect.\,\ref{Sec:profiles} and Appendix\,\ref{Sec:limitations}).

\citet{2014arXiv1404.0559L} have recently reported the existence of these linear trends in the \mui\ -- \hi\ and \muo\ - \ho\ planes of real antitruncated disc galaxies with IR data, which we also find in the $R$-band data by E08 and G11. Our major merger simulations also follow linear trends in these photometric plane, in agreement with observations.

\begin{figure}[t]
\center
\includegraphics[angle=270, width = 0.49\textwidth, bb = 60 50 540 750, clip]{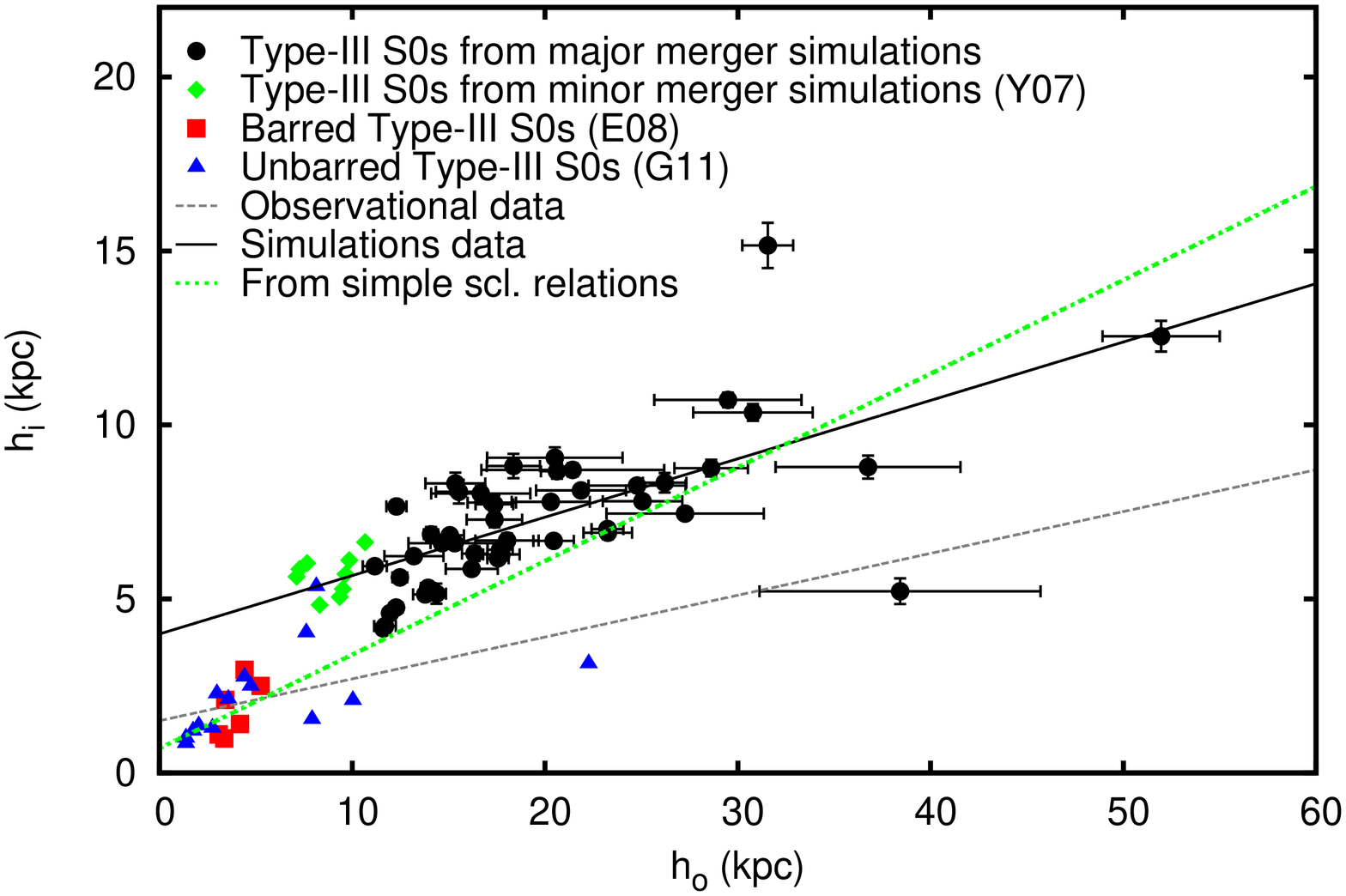}  
\caption{Relation between \hi\ and \ho\ in the antitruncated S0-like remnants and the real Type III S0s from the reference observational sample (E08; G11). Data of the antitruncations formed in the minor merger simulations by Y07 are overplotted for comparison. \emph{Solid line}: Linear fit to our simulations sample. \emph{Dashed line}: Linear fit to the observational data by E08 and G11. \emph{Dotted-dashed line}: Expected relation assuming the simple scaling relations $\hi \propto \rbreak$, $\ho \propto \rbreak$, and $\mubreak \propto \rbreak$ obtained from data (see Sect.\,\ref{Sec:basic_scaling}). Consult the legend in the figure.   [\emph{A colour version of this figure is available in the online edition.}]
}
\label{fig:hiho}
\end{figure}

In Fig.\,\ref{fig:hiho}, we analyse the relation between \hi\ and \ho\ in the real and simulated data. We find a strong linear correlation between the scalelengths of the inner and outer discs in real data and simulations. However, the linear fits to each distribution in the plane differ slightly in the slopes, probably as a result of the large dispersion of the data (see the resulting fits in Table\,\ref{tab:fits_results}). 

\begin{figure*}[ht!]
\center
\includegraphics[angle=270, width = 0.49\textwidth, bb = 60 50 540 750, clip]{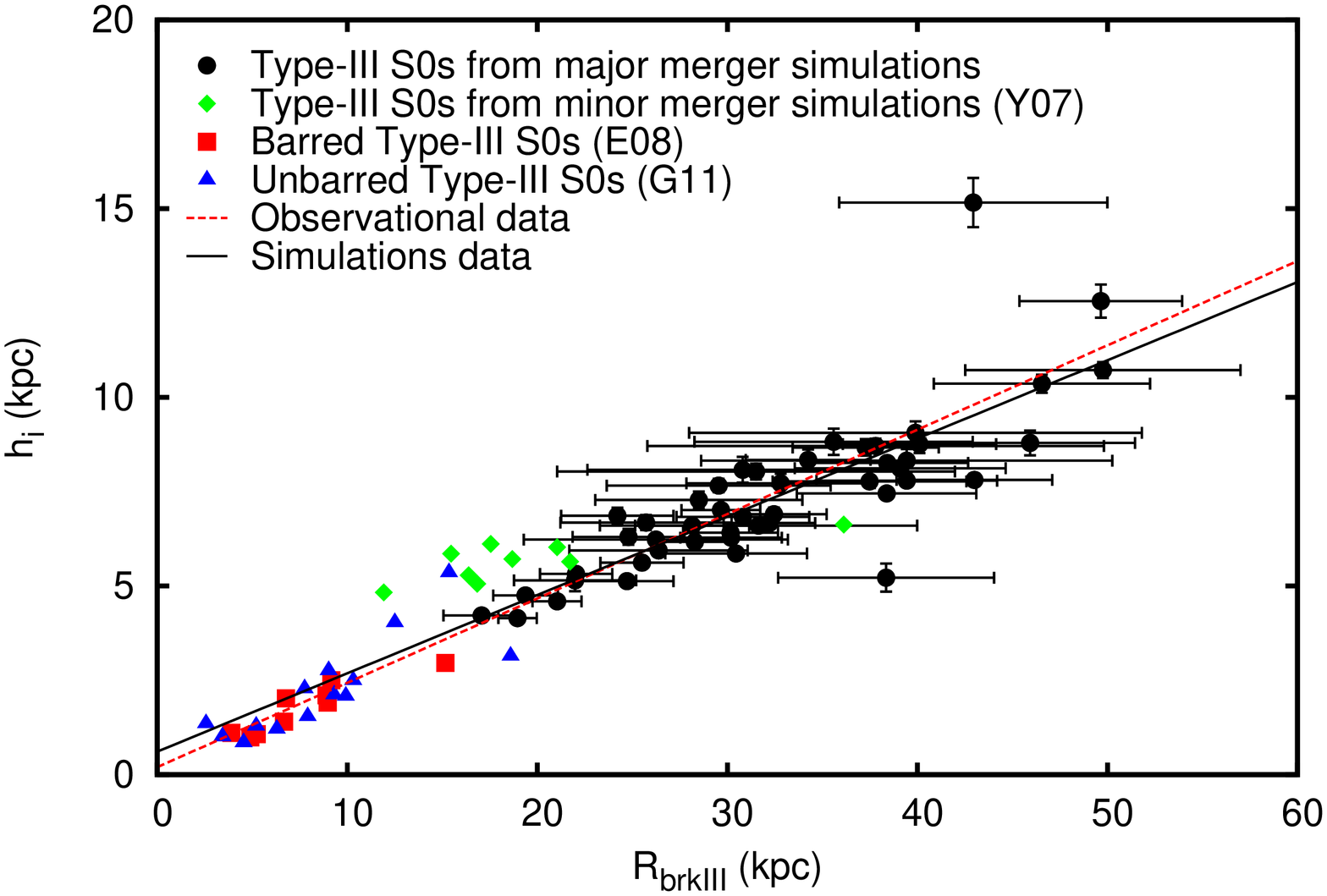}  
\includegraphics[angle=270, width = 0.49\textwidth, bb = 60 50 540 750, clip]{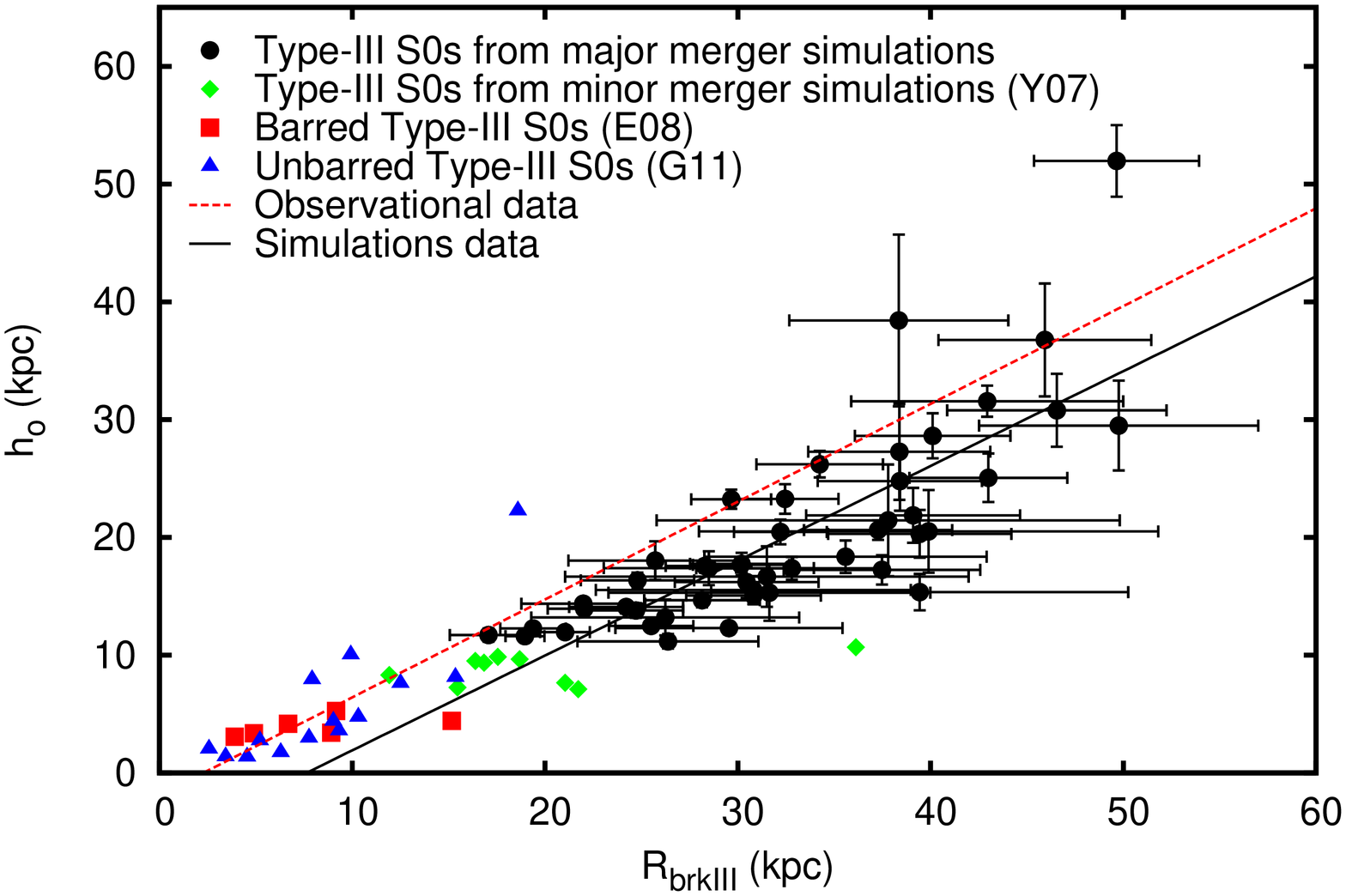}  
\includegraphics[angle=270, width = 0.49\textwidth, bb = 60 50 540 750, clip]{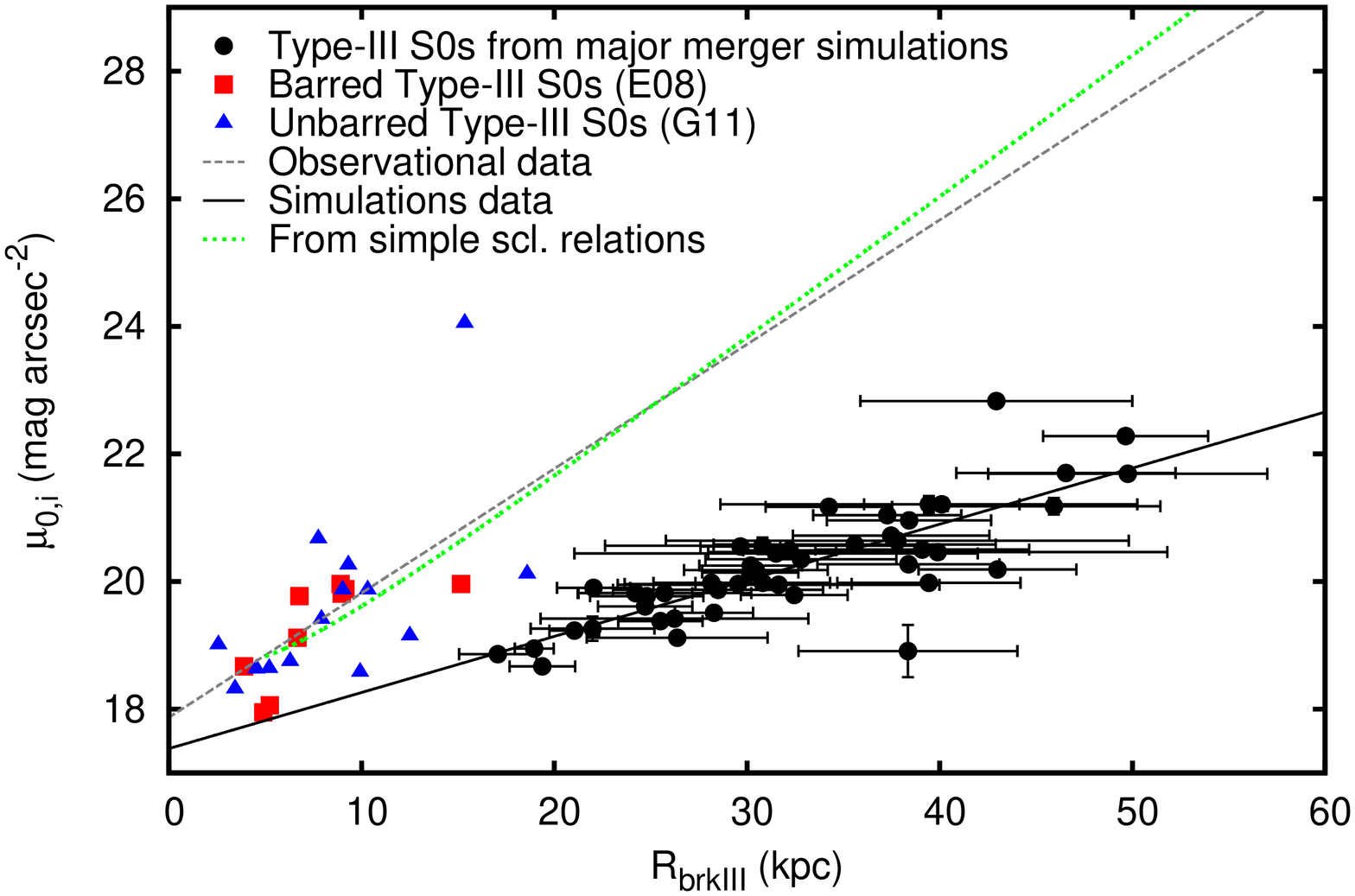}  
\includegraphics[angle=270, width = 0.49\textwidth, bb = 60 50 540 750, clip]{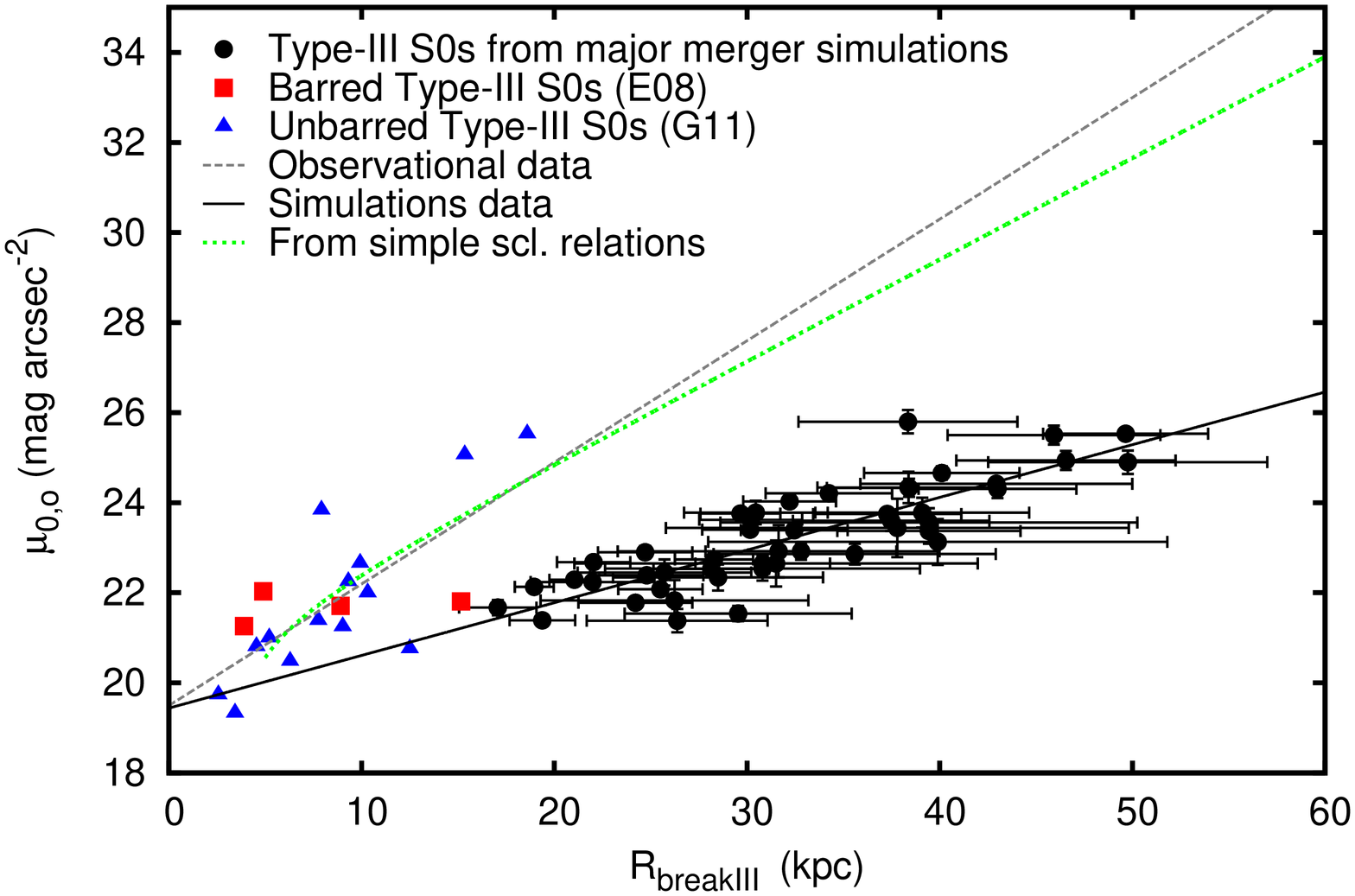}  
\caption{Trends of the photometric parameters of the inner and outer discs with the antitruncation radius, compared to real data on Type III S0 galaxies (E08; G11). \emph{Top}: $R$-band scalelengths of the inner and outer discs vs.\, \rbreak\ (left and right panels, respectively). The fits performed to observations in these planes are two of the three simple scaling relations from which the relations observed in the rest of photometric planes can be derived (see Sect.\,\ref{Sec:basic_scaling}), so they have been remarked in red. Data of the antitruncations formed in the minor merger simulations by Y07 are overplotted for comparison. \emph{Bottom}:  $R$-band central surface brightness of the inner and outer discs vs.\,\rbreak\ (left and right panels). \emph{Solid lines}: Linear fits to our simulations sample. \emph{Dashed lines}: Linear fits to the observational data by E08 and G11. \emph{Dotted-dashed line}: Expected relations assuming the simple scaling relations $\hi \propto \rbreak$, $\ho \propto \rbreak$, and $\mubreak \propto \rbreak$ obtained from data (see Sect.\,\ref{Sec:basic_scaling}). Consult the legend in the figure.  [\emph{A colour version of this figure is available in the online edition.}]
}
\label{fig:trends_rbrk}
\end{figure*}

This linear trend can be expected from the fact that both scalelengths fulfill tight linear trends with \rbreak, as derived from Fig.\,\ref{fig:trends_rbrk}. The top panels show the distributions of our Type III S0-like remnants in the photometric planes relating \hi\ and \ho\ with \rbreak. Real data have been plotted again for comparison, as well as the resulting linear fits to the simulations and to the observations (see Table\,\ref{tab:fits_results}). We also represent the location of the remnants that result from the minor merger simulations by Y07. 

The plots indicate that Type III S0 galaxies present more external \rbreak\ on discs with higher scalelengths. The similarity of the trends followed by simulations and data in the $\hi$ -- \rbreak\ and $\ho$ -- \rbreak\ planes is remarkable: simulations overlay the extrapolation of the linear trends shown by observations towards higher \rbreak\ values. The fact that the minor mergers by Y07 also fulfill these extrapolations, overlapping our major merger simulations at the lowest \rbreak\ values is also very significant. This suggests that the physical processes after the formation of antitruncations in major and minor mergers must be similar, even though the underlying disc structure has a different origin in the two cases: the disc of the main progenitor survives in minor events, whereas the disc is destroyed and rebuilt in major encounters.

The different location of Y07 simulations and ours in these diagrams should not be interpreted as evidence of the minor-merger origin of the breaks in real galaxies, just because they are located at low \rbreak\ values. These diagrams do not account for the different scale in mass, luminosity, and size of the simulated galaxies with respect to the real ones. The remnants of Y07 simulations have stellar masses similar to our less massive simulations ($\sim 10^{11}\Msun$). We will show that, once the scalelengths in these diagrams are normalized to the total galaxy size, both major and minor merger simulations overlap with observations (Sects.\,\ref{Sec:scalefree} and \ref{Sec:scaling}).

The bottom panels of Fig.\,\ref{fig:trends_rbrk} show the location of simulated and observational data in the $\mui$ -- \rbreak\ and $\muo$ -- \rbreak\ diagrams. As already noticed, there is an offset of $\sim 1$\,mag\,arcsec$^{-2}$ between models and real data, but taking this into account, the simulations would lie on top of the extrapolation of the observational trends towards higher \rbreak\ (see Sect.\,\ref{Sec:age_effects}).

Summarizing, we have found that the real Type III S0s present well-defined linear trends of the characteristic parameters of the inner and outer discs with \rbreak\ and between them that had not been reported in previous studies. The antitruncated S0-like remnants resulting from major merger simulations reproduce these linear observational trends. 

\subsection{Scale-free photometric planes of antitruncations}
\label{Sec:scalefreeplanes}

In order to see if our simulated Type III profiles are realistic, we need to compare their characteristic photometric parameters taking into account that our remnants have different mass, luminosity, and size scales to the Type III S0s in the samples by E08 and G11. Therefore, we have also analysed the distributions of real and simulated galaxies in scale-free diagrams.

We have removed the offset in surface brightness between our models and real data by plotting differences of central surface brightness values in the inner and outer discs with respect to \mubreak\ in each galaxy, instead of net values. Typical scalelengths have also been normalized to the total optical size of the galaxy to remove size scaling effects between real data and the simulations. We have considered the radius of the isophote with $\mu _B=25$\,mag\,arcsec$^{-2}$ as an estimate of the total optical radius of the galaxy (\risoph), following E08 and G11 (\risoph\ is thus defined in the $B$ band). The values of $\risoph$ for each S0-like remnant are listed in Table\,\ref{tab:antitrunc}.

\begin{figure*}[ht!]
\center
\includegraphics[angle=270, width = 0.49\textwidth, bb = 60 50 540 750, clip]{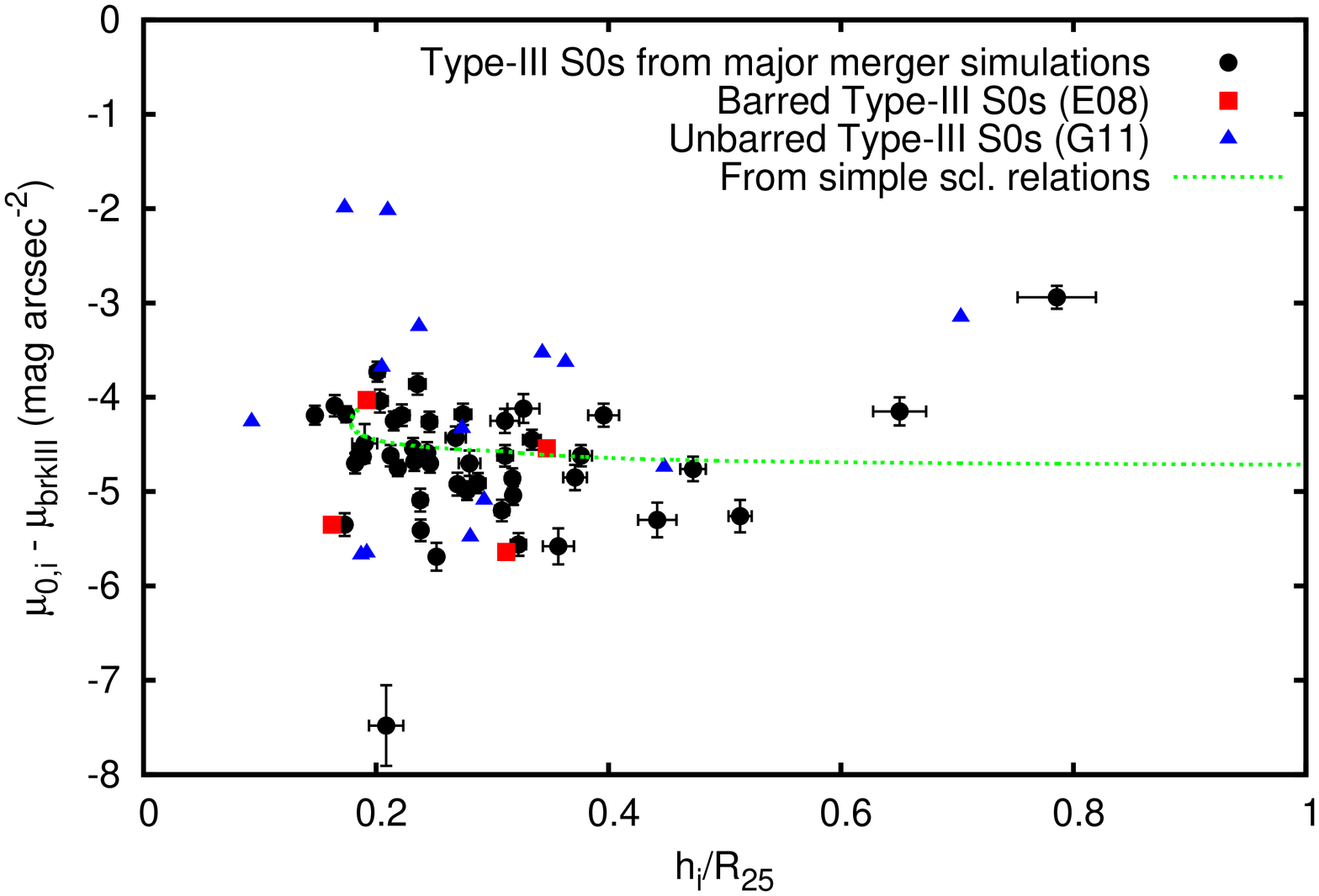} 
\includegraphics[angle=270, width = 0.49\textwidth, bb = 60 50 540 750, clip]{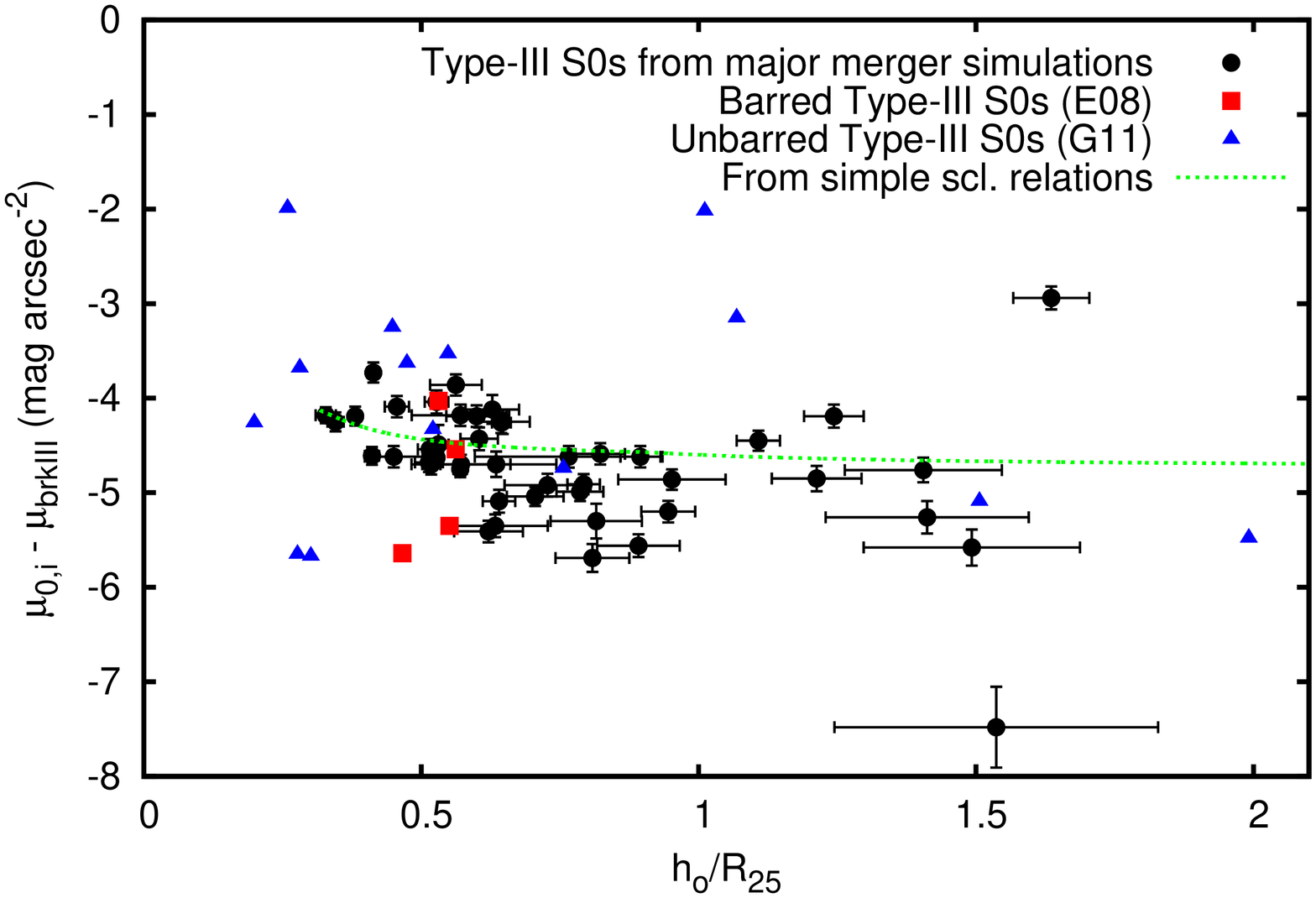}  
\includegraphics[angle=270, width = 0.49\textwidth, bb = 60 50 540 750, clip]{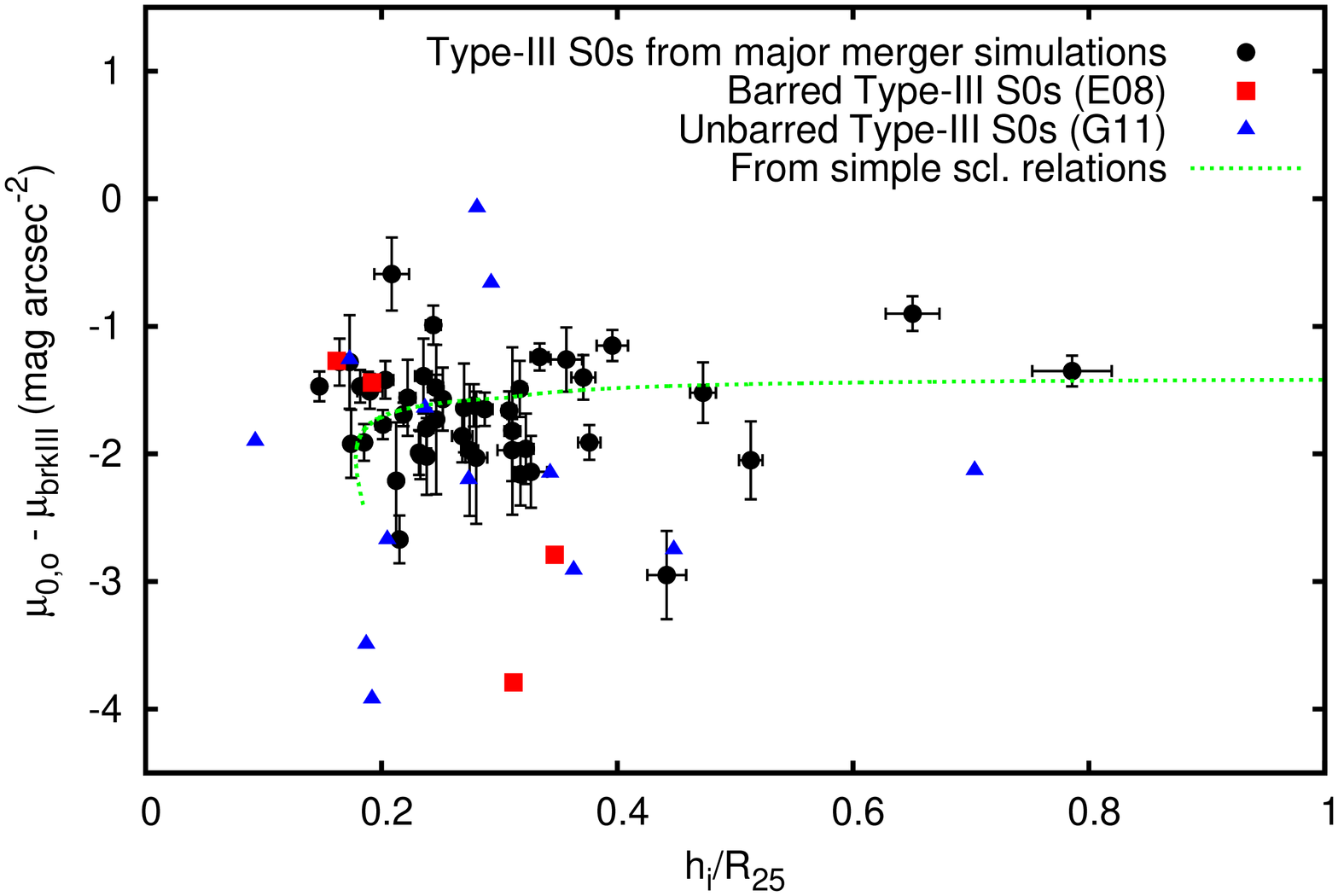}  
\includegraphics[angle=270, width = 0.49\textwidth, bb = 60 50 540 750, clip]{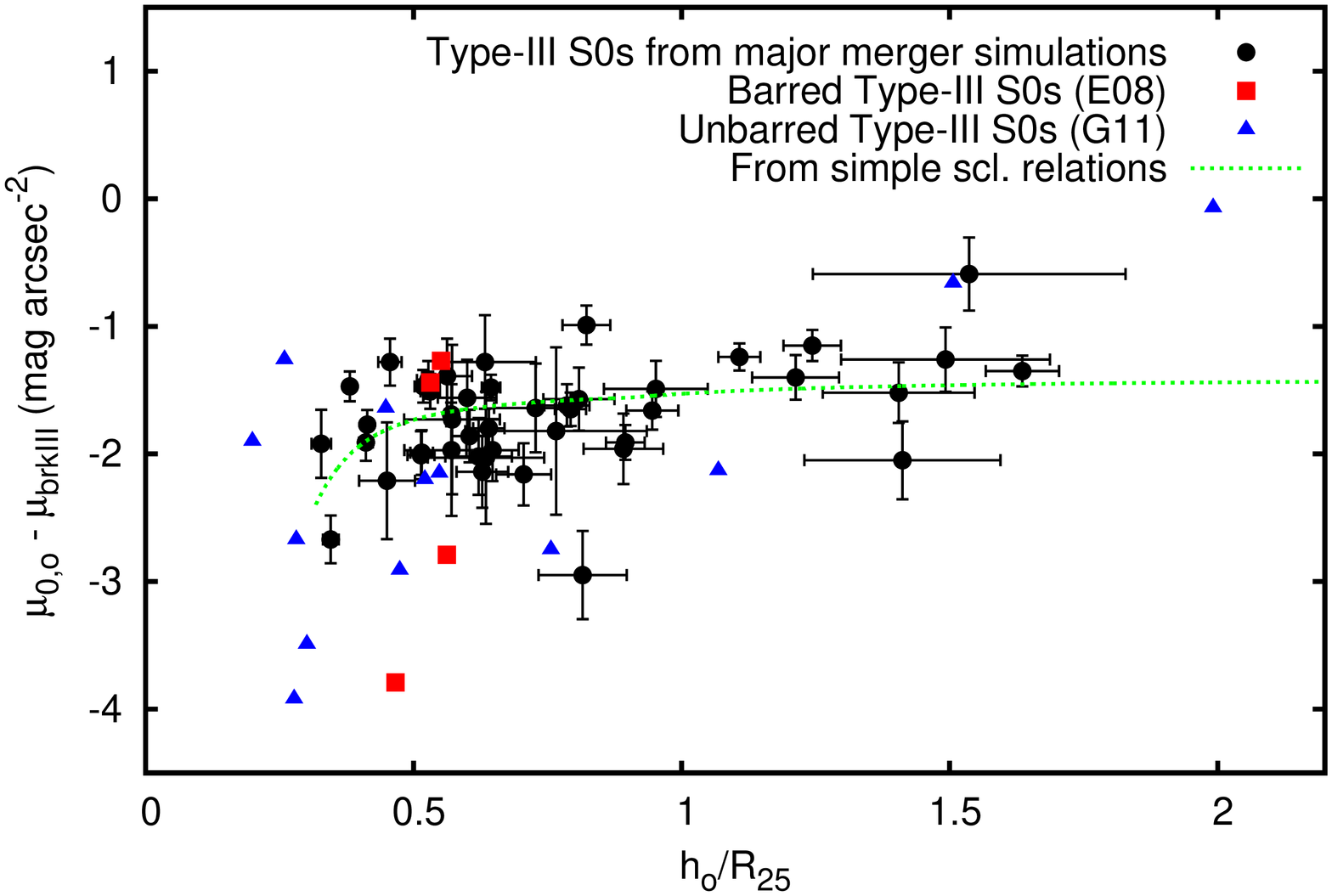}  
\caption{Scale-free photometric planes showing the distributions of the central surface brightness of the inner and outer discs referred to \mubreak\ vs. $\hi/\risoph$ and $\ho/\risoph$, for both simulated and real Type III S0s (E08; G11). \emph{Dotted line}: Expected relations derived from the simple observational scaling relations $\hi \propto \rbreak$, $\ho \propto \rbreak$, and $\mubreak \propto \rbreak$ (see Sect.\,\ref{Sec:basic_scaling}). Consult the legend in the figure.  [\emph{A colour version of this figure is available in the online edition.}] 
}
\label{fig:trends_hiR25_hoR25}
\end{figure*}

\subsubsection{Planes showing no correlation}
\label{Sec:scalefree}

In the top panels of Fig.\,\ref{fig:trends_hiR25_hoR25}, we plot the distributions of our S0 Type III remnants in the diagrams ($\mui -\mubreak$) vs.\,the scalelengths of the inner and outer discs normalized to \risoph. The bottom panels show the same for ($\muo -\mubreak$). The figure shows that the simulations overlie the observational data in all these planes, but neither observations nor major merger simulations exhibit any trends. Therefore, major mergers can build up antitrucated S0-like remnants that reproduce the lack of any correlation in these planes observed in real Type III S0s.

\begin{figure*}[t!h]
\center
\includegraphics[angle=270, width = 0.49\textwidth, bb = 60 50 540 750, clip]{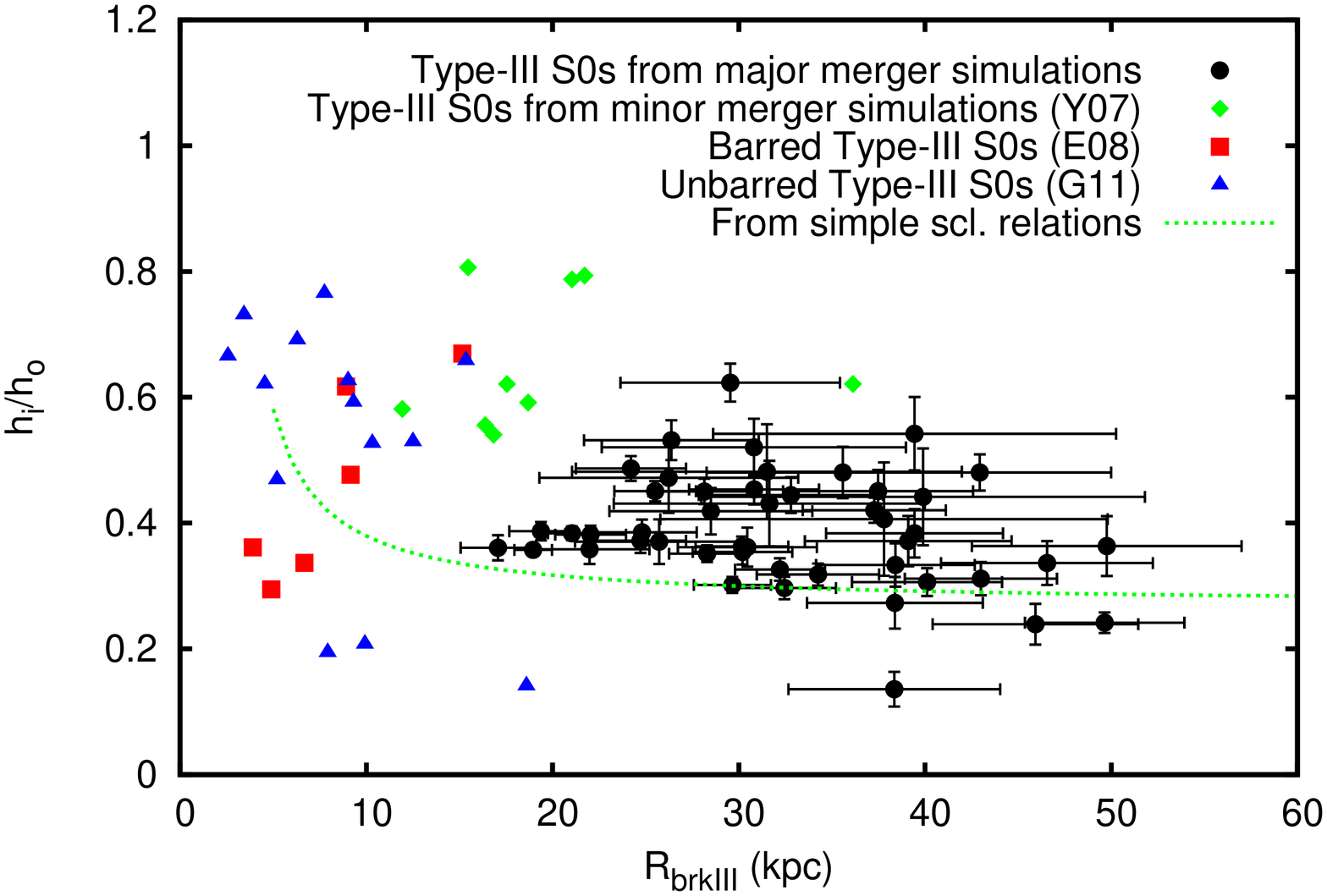}  
\includegraphics[angle=270, width = 0.49\textwidth, bb = 60 50 540 750, clip]{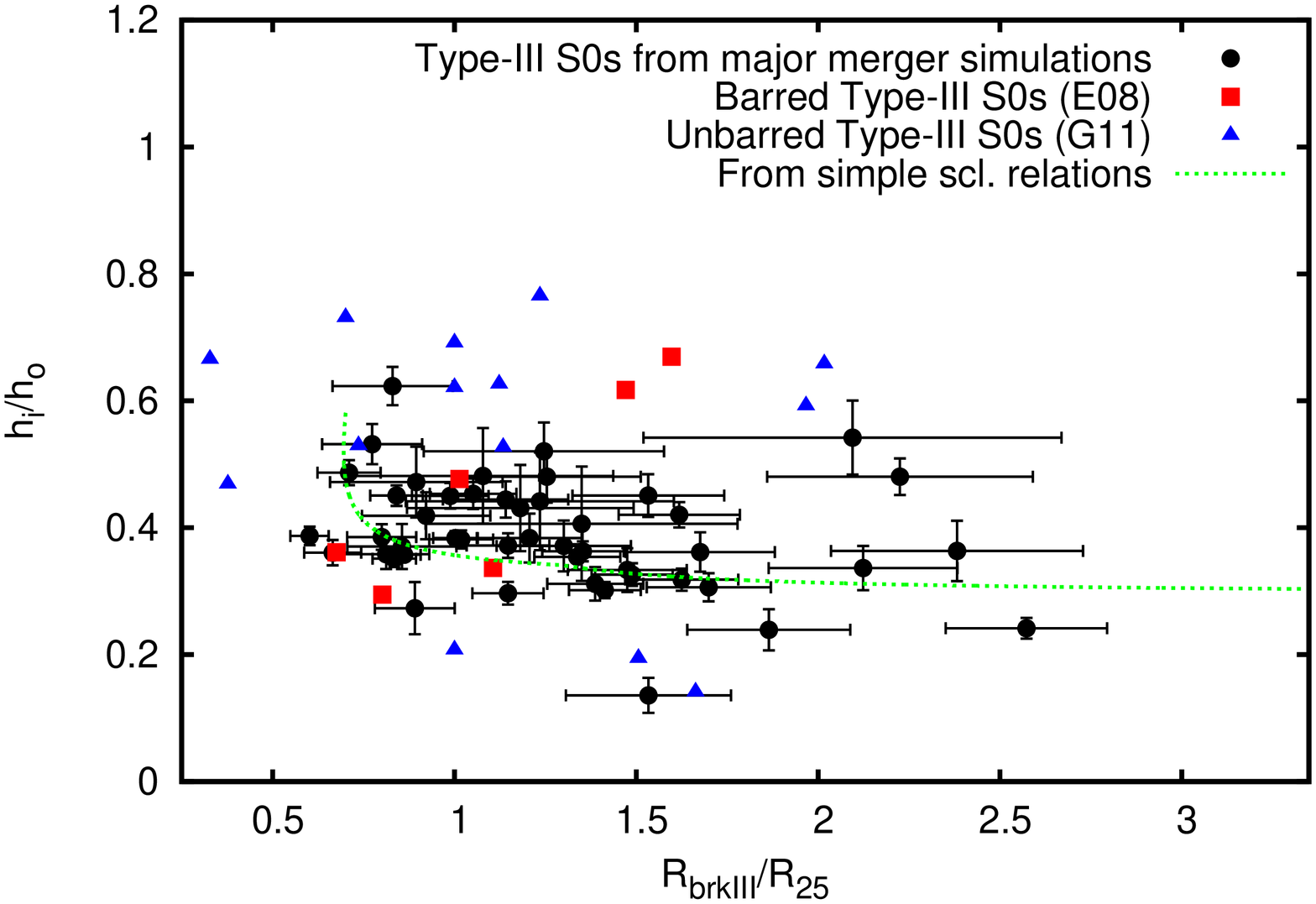}  
\caption{Scale-free photometric planes showing the distributions of $\hi/\ho$ vs. $\rbreak$ and $\rbreak/\risoph$ of our Type III S0-like remnants, compared to real antitruncated S0s (E08; G11). Data of the antitruncations formed in the minor merger simulations by Y07 are overplotted for comparison in the left diagram. \emph{Dotted line}: Expected relations derived from the simple observational scaling relations $\hi \propto \rbreak$, $\ho \propto \rbreak$, and $\mubreak \propto \rbreak$ (see Sect.\,\ref{Sec:basic_scaling}). Consult the legend in the figure.  [\emph{A colour version of this figure is available in the online edition.}]
}
\label{fig:trends_hiho_noscaling}
\end{figure*}

We analyse the relation between the ratio of the scalelengths of the inner and outer discs with the radial location of the antitruncation in Fig.\,\ref{fig:trends_hiho_noscaling}. In the left panel, we show the distribution of our S0-like remnants and of real data without normalizing $\rbreak$, while the right panel plots $\hi/\ho$ as a function of $\rbreak/\risoph$. The minor merger simulations by Y07 have been overplotted in the first panel. Our remnants are clearly displaced towards higher \rbreak\ than the Type III S0s from E08 and G11 (left panel), but once this difference in sizes is accounted for, both distributions overlap (right panel). Again, the fact that simulations reproduce the lack of trends of $\hi/\ho$ with $\rbreak/\risoph$ of real antitruncated S0s supports major mergers as a feasible mechanism to explain the formation of antitruncations in these galaxies. 

Minor merger simulations by Y07 are also displaced towards higher \rbreak\ values than real S0s in the left panel of Fig.\,\ref{fig:trends_hiho_noscaling}, but they also exhibit $\hi/\ho$ ratios similar to observations. Notice that, while major merger simulations satisfy the observational $\hi/\ho$ distribution from $\sim0.2$ up to $\sim0.6$, the minor merger remnants dominate the observed range from $\sim0.6$ to $\sim0.8$. This result strongly supports a combination of major and minor mergers to explain the buildup of antitruncated S0 galaxies.  

\begin{figure*}[th!]
\center
\includegraphics[angle=270, width = 0.45\textwidth, bb = 60 50 540 750, clip]{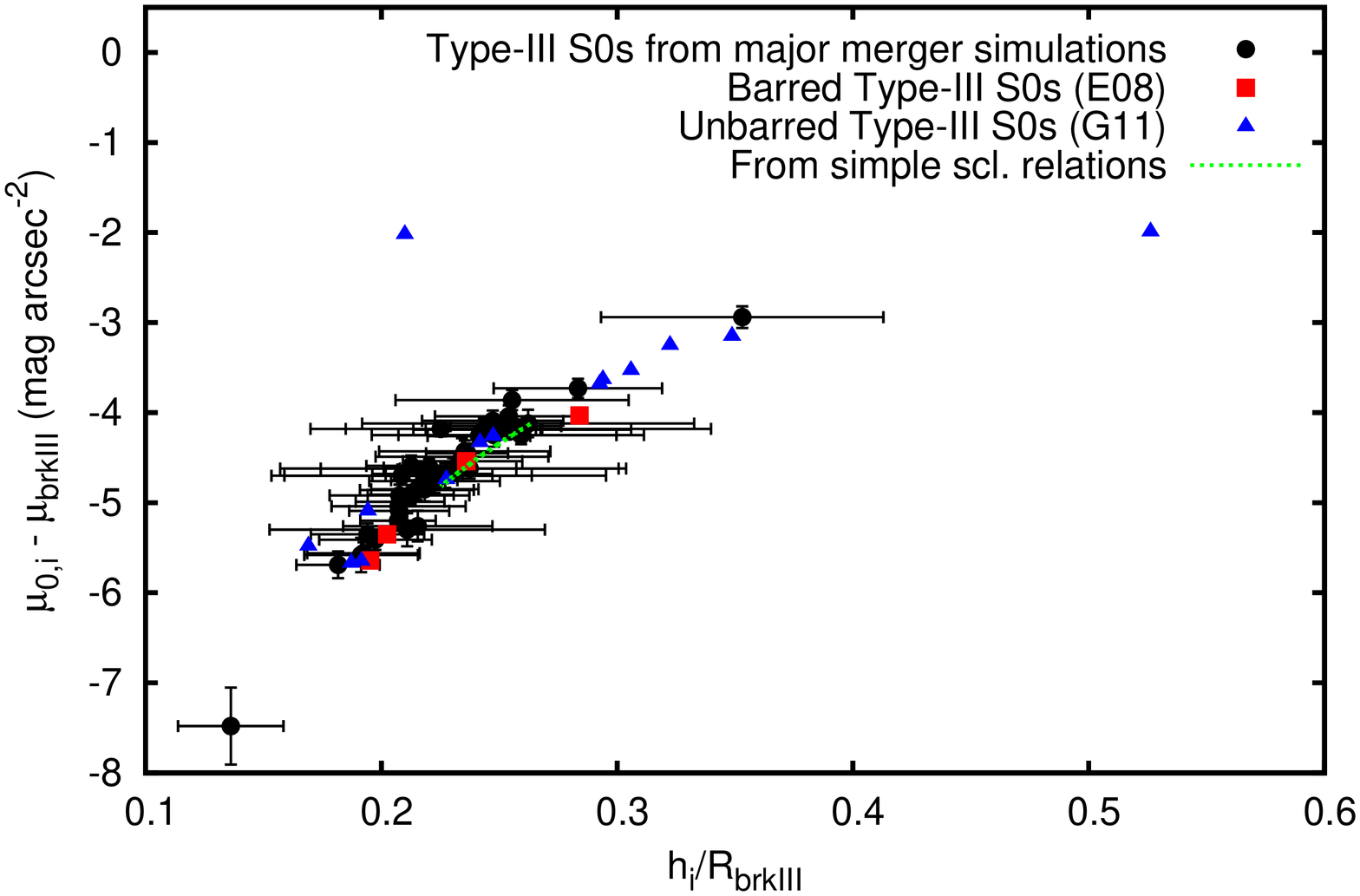}  
\includegraphics[angle=270, width = 0.45\textwidth, bb = 60 50 540 750, clip]{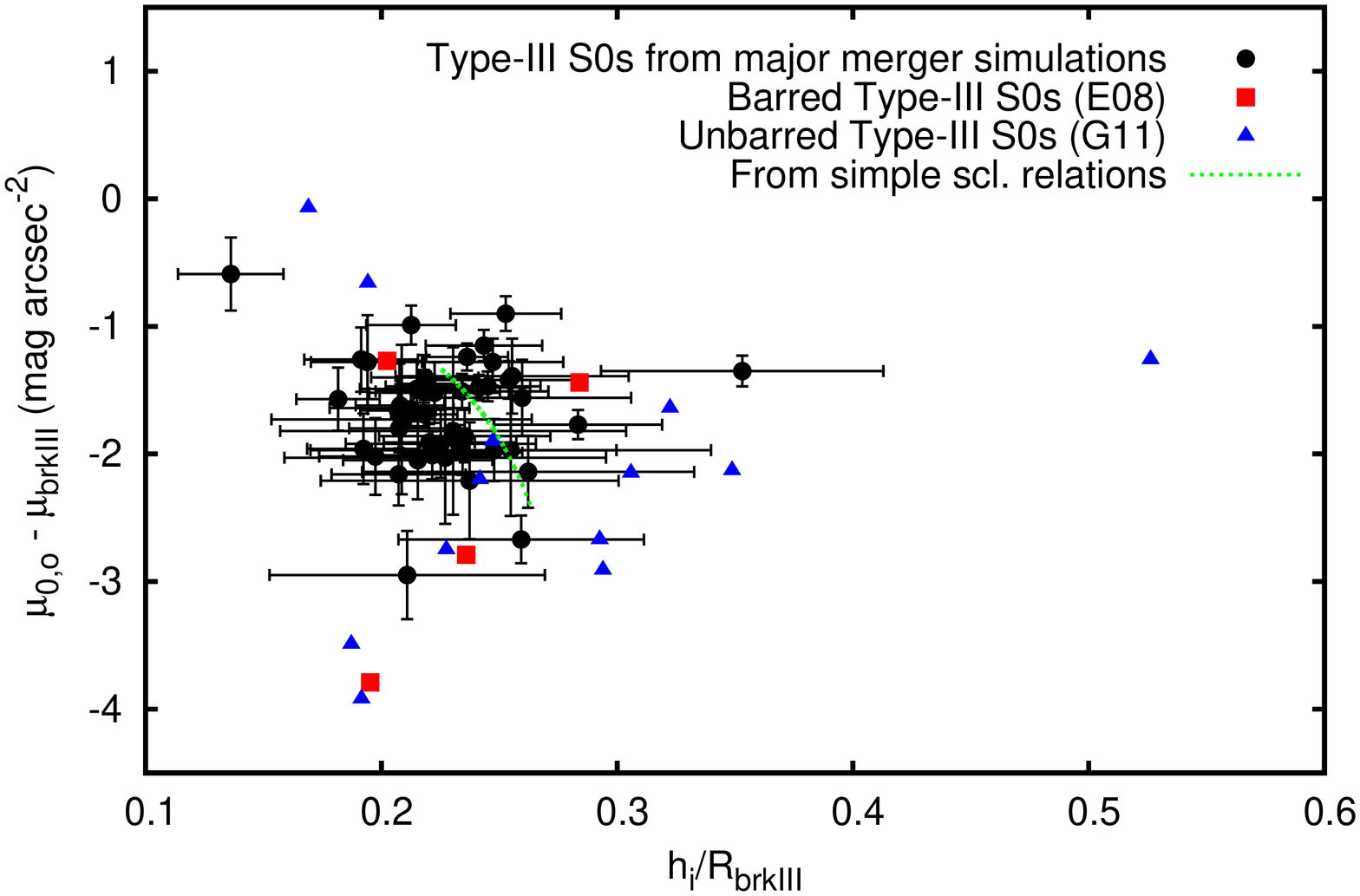}  
\includegraphics[angle=270, width = 0.45\textwidth, bb = 60 50 540 750, clip]{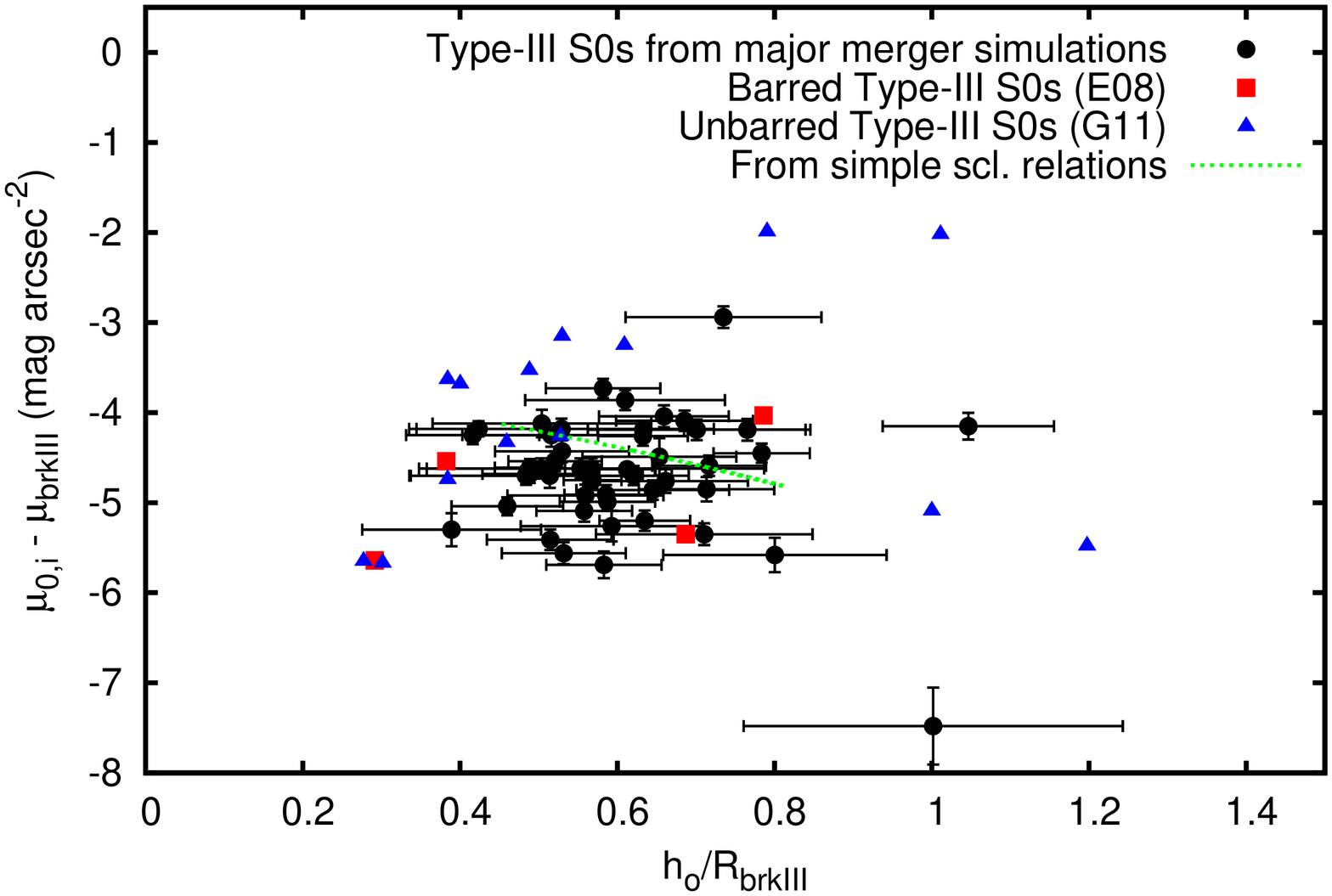}  
\includegraphics[angle=270, width = 0.45\textwidth, bb = 60 50 540 750, clip]{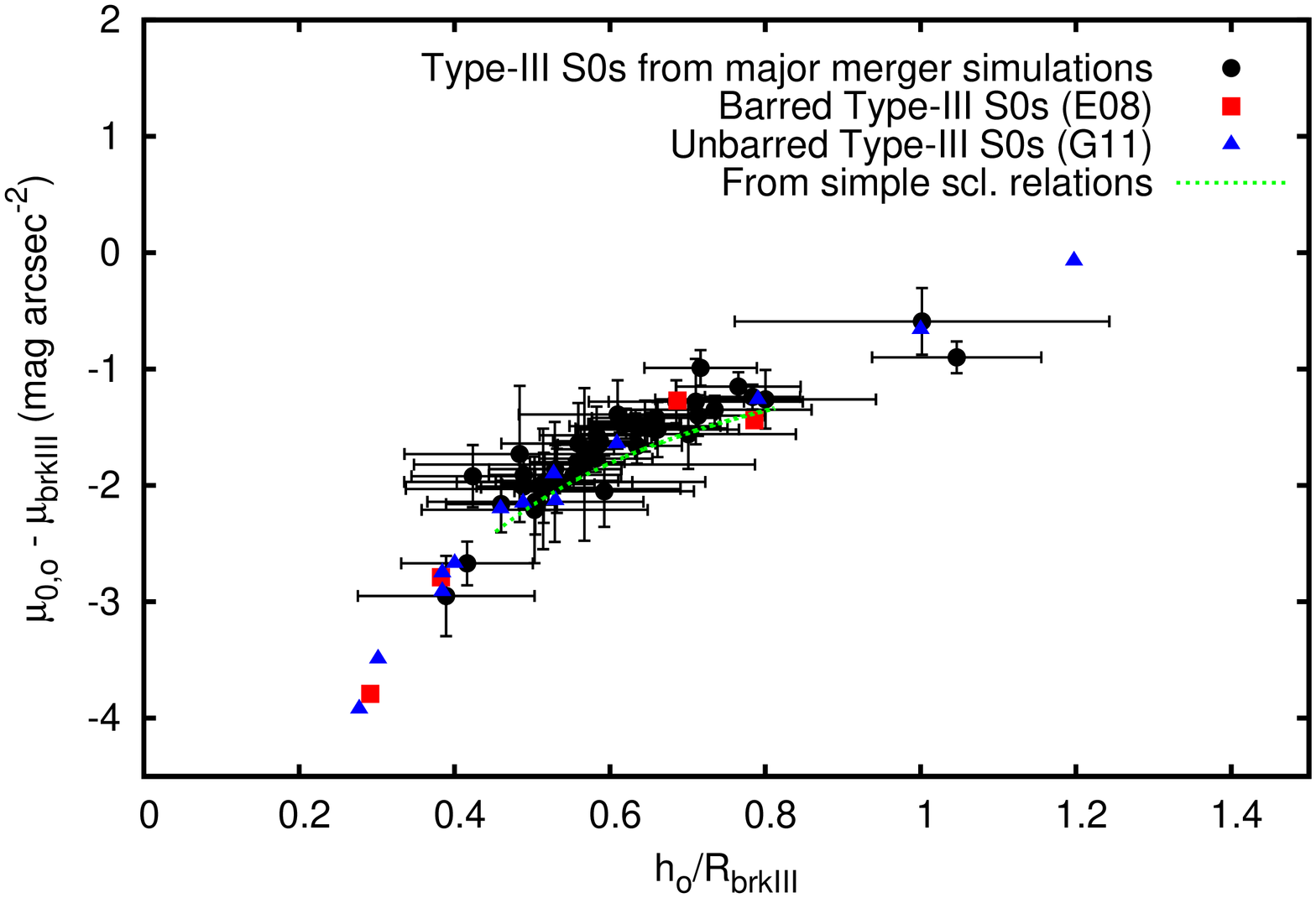}  
\includegraphics[angle=270, width = 0.45\textwidth, bb = 60 50 540 750, clip]{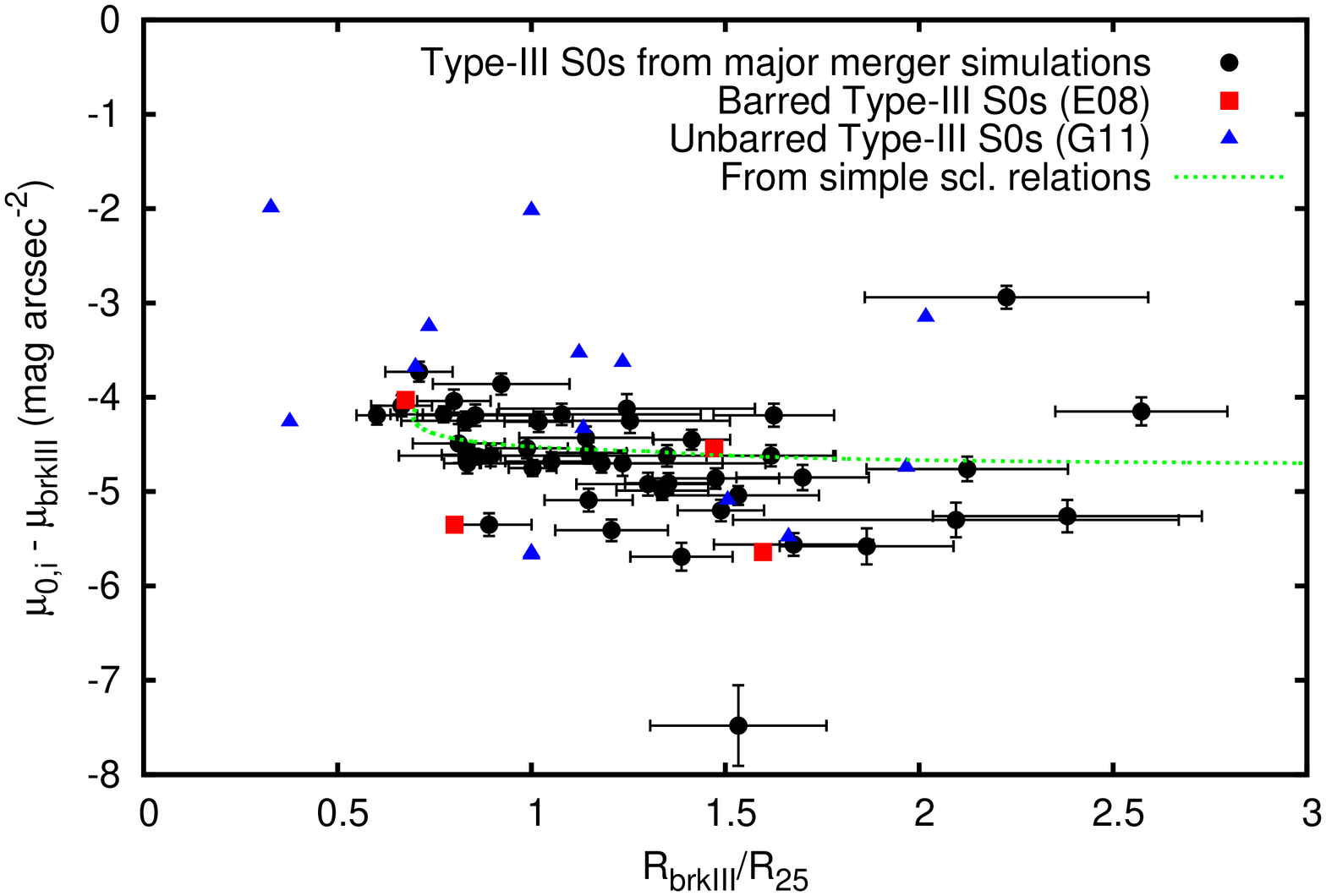}  
\includegraphics[angle=270, width = 0.45\textwidth, bb = 60 50 540 750, clip]{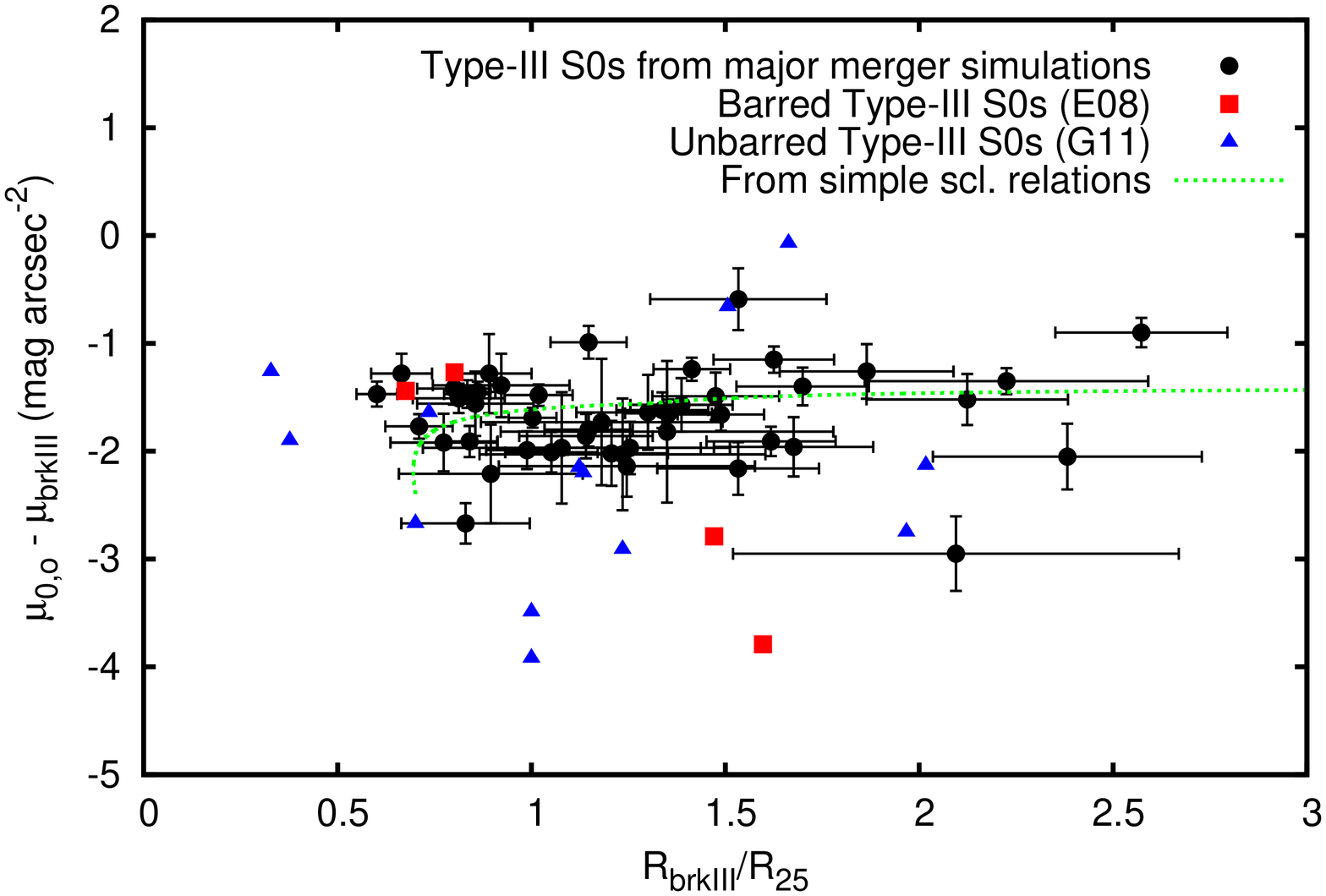}  
\caption{Scale-free photometric planes showing the distribution of the real and simulated Type III S0s in the planes of $(\mui -\mubreak)$ and $(\muo -\mubreak)$ vs.\,$\hi/\rbreak$ (\emph{top panels}), vs.\,$\ho/\rbreak$ (\emph{intermediate panels}), and vs.\,$\rbreak/\risoph$ (\emph{bottom panels}). \emph{Dotted line}: Expected relations derived from the simple observational scaling relations $\hi \propto \rbreak$, $\ho \propto \rbreak$, and $\mubreak \propto \rbreak$ (see Sect.\,\ref{Sec:basic_scaling}). Consult the legend in the figure.  [\emph{A colour version of this figure is available in the online edition.}]
}
\label{fig:trends_hiRbrk_hoRbrk}
\end{figure*}

In Fig.\,\ref{fig:trends_hiRbrk_hoRbrk}, we compare the location of real data and simulations in the planes $(\mui -\mubreak)$ vs.\,$\ho/\rbreak$ (intermediate left panel) and $(\muo -\mubreak)$ vs.\,$\hi/\rbreak$ (top right panel). The two distributions overlap completely, showing no clear trend. In this figure, we also show the relations in the planes $(\mui -\mubreak)$ -- $\hi/\rbreak$ and $(\muo -\mubreak)$ -- $\ho/\rbreak$ (top left and intermediate right panels, respectively). These diagrams follow eq.\,\ref{eq:Freemanmu} for the inner and outer discs, evaluated at the break radius by definition. Therefore, the tight trends observed in these planes were expected. In the bottom panels of Fig.\,\ref{fig:trends_hiRbrk_hoRbrk}, we represent the distributions of real and simulated S0s in the $(\mui - \mubreak)$ -- $\rbreak/\risoph$ and $(\muo - \mubreak)$ -- $\rbreak/\risoph$ diagrams. Real and simulated Type III S0s also overlap in the diagrams, showing flat trends with a noticeable dispersion.

In conclusion, these scale-free photometric diagrams demonstrate that our S0-like remnants develop antitruncations and inner and outer discs with photometric structures consistent with those observed in real Type III S0s, once the differences in mass, luminosity, and size of the real data and the simulations are accounted for.

\subsubsection{Planes with strong photometric scaling relations}
\label{Sec:scaling} 

We have shown that there are tight linear relations between the photometric parameters of the inner and outer discs and the breaks in real Type III S0 galaxies, and that the antitruncated S0-like remnants resulting from major merger simulations lie on top of the extrapolations of the observational trends towards higher values of \hi, \ho, and \rbreak\ (Sects.\,\ref{Sec:rbreak} and \ref{Sec:innerouterdiscs}). Now, we analyse the strong scaling relations underlying real antitruncations in scale-free photometric planes, and how major mergers can explain them. The linear fits to the scaling relations satisfied by real and simulated data shown in this Section are listed in Table\,\ref{tab:fits_results}. 

\begin{figure*}[th!]
\center
\includegraphics[angle=270, width = 0.49\textwidth, bb = 60 50 540 750, clip]{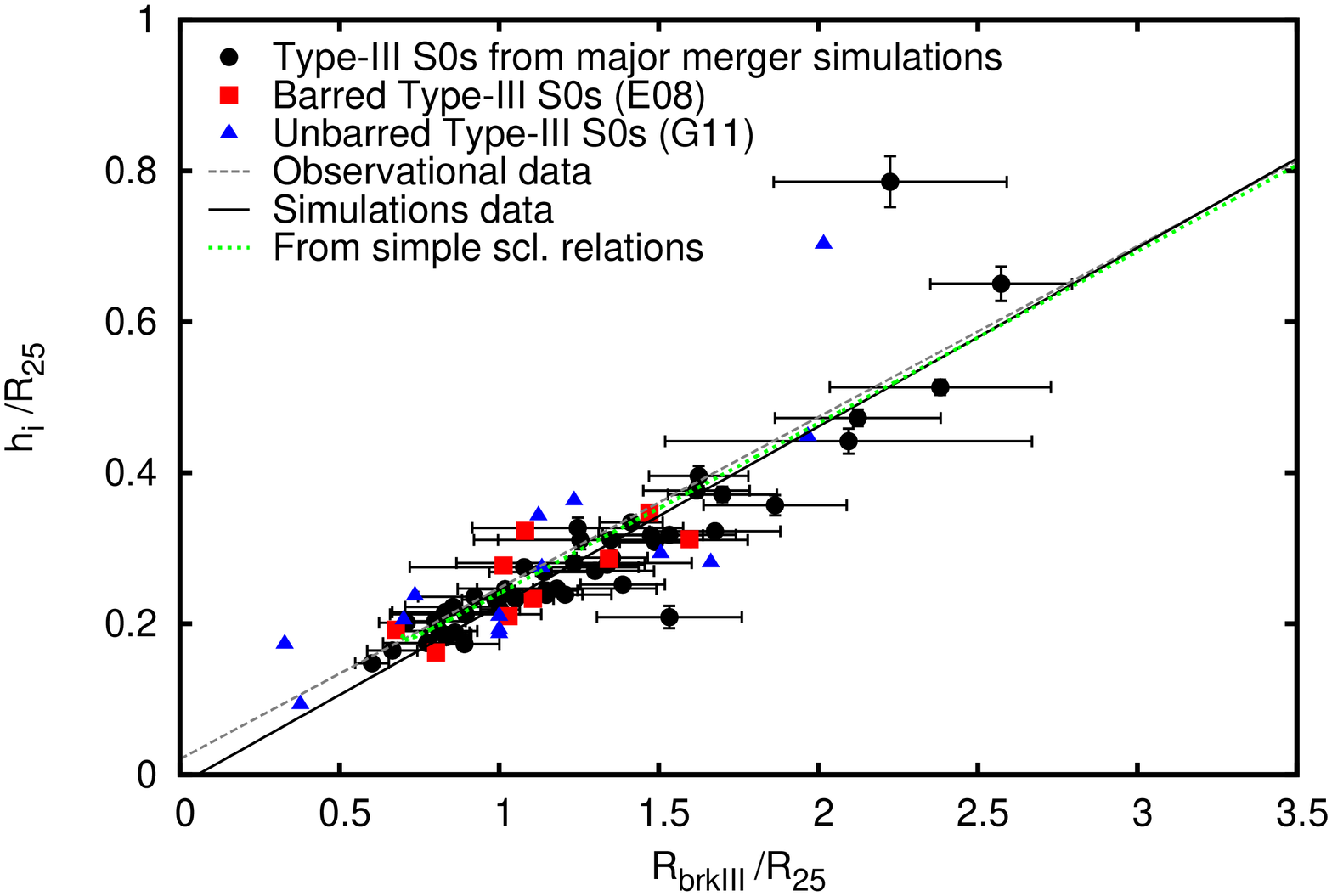}  
\includegraphics[angle=270, width = 0.49\textwidth, bb = 60 50 540 750, clip]{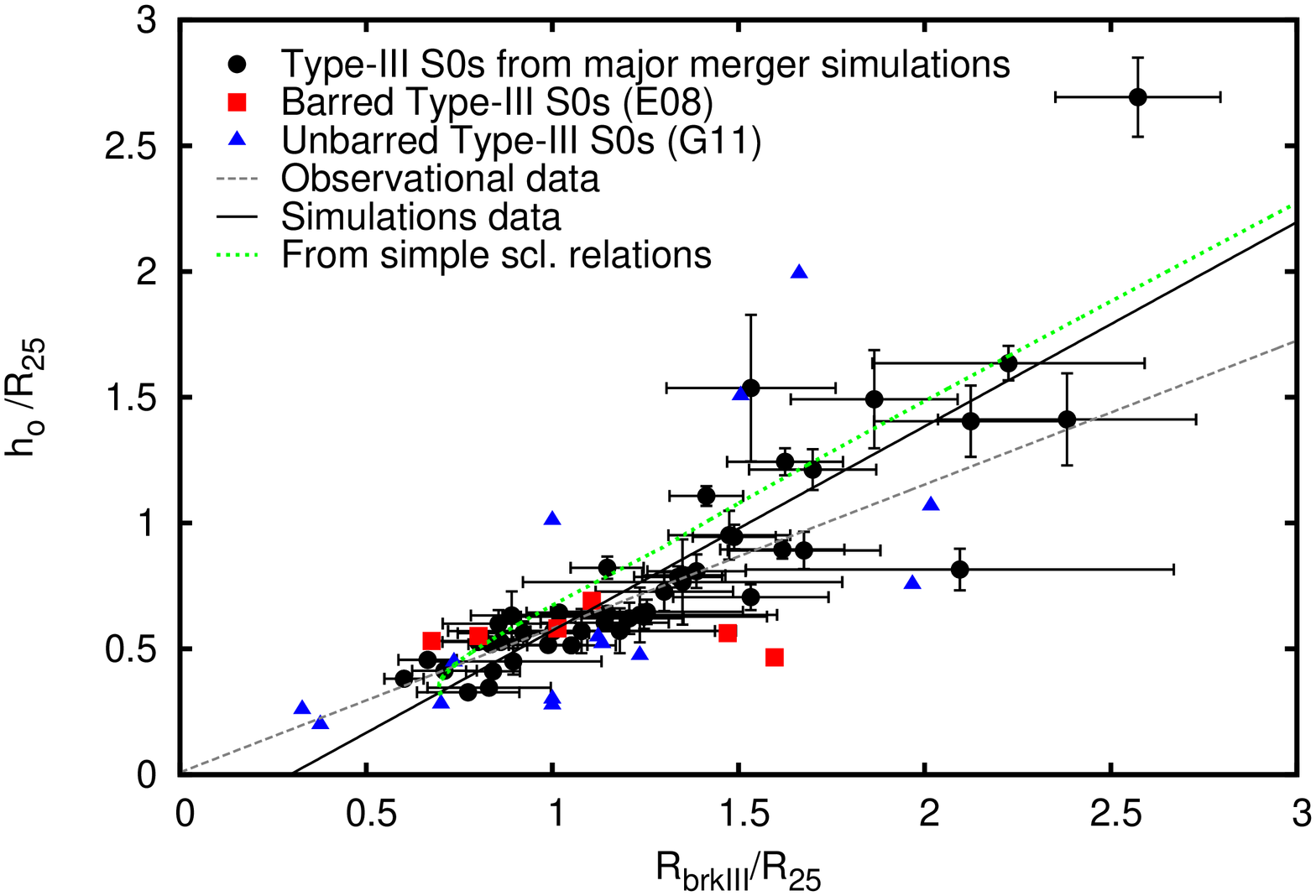}  
\includegraphics[angle=270, width = 0.49\textwidth, bb = 60 50 540 750, clip]{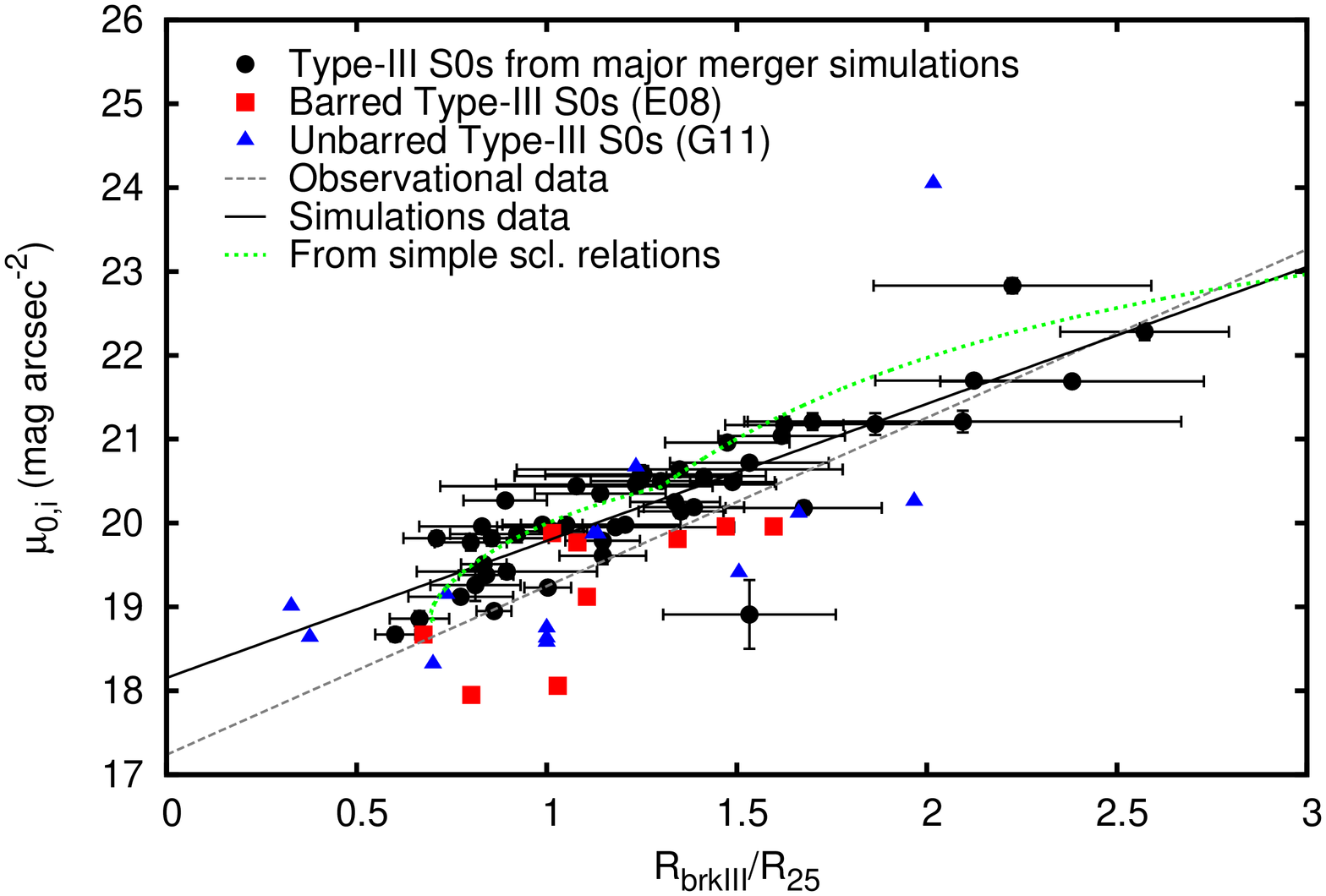}  
\includegraphics[angle=270, width = 0.49\textwidth, bb = 60 50 540 750, clip]{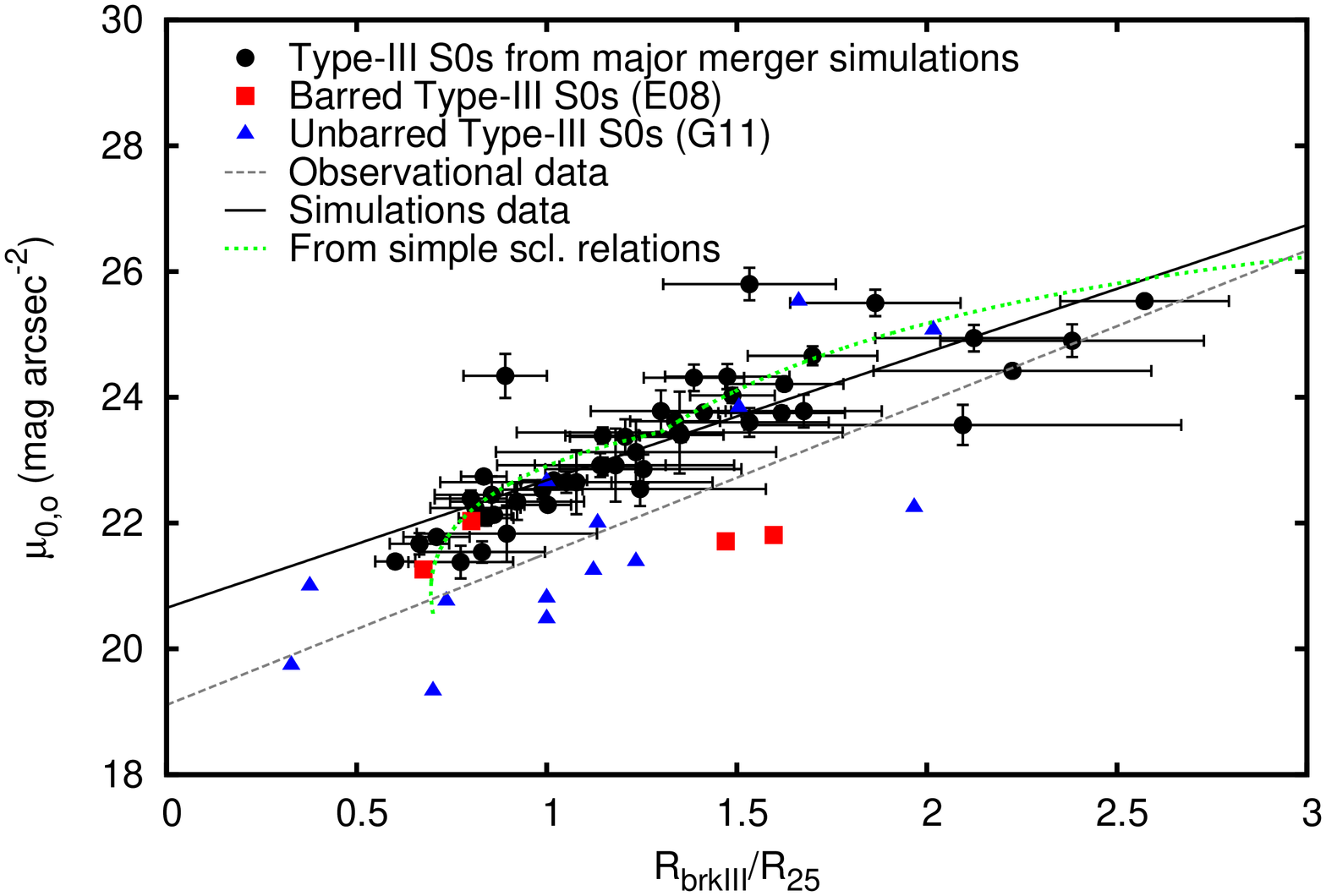}  
\caption{Photometric scaling relations of the characteristic parameters of the inner and outer discs with $\rbreak/ \risoph$, in our Type III S0-like remnants and in real Type III S0 galaxies (E08; G11). \emph{Solid line}: Linear fit to our simulations. \emph{Dashed line}: Linear fit to the observational data by E08 and G1. \emph{Dotted line}: Expected relations derived from the simple observational scaling relations $\hi \propto \rbreak$, $\ho \propto \rbreak$, and $\mubreak \propto \rbreak$ (see Sect.\,\ref{Sec:basic_scaling}).  Consult the legend in the figure.  [\emph{A colour version of this figure is available in the online edition.}]
}
\label{fig:trends_rbrkr25}
\end{figure*}

In the top panels of Fig.\,\ref{fig:trends_rbrkr25}, we show the distributions of our Type III S0-like remnants in the $\hi/\risoph$ -- $\rbreak/\risoph$ and $\ho/\risoph$ -- $\rbreak/\risoph$ planes. Our simulations show clear linear correlations with $\rbreak/\risoph$, overlapping with the observational sample. In fact, the linear fits to simulations and data in the two diagrams (lines in the figure) are very similar (see Table\,\ref{tab:fits_results}). 

In the bottom panels of Fig.\,\ref{fig:trends_rbrkr25}, we analyse the trends of \mui\ and \muo\ with $\rbreak/\risoph$ of the S0-like remnants, in comparison with observational data. The central surface brightness of the inner and outer discs become fainter as $\rbreak/\risoph$ increases in the real and simulated S0s. We know that the observations and the simulated remnants present an offset of $\sim 1$\,mag\,arcsec$^{-2}$ in the surface brightness values (see Fig.\,\ref{fig:trends_rbrk}) due to the youth of the starbursts induced by the encounters in the remnants and by the assumptions adopted to estimate $M/L$ (see Sect.\,\ref{Sec:profiles} and Appendix\,\ref{Sec:limitations}). Considering that this offset might be corrected allowing the remnants to relax for a few Gyrs more or just changing these assumptions, the simulations would be displaced towards fainter \mui\ and \muo\ in the bottom panels of Fig.\,\ref{fig:trends_rbrkr25}. However, \risoph\ would also become smaller, and thus the simulations would also move towards higher $\rbreak/\risoph$ values in these planes, meaning that real and simulated data would still be consistent in these diagrams. Note also that this global dimming of the remnants would move simulations slightly diagonally in the planes shown at the top panels of the same figure, so simulations and observations would still overlap in these planes.

\begin{figure}[!ht]
\center
\includegraphics[angle=270, width = 0.45\textwidth, bb = 60 50 540 750, clip]{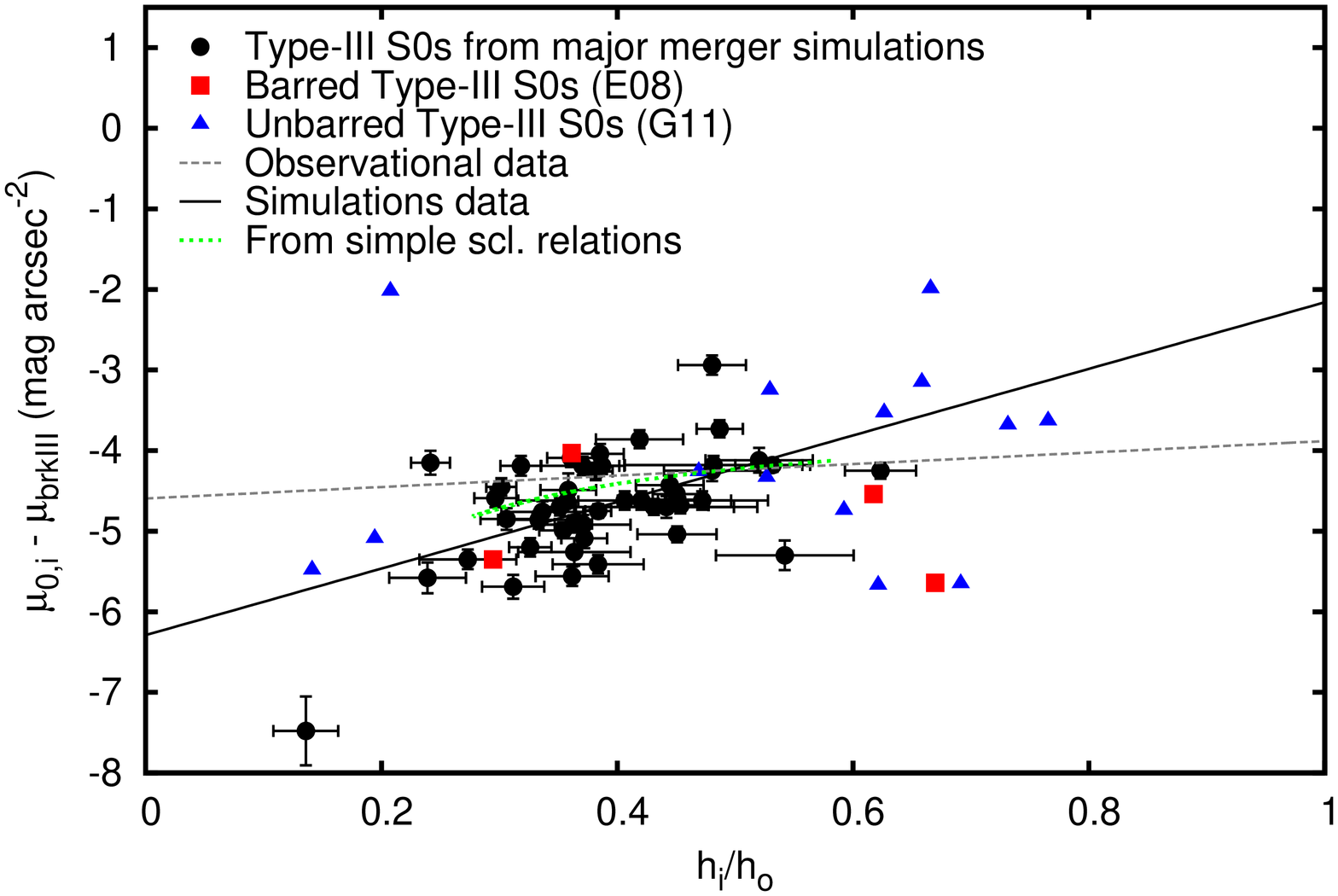}  
\includegraphics[angle=270, width = 0.45\textwidth, bb = 60 50 540 750, clip]{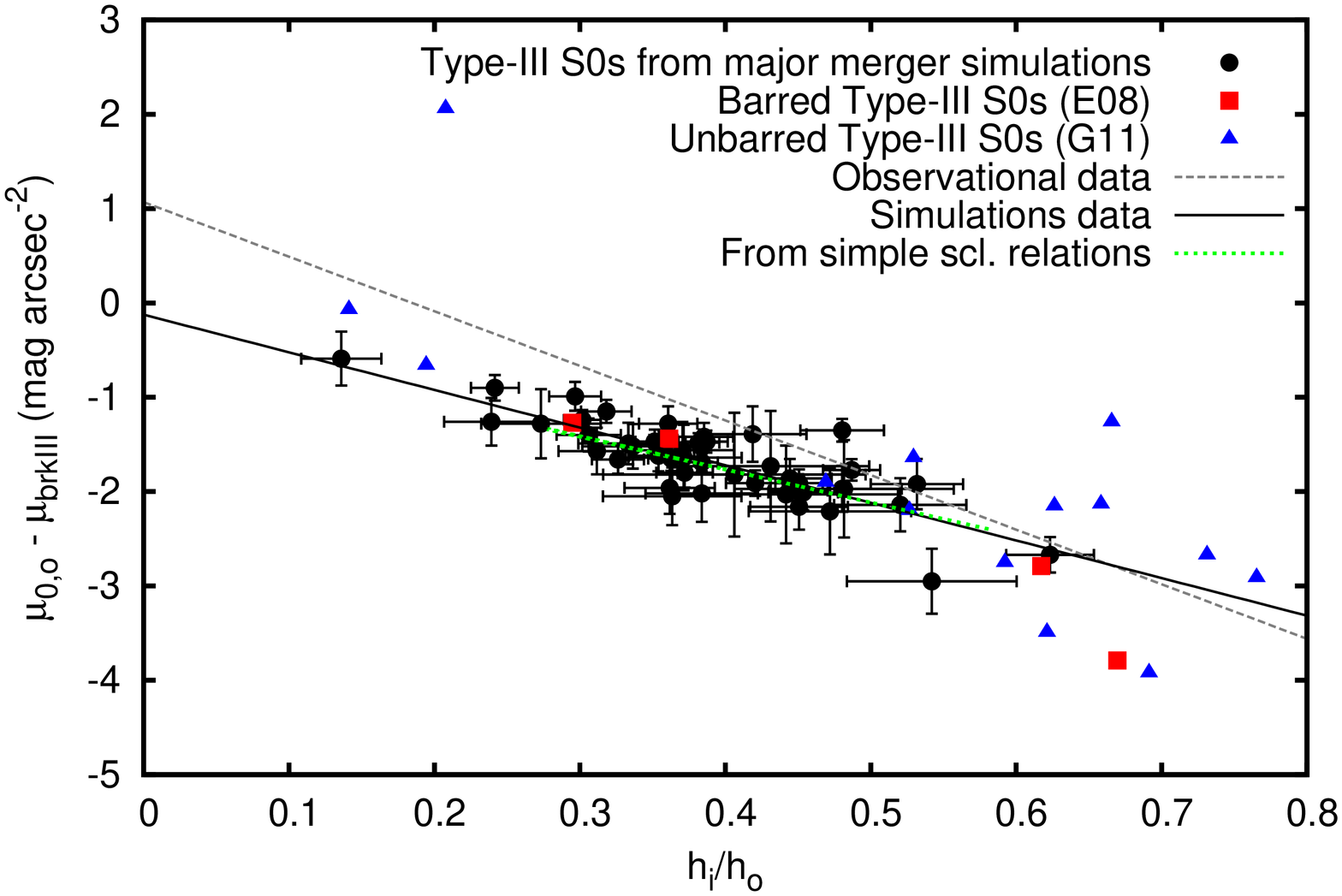}  
\includegraphics[angle=270, width = 0.45\textwidth, bb = 60 50 540 750, clip]{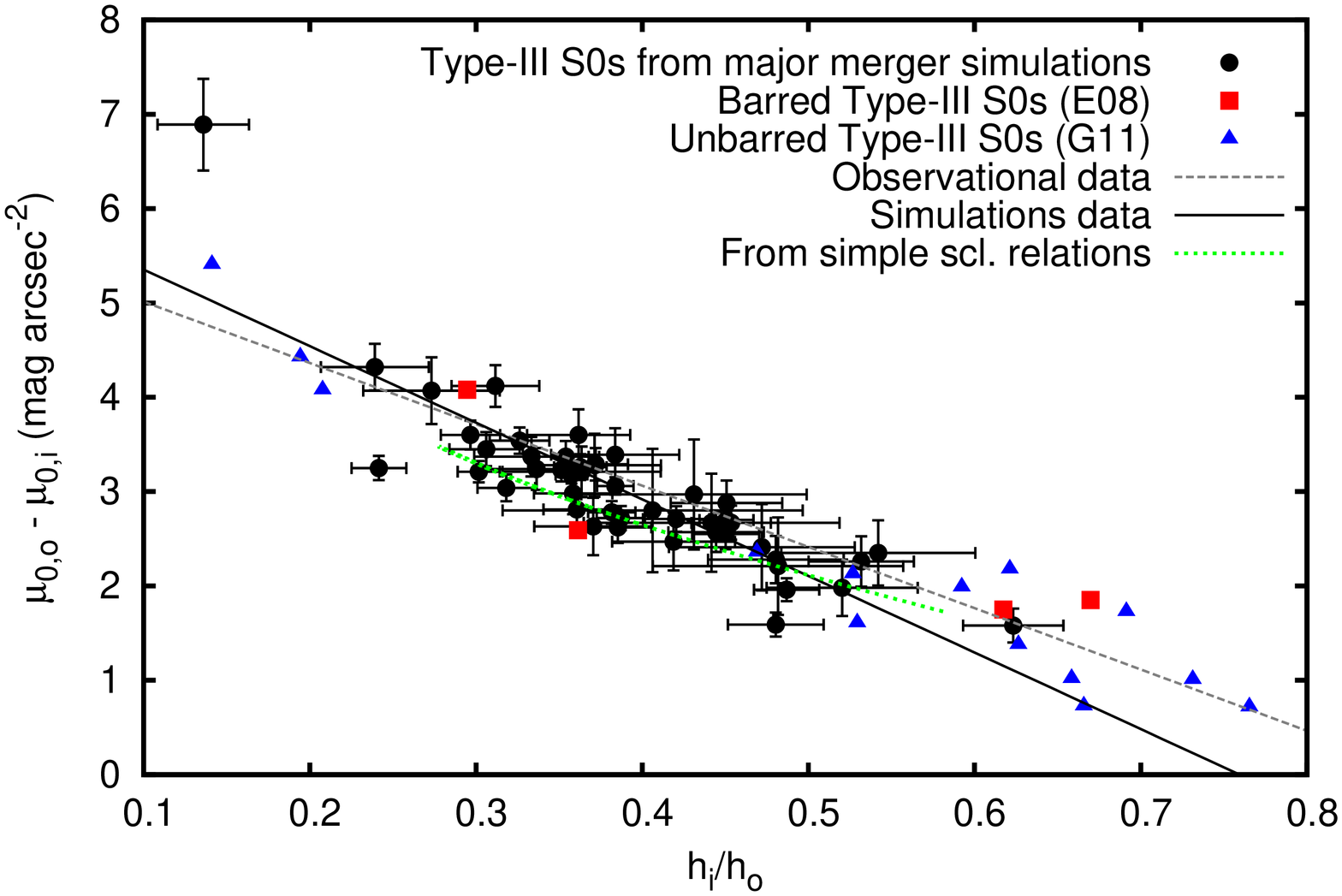}  
\caption{Photometric scaling relations in scale-free diagrams of the characteristic surface brightness of the inner and outer discs and breaks (referred to \mubreak) with $\hi/\ho$, for real and simulated Type III S0s. \emph{Solid line}: Linear fit to our simulations sample. \emph{Dashed line}: Linear fit to the observational data by E08 and G11. \emph{Dotted line}: Expected relations derived from the simple observational scaling relations $\hi \propto \rbreak$, $\ho \propto \rbreak$, and $\mubreak \propto \rbreak$ (see Sect.\,\ref{Sec:basic_scaling}). Consult the legend in the figure.  [\emph{A colour version of this figure is available in the online edition.}]
}
\label{fig:trends_hiho}
\end{figure}

In Fig.\,\ref{fig:trends_hiho}, we show the photometric relations of $(\mui - \mubreak)$, $(\muo - \mubreak)$, and $(\muo - \mui)$ vs. the ratio $\hi / \ho$ for our remnants and the reference observational data. Once we account for the different scales in length, mass, and luminosity of the simulated and real galaxies, they overlap in the photometric planes, as observed here again. The major merger simulations tightly reproduce the trends shown by observations in these planes, revealing the existence of strong scaling relations between the characteristic surface brightness of the breaks and of their inner and outer discs with $\hi / \ho$ in both real and simulated Type III S0s. The $(\mui - \mubreak)$ -- $\hi / \ho$ relation is weaker than the other two, for both observational and simulated data (see Table\,\ref{tab:fits_results}). 

Figure\,\ref{fig:trends_Rbrkho_hiho} represents the photometric planes $\hi/\ho$ -- $\rbreak/\hi$ and $\hi/\ho$ -- $\rbreak/\ho$  for the major merger simulations and the reference observational sample. The left panel indicates that $\hi/\ho$ decreases as the $\rbreak / \hi$ ratio increases in both real and simulated datasets, according to the observational results by \citet{2014arXiv1404.0559L} (see their Fig.\,10). The trend in this photometric plane is not significant, so there is no tight scaling relation in this diagram. However, the trend between $\hi/\ho$ and $\rbreak/\ho$ is better defined for both simulations and data and can be considered as a scaling relation (right panel). Moreover, major merger simulations fulfill the observational $\hi/\ho$ distribution in the zones where minor mergers by Y07 do not fall, and viceversa. 

Summarizing, we have found two important results: 1) that real Type III S0 galaxies present well-defined scaling relations between the photometric parameters of their breaks and their inner and outer discs, and 2) that the Type III S0-like remnants which result from major merger simulations reproduce these observational scaling relations. This means that the structures of the inner and outer discs in the antitruncation phenomena are tightly linked in both real and simulated galaxies, supporting major mergers as a feasible mechanism to explain the formation of Type III S0s.

\subsection{Influence of bars in the photometric relations}
\label{Sec:bars} 

We find no differences between the trends shown by the barred and unbarred Type III S0s of the observational samples (E08 and G11, respectively) in any diagrams studied in Sects.\,\ref{Sec:rbreak}--\ref{Sec:scaling}. This result supports that bars are structurally unrelated to antitruncations, as already argued by \citet{2009IAUS..254..173S} and G11 from the relative frequency of antitruncated discs in barred and unbarred S0s (see Sect.\,\ref{Sec:introduction}). Some of our major merger remnants have ovals and small inner bars, but none has developed a strong bar (see Paper I).

\begin{table*}
\caption{Linear fits performed to the trends in several photometric planes of real Type III S0s and the Type III S0-like remnants}
\label{tab:fits_results}
\begin{center}
{\footnotesize
\begin{tabular}{llccrrlccrr}
\toprule
 &  & \multicolumn{4}{c}{Observational data} & & \multicolumn{4}{c}{Major merger simulations}\\
\vspace{-0.3cm}\\\cline{3-6}\cline{8-11}\vspace{-0.3cm}\\
\multicolumn{1}{l}{No.}& \multicolumn{1}{l}{Photometric relation} & $m$ & $c_0$ & $\chi^{\mathrm{2}}$ & $\rho$ & & $m$ & $c_0$ & $\chi^{\mathrm{2}}$ & $\rho$\\
(1) & \multicolumn{1}{c}{(2)} & (3)&(4)&(5)&(6)& & (7)&(8)&(9)&(10)\\
\midrule
1& $\hi$ vs.\,$\rbreak$ &0.224$\pm$0.032 & 0.20$\pm$0.30 & 7.807 & 0.835 & & 0.207$\pm$0.022 & 0.61$\pm$0.73 & 64.654 & 0.818\\
2& $\ho$ vs.\,$\rbreak$ & 0.83$\pm$0.17 & -1.9$\pm$1.6 & 175.896 & 0.805 & & 0.805$\pm$0.090 & -6.1$\pm$3.0 & 1105.918  & 0.800\\
3& $\mui$ vs.\,$\rbreak$ & 0.195$\pm$0.052 & 17.87$\pm$0.48 & 20.340 & 0.634  & & 0.0879$\pm$0.0095 & 17.38$\pm$0.32 & 12.178 & 0.811\\
4& $\muo$ vs.\,$\rbreak$  & 0.270$\pm$0.061 & 19.49$\pm$0.59 & 20.368  & 0.742 & & 0.117$\pm$0.011& 19.44$\pm$0.37& 16.906 & 0.843 \\
5& $\mubreak$ vs.\,$\rbreak$  & 0.224$\pm$0.074 & 21.84$\pm$0.72 &  30.036 & 0.603 &  & 0.1150$\pm$0.0080  & 21.18$\pm$0.27 & 8.705 &  0.906\\
\vspace{-0.2cm}\\\hline\vspace{-0.2cm}\\
6& $\mui$ vs.\,$\hi$ & 0.94$\pm$0.14 & 17.55$\pm$0.33 & 11.074 & 0.821 & & 0.396$\pm$0.024 & 17.33$\pm$0.18 & 4.968 & 0.928\\
7& $\muo$ vs.\,$\ho$  & 0.261$\pm$0.051 & 20.45$\pm$0.36 & 17.10 & 0.79 & & 0.1225$\pm$0.0094 & 20.79$\pm$0.20 & 12.300 & 0.888\\
\vspace{-0.2cm}\\\hline\vspace{-0.2cm}\\
8& $\hi/\risoph$ vs.\,$\rbreak/\risoph$ & 0.226$\pm$0.035 & 0.021$\pm$0.043 & 0.112 & 0.813 & & 0.237$\pm$0.017 & -0.013$\pm$0.023 & 0.132 & 0.899 \\
9& $\ho/\risoph$ vs.\,$\rbreak/\risoph$ & 0.57$\pm$0.18 & 0.01$\pm$0.22 & 2.403 & 0.602 & & 0.812$\pm$0.069 & -0.240$\pm$0.093 & 2.108 & 0.870 \\
10& $\mui$ vs.\,$\rbreak/\risoph$ & 2.01$\pm$0.44 & 17.24$\pm$0.53 & 16.903 & 0.709 & & 1.63$\pm$0.14 & 18.15$\pm$0.19 & 9.045 & 0.864\\
11& $\muo$ vs.\,$\rbreak/\risoph$  & 2.41$\pm$0.56 & 19.11$\pm$0.69 & 21.133 & 0.731 & &  2.03$\pm$0.20 & 20.65$\pm$0.27 & 17.395 & 0.838\\
12& $\mubreak$ vs.\,$\rbreak/\risoph$  & 2.77$\pm$0.48 & 20.64$\pm$0.58 & 15.135 &  0.824& &  1.96$\pm$0.15 & 21.41$\pm$0.21 & 10.395  & 0.887\\
\vspace{-0.2cm}\\\hline\vspace{-0.2cm}\\
13 &  $\hi/\ho$ vs.\,$\rbreak / \hi$      & -0.094$\pm$0.039 & 0.90$\pm$0.17 & 0.522 & -0.496 & & -0.079$\pm$0.016 &0.016$\pm$0.075 & 0.236 & -0.583\\
14 &  $\hi/\ho$ vs.\,$\rbreak / \ho$      & 0.177$\pm$0.035 &  0.151$\pm$0.077 & 0.289 & 0.763 & & 0.223$\pm$0.021 &0.007$\pm$0.037 & 0.103 & 0.884\\
15 & $(\mui - \mubreak)$ vs.\,$\hi/\ho$     & 0.7$\pm$1.5 & -4.57$\pm$0.82 & 22.927 & 0.113 & & 4.13$\pm$0.96 & -6.29$\pm$0.38 & 14.822 & 0.540 \\
16 & $(\muo - \mubreak)$ vs.\,$\hi/\ho$   & -5.9$\pm$1.1 & 1.10$\pm$0.60 & 12.502 & -0.803 & & -3.99$\pm$0.40 & -0.12$\pm$0.16 & 2.592 & -0.829\\
17 &  $(\muo - \mui)$ vs.\,$\hi/\ho$     &  -6.53$\pm$0.54 & 5.68$\pm$0.30 & 3.092 & -0.949 & & -8.12$\pm$0.71 & 6.17$\pm$0.28 & 8.057 & -0.863\\
\bottomrule 
\end{tabular}
\begin{minipage}[t]{\textwidth}{\vspace{0.2cm}
\footnotesize
\emph{Columns}: (1) ID number of the photometric relation. (2) Fitted photometric relation. (3) Slope of the linear fit to the observational data (E08; G11). (4) Y-intercept of the linear fit to the observational data. (5) Pearson's cumulative test statistic for the observational linear fit. (6) Pearson product-moment correlation coefficient for the observational linear fit. (7) Slope of the linear fit to our major merger S0-like remnants. (8) Y-intercept of the linear fit to the major merger S0-like remnants. (9) Pearson's cumulative test statistic for the linear fit to the major merger simulations data. (10) Pearson product-moment correlation coefficient for the  major merger simulations data.}
\end{minipage}
}
\end{center}
\end{table*}

\subsection{Kolmogorov-Smirnov tests to simulations and real data}
\label{Sec:KS}

We have shown that the antitruncated S0-like remnants of major merger simulations follow analogous trends and exhibit similar values to real Type III S0s in several photometric planes. We have tested whether the similarity of the distributions of real and simulated data is statistically significant in some photometric planes, using the Kolmogorov-Smirnov (KS) test in two dimensions. We have centred on those diagrams where both data and simulations overlap, as well as in those where the characteristic surface brightness in the simulations are displaced from the observations just by an offset of $\sim 1$\,mag\,arcsec$^{-2}$. 

The KS test checks whether the two samples in each diagram can have been shown from the same parent distribution at a given significance level $\alpha$ (typically, $\alpha = 0.05$). We have used the implementation of the test made by P.\,Yoachim (\texttt{KS2D}, programmed in IDL), who has kindly made it publicly available in his web page\footnote{\texttt{KS2D} can be downloaded from: http://www.astro.washington.edu/ \-users/\-yoachim/ \-code.php}. The probability ($p$) of obtaining the observational and simulated samples considering that the null hypothesis is true (i.e. that they have been taken from the same parent distribution) are listed in Table\,\ref{tab:KS2D}. If $p> \alpha = 0.05$, the null hypothesis is accepted, meaning that the observational and simulated samples come from the same parent distribution at 95\% of confidence. Otherwise, the null hypothesis is rejected and then both samples are different at 95\% of confidence level.

In Sect.\,\ref{Sec:basic_scaling} we demonstrated that most parameters tightly correlate with \rbreak. In these planes, real and simulated data do not overlap, because the S0s in the observational sample have $\rbreak < 20$\,kpc and our remnants are larger ($\rbreak> 20$\,kpc typically). Therefore, no KS test has been carried out in the planes involving non-normalized scalelengths on one axis (such as \rbreak, \hi, or \ho), as we obviously expect $p< \alpha = 0.05$ in them. The KS test in the diagrams involving \mubreak, \mui, or \muo\ on one axis also rejects the null hypothesis at $\alpha = 0.05$, because the two distributions do not overlap due to the offset of $\sim 1$\,mag\,arcsec$^{-2}$ in surface brightness that we have found between real data and simulations. As this offset may be a consequence of the extreme youth of the starbursts induced by the merger and/or to the assumptions adopted to convert mass into light (see Sect.\,\ref{Sec:profiles} and Appendix\,\ref{Sec:limitations}), we have tested whether real and simulated Type III S0s would be similar in these diagrams if the S0-like remnants and the real S0s are overlapped by considering the difference between the $y$-intercepts of the linear fits to each sample in Table\,\ref{tab:fits_results}. 

This has been done for the planes  $\mui$, $\muo$, and $\mubreak$ vs.\, $\rbreak/\risoph$. The results are shown in Table\,\ref{tab:KS2D}. In the \mui\ and \muo\ planes, the KS test indicates that the real and simulated samples are similar at 95\% of confidence level. In the plane $\mubreak$ vs.\, $\rbreak/\risoph$, the $p$ value increases noticeably when this offset in surface brightness is accounted for, but not enough to consider both distributions as similar at a 95\% confidence level.

The majority of the relations involving $\rbreak/\risoph$ pass the test, indicating that the real and simulated samples can be considered as similar at a 95\% confidence level (Table\,\ref{tab:KS2D}). However, no photometric plane involving $\hi/\ho$ passes the KS test. This is because our remnants span the range of $\hi/\ho$ values observed in real data only partially, biassing the test (see, e.g. Fig.\,\ref{fig:trends_hiho_noscaling}). Our simulations span a limited region of the initial conditions space (Sect.\,\ref{Sec:generallimitations}), so it is normal that the remnants do not reproduce the whole observational range of all photometric parameters. In fact, it is already quite surprising that they span so widely. Moreover, if we include Y07 results in the KS test as part of our simulated sample, the $p$ value of real and simulated data in the planes $\hi/\ho$ -- $\rbreak/\hi$ and $\hi/\ho$ -- $\rbreak/\ho$ increases significantly, although insufficiently to consider the real and simulated samples to have been taken from the same parent distribution (see Table\,\ref{tab:KS2D}). This is because the minor merger models span a range of $\hi/\ho$ different to our major merger simulations(see Figs.\,\ref{fig:trends_hiho_noscaling} and \ref{fig:trends_Rbrkho_hiho}). 

We can thus conclude that the similarity of the real and simulated distributions of Type III S0s is statistically significant in all photometric planes where the two samples overlap visually or accounting for the offset in the characteristic surface brightness observed between them.

\begin{table}[h]
\begin{threeparttable}
\caption{Kolmogorov-Smirnov test to the 2D distributions of real Type III S0s data and of Type III S0-like remnants in several photometric planes related to the antitruncations}
\label{tab:KS2D}
{\footnotesize
\begin{tabular}{rllccc}
\toprule
No. & Photometric plane &  \multicolumn{2}{c}{$p$-value\tnote{a}} & KS test\tnote{b} &  \multicolumn{1}{c}{Fig.} \\
\midrule
%1 & $\mubreak$ vs.\,$\rbreak/\risoph$ & \multicolumn{2}{l}{0.0013} & NS & \ref{fig:mubreak_Rbrk} \\
1 & $\mubreak$ vs.\,$\rbreak/\risoph$ (displaced\tnote{c} ) & \multicolumn{2}{l}{0.021} & S & \ref{fig:mubreak_Rbrk} \\
\midrule
2 & $V_\mathrm{rot}$ vs.\,$\rbreak/\risoph$ & \multicolumn{2}{l}{0.068}  & S & \ref{fig:Rbrk_R25_vrot}  \\
\midrule
3 & $(\mui - \mubreak)$ vs.\,$\hi/\risoph$                 & \multicolumn{2}{l}{0.018}  & NS & \ref{fig:trends_hiR25_hoR25}  \\
4 & $(\mui - \mubreak)$ vs.\,$\ho/\risoph$                 & \multicolumn{2}{l}{0.0054}  & NS & \ref{fig:trends_hiR25_hoR25}  \\
5 & $(\muo - \mubreak)$ vs.\,$\ho/\risoph$                & \multicolumn{2}{l}{0.017} & NS &\ref{fig:trends_hiR25_hoR25}   \\
6 & $(\muo - \mubreak)$ vs.\,$\hi/\risoph$                & \multicolumn{2}{l}{0.035}   & NS &\ref{fig:trends_hiR25_hoR25} \\
\midrule
7 & $\hi/\ho$ vs.\,$\rbreak / \risoph$                      & \multicolumn{2}{l}{0.0013} & NS & \ref{fig:trends_hiho_noscaling}   \\
\midrule
8 & $(\mui - \mubreak)$ vs.\,$\hi/\rbreak$                 & \multicolumn{2}{l}{0.044}  & NS &  \ref{fig:trends_hiRbrk_hoRbrk}  \\
9 & $(\mui - \mubreak)$ vs.\,$\ho/\rbreak$                 & \multicolumn{2}{l}{0.0009}  & NS &  \ref{fig:trends_hiRbrk_hoRbrk}  \\
10 & $(\muo - \mubreak)$ vs.\,$\hi/\rbreak$                 & \multicolumn{2}{l}{0.0008}  &  NS & \ref{fig:trends_hiRbrk_hoRbrk}   \\
11 & $(\muo - \mubreak)$ vs.\,$\ho/\rbreak$                 & \multicolumn{2}{l}{0.023}  &  NS & \ref{fig:trends_hiRbrk_hoRbrk}   \\
12 & $(\mui - \mubreak)$ vs.\,$\rbreak/\risoph$                 & \multicolumn{2}{l}{0.026}  & NS & \ref{fig:trends_hiRbrk_hoRbrk} \  \\
13 & $(\muo - \mubreak)$ vs.\,$\rbreak/\risoph$                 & \multicolumn{2}{l}{0.0086}  & NS & \ref{fig:trends_hiRbrk_hoRbrk} \  \\
\midrule
14 & $\hi/\risoph$ vs.\,$\rbreak/\risoph$ &  \multicolumn{2}{l}{0.71} &  S &\ref{fig:trends_rbrkr25} \\
15 & $\ho/\risoph$ vs.\,$\rbreak/\risoph$ & \multicolumn{2}{l}{0.066} &  S &\ref{fig:trends_rbrkr25} \\
16 & $\mui$ vs.\,$\rbreak/\risoph$ (displaced\tnote{c} ) &  \multicolumn{2}{l}{0.23} & S & \ref{fig:trends_rbrkr25} \\
17 & $\muo$ vs.\,$\rbreak/\risoph$ (displaced\tnote{c} ) &  \multicolumn{2}{l}{0.67} & S & \ref{fig:trends_rbrkr25} \\
\midrule
18 & $(\mui - \mubreak)$ vs.\,$\hi/\ho$                 & \multicolumn{2}{l}{0.00048} &  NS &\ref{fig:trends_hiho}   \\
19 & $(\muo - \mubreak)$ vs.\,$\hi/\ho$                & \multicolumn{2}{l}{0.00073} &  NS &\ref{fig:trends_hiho}   \\
20 & $(\muo - \mui)$ vs.\,$\hi/\ho$                      & \multicolumn{2}{l}{0.00030}  &  NS &\ref{fig:trends_hiho}  \\
\midrule
21 & $\hi/\ho$ vs.\,$\rbreak / \hi$                      & \multicolumn{2}{l}{0.00079}  &  NS &\ref{fig:trends_Rbrkho_hiho}  \\
22 & $\hi/\ho$ vs.\,$\rbreak / \hi$ (with Y07)                     & \multicolumn{2}{l}{0.024}  & NS  &\ref{fig:trends_Rbrkho_hiho}  \\
23 & $\hi/\ho$ vs.\,$\rbreak / \ho$                      & \multicolumn{2}{l}{0.0012}  &  NS &\ref{fig:trends_Rbrkho_hiho}  \\
24 & $\hi/\ho$ vs.\,$\rbreak / \ho$ (with Y07)                     & \multicolumn{2}{l}{0.030}  & NS  &\ref{fig:trends_Rbrkho_hiho}  \\
\bottomrule \vspace{0.1cm}
\end{tabular}
\begin{minipage}{0.48\textwidth}
\footnotesize
\begin{itemize}
 \item[a] The $p$-values have been computed using the implementation of the KS test in 2D by P.\,Yoachim (\texttt{KS2D}, see the text).
\item[b] According to the KS test, the distributions of observational data and of simulations are similar (S) in a 95\% confidence level if $p > \alpha = 0.05$. If not, the null hypotesis is rejected, meaning that both samples have been taken from different parent distributions and cannot be considered as similar (NS). 
\item[c] Displaced relations assume an offset in mag\,arcsec$^{-2}$ to overlap the distributions of real and simulated data in the photometric plane. The offset has been set to the difference between the $y$-intercepts of the linear fits performed to each sample in the plane (see Table\,\ref{tab:fits_results}).
\end{itemize}
\end{minipage}
}
\end{threeparttable}
\end{table}

\subsection{Modelling from basic photometric scaling relations}
\label{Sec:basic_scaling} 

In this section, we show that these relations, as well as the distributions in all photometric planes studied in Sects.\,\ref{Sec:photplanes} -- \ref{Sec:scalefreeplanes}, can be derived from eq.\,\ref{eq:Freemanmu} considering only three simple scaling relations derived from observations as basis: $\hi \propto \rbreak$ (Fig.\,\ref{fig:trends_rbrk}), $\ho \propto \rbreak$ (Fig.\,\ref{fig:trends_rbrk}), and $\mubreak \propto \rbreak$ (Fig.\,\ref{fig:mubreak_Rbrk}). 

We have assumed the values of the constants obtained from the linear fits performed to these relations using observational data (see Table\,\ref{tab:fits_results}), because of the $\sim 1$\,mag\,arcsec$^{-2}$ offset of the simulations with respect to observations (see Sect.\,\ref{Sec:profiles} and Appendix\,\ref{Sec:limitations}). Therefore, the inner and outer discs of real antitruncated S0s fulfill the simple scaling relations,

\begin{eqnarray}
 \hi = 0.22\,\rbreak + 0.20,     \label{eq:scalingrelation1} \\
  \ho = 0.83\, \rbreak -1.9, \label{eq:scalingrelation2}\\
  \mubreak =  0.22\,\rbreak + 21.84, \label{eq:scalingrelation3}
\end{eqnarray}

\noindent where \rbreak\ is in kpc and \mubreak\ is in mag\,arcsec$^{-2}$, both for the $R$ band. These simple scaling relations have been highlighted in their corresponding diagrams over the other scaling relations by using red lines. The Type III S0-like remnants resulting from our major merger simulations show analogous trends, except for the offset in \mubreak\ widely commented in previous sections (see Table\,\ref{tab:fits_results}). We have selected the scaling relation in eq.\,\ref{eq:scalingrelation3} instead of $\mui \propto \rbreak$ or $\muo \propto \rbreak$ (see Fig.\,\ref{fig:trends_rbrk}) to avoid considering a region of the disc as privileged. 

Additionally, the photometric parameters of the inner and outer discs relate to those of the break through eq.\,\ref{eq:Freemanmu}, because \rbreak\ and \mubreak\ are defined at the radius in the galaxy where the linear fits performed to both discs cross. Therefore, the photometric parameters are also connected through the following two relations:

\begin{eqnarray}
  \mubreak = \mui  + \frac{2.5}{\ln 10}\frac{\rbreak}{\hi}, \label{eq:inner_disc}\\
  \mubreak = \muo  + \frac{2.5}{\ln 10}\frac{\rbreak}{\ho}. \label{eq:outer_disc}
\end{eqnarray}

Thus, we have six unknown parameters (\hi, \mui, \ho, \muo, \rbreak, and \mubreak) related through five equations (eqs.\,\ref{eq:scalingrelation1} -- \ref{eq:outer_disc}). Considering \rbreak\ as the free parameter, we can estimate the expected values of the other five parameters just by solving the previous system of equations. 

We have derived the theoretical trends expected from eqs.\,\ref{eq:scalingrelation1} -- \ref{eq:outer_disc} in all the photometric planes analysed in this study. They are overplotted in each photometric plane shown in Sects.\,\ref{Sec:photplanes} -- \ref{Sec:scalefreeplanes} with green dotted lines. The fit performed to the relation $\ho \propto \rbreak$ shown by real data implies $\ho<0$ for $\rbreak \lesssim 5$\,kpc (eq.\,\ref{eq:scalingrelation2}), so we have limited the predictions in all planes to $\rbreak>5$\,kpc to avoid singularities of the equations or unrealistic predicted values of the photometric parameters, up to $\rbreak = 100$\,kpc. In the photometric planes where the scalelengths are normalized by \risoph, we have derived it by estimating $R_{25}(B)$ first. We have approximated $\risoph(R)$ by the radius at which the  fit to the inner disc gets $\mu(R) = 25$\,mag\,arcsec$^{-2}$ in the case that $\mubreak>25$\,mag\,arcsec$^{-2}$. In case that $\mubreak\leq 25$\,mag\,arcsec$^{-2}$, the outer disc profile has been used instead. Finally, we have transformed $\risoph (R)$ into \risoph\ (defined in the $B$ band) by using a scale factor equal to $0.77$, which corresponds to the median of the $\risoph(B)/\risoph(R)$ ratios exhibited in our models.

\begin{figure*}[th!]
\center
\includegraphics[angle=270, width = 0.49\textwidth, bb = 60 50 540 750, clip]{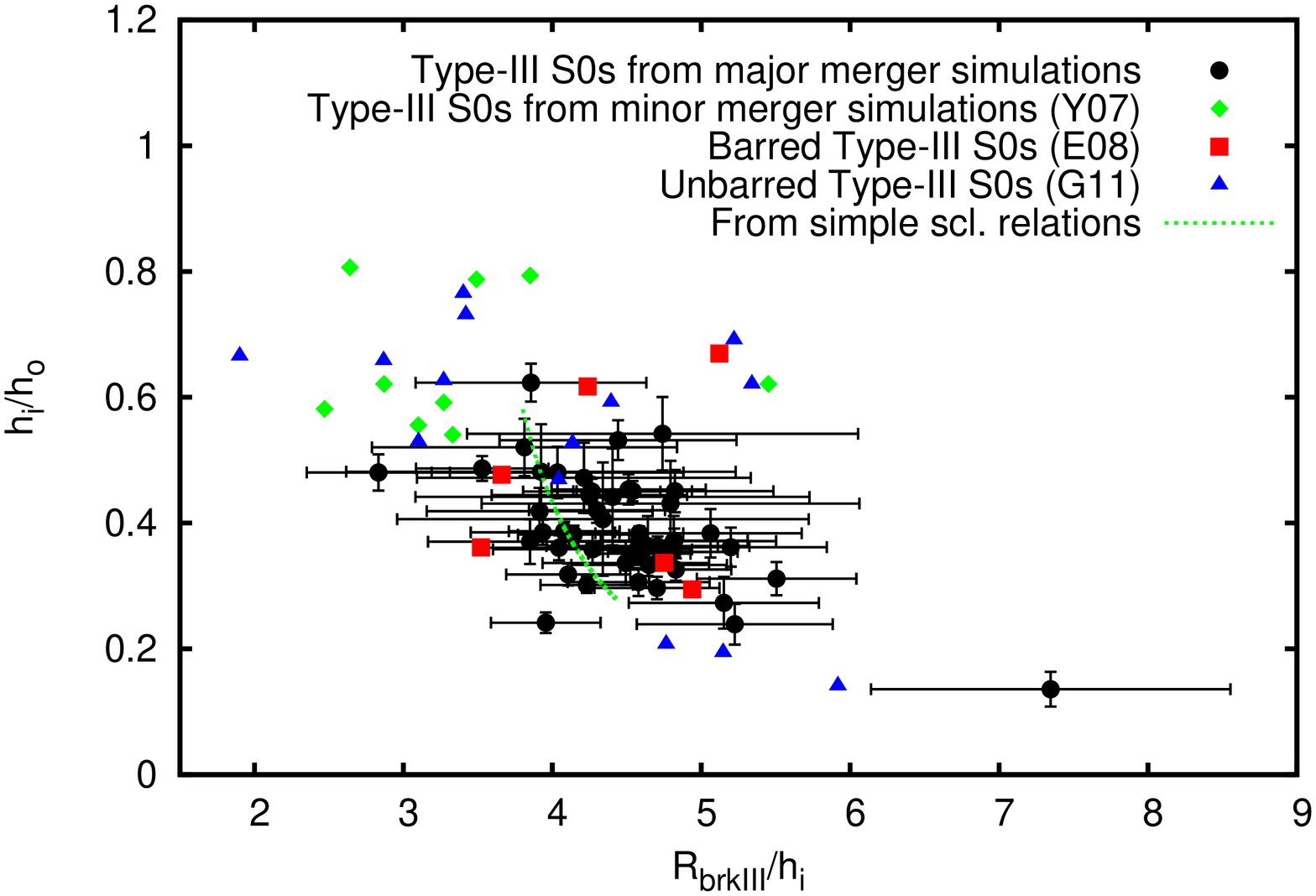}  
\includegraphics[angle=270, width = 0.49\textwidth, bb = 60 50 540 750, clip]{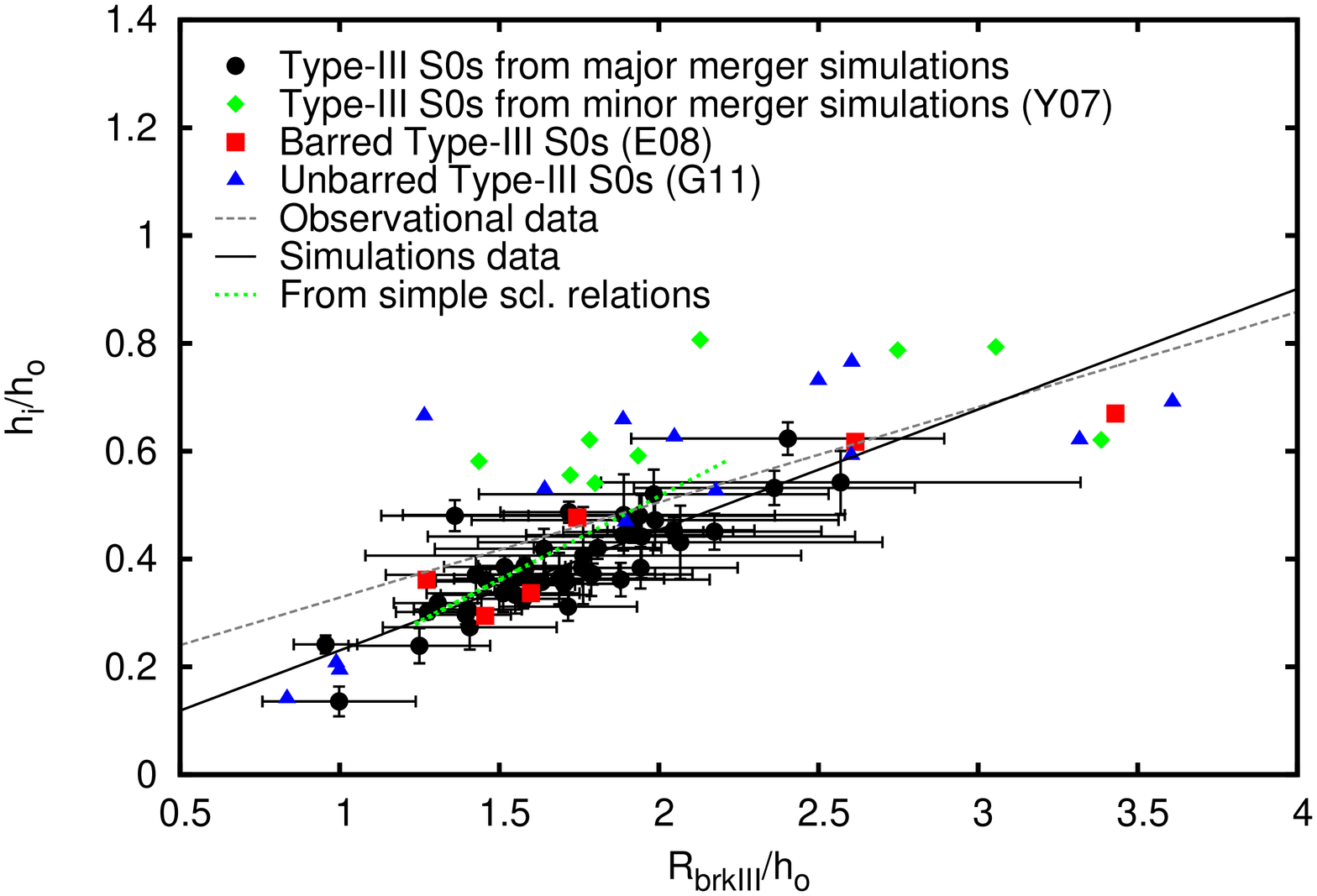}  
\caption{Photometric scaling relations $\hi/\ho$ -- $\rbreak/\hi$ and $\hi/\ho$ -- $\rbreak/\ho$ for our Type III S0-like remnants and the antitruncated S0s in the reference observational sample (E08; G11). Data of the antitruncations formed in the minor merger simulations by Y07 are overplotted for comparison. \emph{Solid line}: Linear fit to our simulations sample. \emph{Dashed line}: Linear fit to the observational data by E08 and G11. \emph{Dotted line}: Expected relations derived from the simple observational scaling relations $\hi \propto \rbreak$, $\ho \propto \rbreak$, and $\mubreak \propto \rbreak$ (see Sect.\,\ref{Sec:basic_scaling}). Consult the legend in the figure.  [\emph{A colour version of this figure is available in the online edition.}]
}
\label{fig:trends_Rbrkho_hiho}
\end{figure*}

Figures\,\ref{fig:mubreak_Rbrk} -- \ref{fig:trends_Rbrkho_hiho} demonstrate that the trends in each photometric plane expected from these three simple scaling relations reproduce those obtained in real and simulated data very well (in particular, see Figs.\,\ref{fig:trends_hiR25_hoR25} -- \ref{fig:trends_Rbrkho_hiho}). In some diagrams, the trends expected from the three simple scaling relations are even more defined than those shown by real Type III S0s and by our antitruncated S0-like remnants (see, e.g. Figs.\,\ref{fig:trends_hiR25_hoR25} -- \ref{fig:trends_hiRbrk_hoRbrk}). 

\begin{figure*}[ht!]
\center
\includegraphics[width = 0.87\textwidth]{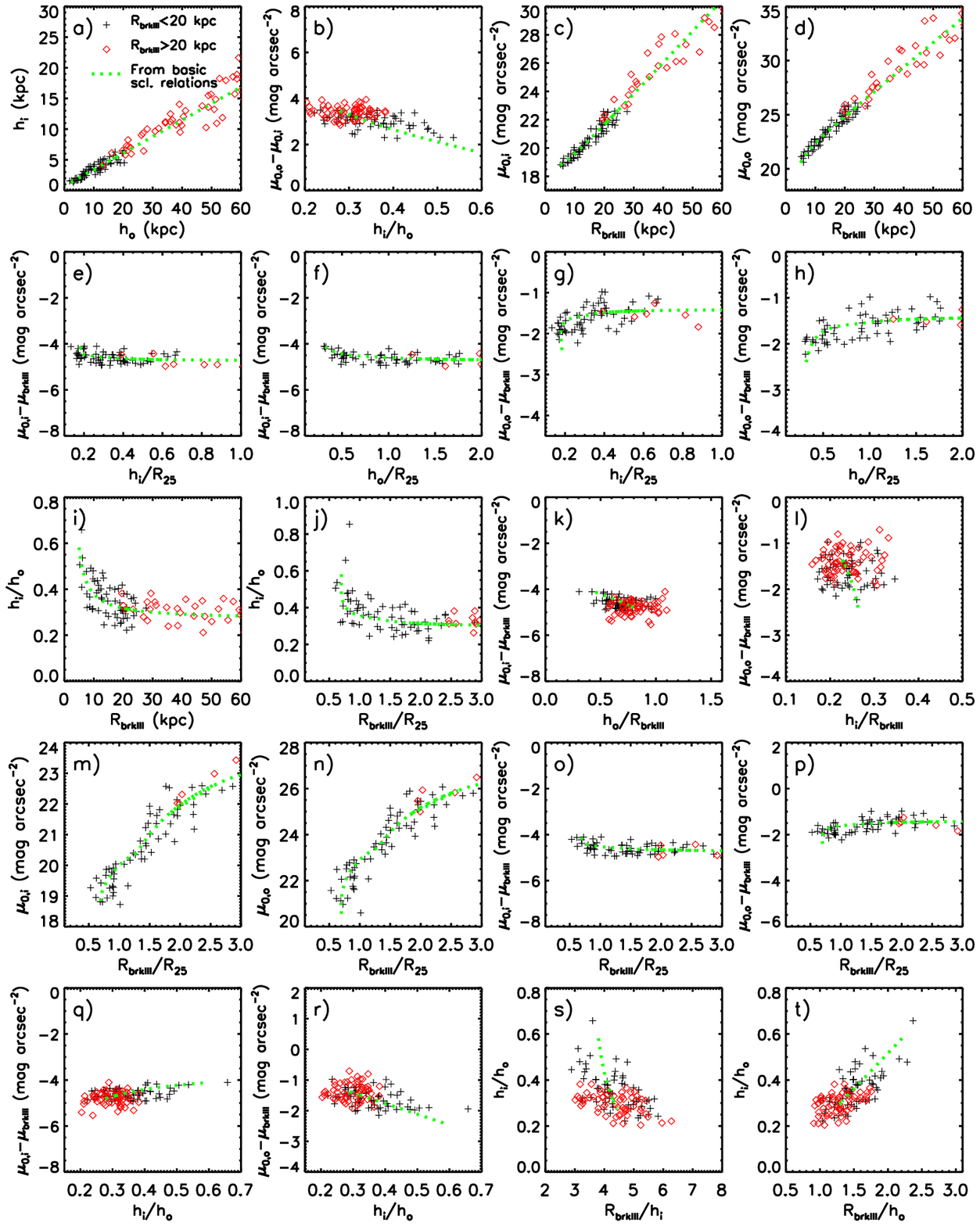}   % , bb = 60 50 540 750, clip
\caption{Trends expected in several photometric planes from the modelling obtained from three basic scaling relations fulfilled by real and simulated Type III S0s  (\emph{dashed lines}): $\hi \propto \rbreak$, $\ho \propto \rbreak$, and $\mubreak \propto \rbreak$. The observational relations have been assumed for the modelling (see Sect.\,\ref{Sec:basic_scaling}). We have overplotted the locations expected from this modelling for 60 values of \rbreak, but inserting typical observational errors in the scalelengths and surface brightness values at random (\emph{symbols}). We have distinguished the predictions with $\rbreak < 20$\,kpc (\emph{crosses}) from those with $\rbreak>20$\,kpc (\emph{diamonds}), to compare them with the distributions of the observational data (typically with  $\rbreak < 20$\,kpc) and of our S0-like remnants (with $\rbreak>20$\,kpc) in the figures above. Accounting for errors, the modelling is capable of reproducing the distributions of real data in these planes (compare with the corresponding diagrams in Figs.\,\ref{fig:mubreak_Rbrk} -- \ref{fig:trends_Rbrkho_hiho}).  [\emph{A colour version of this figure is available in the online edition.}]}
\label{fig:trends_expected}
\end{figure*}

Moreover, the dispersion of real and simulated data in these diagrams with respect to the expected relations is basically due to the observational uncertainties. To show this, we have simulated the distribution in these planes of the relations expected from these simple scaling relations, but inserting random errors in the estimates spanning the typical observational ranges (errors of up to $\sim 1$\% for the characteristic surface brightness parameters, and up to $\sim 20$\% for the scalelengths and \rbreak). The results of this simulation of the effects of observational errors in the predictions of this modelling for the majority of the photometric planes studied are plotted all together in Fig.\,\ref{fig:trends_expected}. The theoretical trends without accounting for observational errors have also been overplotted for comparison.

Figure\,\ref{fig:trends_expected} demonstrates that the distribution of real and simulated data, as well as their trends, can be explained in all the photometric planes just considering these three simple scaling relations and accounting for the typical observational errors. The tight theoretical trends derived from eqs.\,\ref{eq:scalingrelation1} -- \ref{eq:outer_disc} disperse in the planes in a similar way to observations and simulations if we just account for typical observational errors (compare each panel in Fig.\,\ref{fig:trends_expected} with the corresponding photometric plane in the figures above). We can even explain the existence of forbidden regions in the planes for real S0s and our remnants, which correspond to the singularities of the system of equations above or to unphysical values of the parameters (as, e.g. negative scalelengths). 

In Fig.\,\ref{fig:trends_expected}, we have represented with different symbols the values expected for $\rbreak \leq 20$\,kpc and for $\rbreak > 20$\,kpc. The former range represents the typical range of \rbreak\ values exhibited by real S0s in the samples by E08 and G11, whereas the latter one is the range covered by our major merger simulations. We have compared the distributions expected from the simple scaling relations in each range of \rbreak\ with those exhibited by real S0s and our antitruncated S0-like remnants in the corresponding photometric planes above. We find that the distributions of real S0s and of Type III S0-like remnants in the planes that do not involve \risoph\ are well reproduced by the expectations of the modelling performed for $\rbreak < 20$ and $\rbreak> 20$\,kpc, respectively. However, real and simulated data overlap in some photometric planes involving \risoph\ (such as $\hi / \ho$ vs.\,$\rbreak / \risoph$, see the right panel of Fig.\,\ref{fig:trends_hiho_noscaling}), whereas the modelled data for $\rbreak < 20$ and $\rbreak> 20$\,kpc from the basic scaling relations are displaced in it (see panel j in Fig.\,\ref{fig:trends_expected}). This may be because of the approximations performed to estimate \risoph\ in the modelling.

In order to test this, we have performed a similar modelling to the one exposed here by replacing the basic scaling relation in eq.\,\ref{eq:scalingrelation3} by the observational $\mubreak \propto \rbreak/\risoph$ relation (middle panel of Fig.\,\ref{fig:mubreak_Rbrk}). This procedure avoided the approximation of $\risoph(R)$ by the profile of the inner or the outer disc since the beginning of the formulation. We found that, with this change, the modelling reproduced the mix of observations and simulations in the planes involving \risoph\ with the new predictions for $\rbreak < 20$ and $\rbreak> 20$\,kpc, but that it did not reproduce the observed \mui\ -- \rbreak, \muo\ -- \rbreak, and \mubreak\ -- \rbreak\ relations. Therefore, although it seems that the approximations performed to estimate \risoph\ are really after the inappropriate mixing of the predictions in some photometric planes, the modelling works better with the \mubreak\ -- \rbreak\ trend than with the \mubreak\ -- $\rbreak/\risoph$ relation. 

Summarizing, the scaling relations, trends, and distributions of real and simulated Type III S0s in the photometric planes involving the characteristic parameters of the breaks and of their inner and outer discs can be predicted just accounting for three basic scaling relations derived for observations (which are also reproduced by our simulations): $\hi \propto \rbreak$, $\ho \propto \rbreak$, and $\mubreak \propto \rbreak$.

\section{Discussion}
\label{Sec:discussion}

Any formation scenario of antitruncated discs must be capable of explaining the following observational results (see Sect.\,\ref{Sec:introduction}): 1) why S0 galaxies present the highest percentage of antitruncated profiles of all Hubble types, and 2) why the relative frequency of these features seems to be independent of the environment. 

Concerning the first issue, it is known both observationally and theoretically that mergers tend to reduce the gas level in a galaxy by inducing starbursts. In fact, major events tend to exhaust it completely for typical gas contents in the progenitors, typically giving rise to a gas-poor remnant \citep[see, e.g.][]{2007A&A...468...61D, 2008A&A...492...31D,2008MNRAS.384..386C,2010MNRAS.404..590L,2011ApJ...730....4B,2011MNRAS.411.2148K,2014arXiv1402.1166K,2014MNRAS.437L..41K,2014arXiv1403.1817M}. The more intense the  merger history of a galaxy is, the more gas-poor it becomes. So, attending to this and the high probability of major mergers to form antitruncations (as indicated by the present study), a major merger scenario would be consistent with the higher percentage of Type III profiles in S0s than in other types, because this mechanism would explain the formation of the antitruncated stellar disc and the assembly of the host S0 galaxy simultaneously. 

The question on the role of environment in the formation of antitruncations is more puzzling. As commented before, \citet{2012ApJ...744L..11E} found that the relative presence of antitruncated S0 galaxies was similar in the Virgo cluster and the field, in contrast with the complete absence of Type-II S0 galaxies in Virgo. These results suggest the existence of environmental processes that prevent Type-II profiles in S0s from forming or that dissipate truncations in the clusters. However, these mechanisms do not affect the fraction of S0s harbouring antitruncated discs. This might be reconciled with the merger scenario if the cluster and group-field environments presented similar relative merger rates. Although it is controversial whether the density of the environment enhances the merger rate or not \citep{2012ApJ...754...26J}. 

The role of mergers in the shaping of S0s in clusters may be underestimated at present. Many studies support other environmental-driven processes (such as gas stripping and strangulation) as the most important drivers of the observed transformation of spirals into S0s in clusters since  $z\sim 0.8$ \citep[see][]{2001ApJ...563..118P,2007A&A...470..173B,2007ApJ...660.1151D,2009ApJ...693..112P}. However, recent observational and theoretical studies suggest that cluster galaxies have undergone a ''pre-processing'' through mergers in groups and filaments before falling into a cluster, and that even the gravitational binding of the group is preserved during the infalling for up to one orbital period (this is known as galaxy ''post-processing''). Besides this, the galaxy-galaxy collisions are enhanced during the pericentre passage of the group in the cluster \citep[][]{2013MNRAS.435.2713V}. Thus, all these mergers and interactions must have been complemented with other evolutionary processes in the clusters to evolve spirals into S0s in the last $\sim 7$--8\,Gyr, and it would be difficult to disentangle the contribution of each process to their buildup. 

Mergers, galaxy harassment, and tidal interactions may have triggered an even more dramatic evolution of spirals into S0s in the field during the same time period \citep[always within groups, see][]{2007ApJ...671.1503M,2009ApJ...692..298W,2011MNRAS.415.1783B}. Considering that $\sim 50$\% of S0s reside in groups \citep{1982ApJ...257..423H,2006ApJS..167....1B,2007ApJ...655..790C,2009ApJ...692..298W}, the relative relevance of mergers in the assembly of S0s may have been similar in both clusters and in the field. This may explain the similar relative percentages of Type III S0s found in both environments. Some observational and theoretical studies support this scenario, suggesting that E-S0 galaxies have undergone at least one major merger during the last $\sim 9$\,Gyr regardless of the environment \citep{2010A&A...519A..55E,2013MNRAS.428..999P}. 

Moreover, the efficiency of the mechanism that produces antitruncations in discs may be affected by gas-removal processes, which take place in both low- and high-density environments (such as ram pressure stripping or thermal evaporation), or even by preferential orbits within the clusters \citep{2008ApJ...672L.103K,2013MNRAS.435.2713V}. This means that the evolutionary scenario is so complex that it will be extremely difficult to quantify the relative role played by mergers in the buildup of Type III S0s in any sort of environment. 

Minor mergers alone seem not to be enough to explain the properties of Type III S0s. Although Y07 showed that they are a plausible mechanism to produce Type III profiles from pure-exponential S0 galaxies, it has not been analysed if these breaks would be detectable taking into account the limitations inherent to observational data and if they really are analogous to those observed in Type III galaxies. We have compared Y07 results with ours when possible, overriding the fact that these authors directly measure antitruncations in mass density profiles (see Figs.\,\ref{fig:trends_rbrk}, \ref{fig:trends_hiho_noscaling}, and \ref{fig:trends_Rbrkho_hiho}). The distributions of minor merger simulations are complementary to those of major mergers in these diagrams, so both sorts of mechanisms would be required to explain these observations. 

The kinematic properties of nearby ellipticals and S0s suggest that major merging may have been relevant for establishing the current dynamical state of $\sim 40$ -- 50\% of present-day S0s at maximum \citep{2014A&A...565A..31T}. Some properties of S0s seem to be more coherent with a buildup driven by minor mergers than by major encounters too \citep[see][]{2006A&A...457...91E,2011A&A...533A.104E,2012A&A...547A..48E,2013A&A...552A..67E}. Therefore, a hybrid scenario in which both major and minor merger events have contributed noticeably to form antitruncated S0s may be more coherent with observations. Nevertheless, the relative fractions of antitruncations in S0 and spiral galaxies, their relation with the galaxy subcomponents, and their dependence on the environment require more observational efforts to discern between different formation scenarios \citep{2012ApJ...759...98C,2012MNRAS.420.2475M,2014arXiv1404.0559L}.

The tight scaling relations that we have found between the photometric parameters of the antitruncations and of the inner and outer discs in real S0s indicate that the physical process after their formation is highly scalable. This scalability ensures that the results obtained in this study can be extrapolated to any mass range, and has allowed a fair comparison with observational S0s that are $\sim 2$ times smaller and fainter than the Type III S0 remnants of our simulations. The inner and outer disc structures in antitruncated S0s are so connected that the values of \mui, \muo, \mubreak, \hi, and \ho\ can be derived just by knowing \rbreak. More significantly, we have found that major mergers are capable of reproducing the distribution of real Type III S0s in all photometric planes. 

\section{Conclusions} 
\label{Sec:conclusions}

We have studied whether major mergers can produce S0-like remnants with realistic antitruncated stellar discs or not. We have analysed 67 E/S0 and S0 relaxed remnants that result from the major merger simulations available in the GalMer database \citep{2010A&A...518A..61C}. In order to ensure a fair comparison with observational data, we have simulated realistic $R$-band surface brightness profiles of the S0-like remnants reproducing the observing conditions of the studies by E08 and G11. We have visually identified the final remnants with antitruncated stellar discs and fitted exponential profiles to their inner and outer discs to obtain their characteristic photometric parameters, as well as define the breaks. The results have been compared with observations in several photometric planes. 

We have found that $\sim 70$\% of the relaxed S0-like remnants in our sample of simulations have clear antitruncated stellar discs. The photometric parameters of the breaks and of the inner and outer discs are compatible with those observed in real Type III S0 galaxies. Our remnants lie on top of the extrapolations of the trends shown by real data towards brighter magnitudes and higher break radii, because of the higher luminosities and sizes of the simulated remnants compared to the data in the reference observational samples. However, simulations and observations overlap completely in scale-free photometric diagrams, satisfying similar trends. 

We report the existence of strong photometric scaling relations between the parameters of the breaks and of their inner and outer discs in real antitruncated S0s, which have not been reported in previous studies. These scaling relations imply that the structures of the inner and outer discs in antitruncated S0s are strongly linked. We also find that the scaling relations, trends, and distributions of real Type III S0s in all photometric planes can be predicted from just three basic scaling relations: $\hi \propto \rbreak$, $\ho \propto \rbreak$, and $\mubreak \propto \rbreak$. The Type III S0-like remnants resulting from our major merger simulations reproduce the distributions, trends, and scaling relations of real antitruncated S0s in all photometric planes analysed. 

The tight scaling relations found between the photometric parameters of the breaks and of the inner and outer discs in both real and simulated S0s indicate that the physical mechanism after their formation is highly scalable, ensuring that the results obtained here for our remnants (with stellar masses of $\sim 1$ -- $3\times 10 ^{11}\Msun$) can be extrapolated to other mass ranges. This scalability has also allowed a direct comparison with observational S0s that are $\sim 2$ times smaller and fainter than the S0 remnants. The offsets of $\sim 1$\,mag\,arcsec$^{-2}$ in the characteristic surface brightness values and of $\sim 1$ -- 2\,mag in total absolute magnitudes in several bands can be explained in terms of the assumptions performed to convert mass into light in the simulations (ages of old stellar particles, SFHs, IMFs, dust extinction) and the blue colours of the remnants due to the recent starbursts induced by the encounters. 

We also find that the existence of bars in real Type III S0s does not affect to the photometric trends of the antitruncations. This supports that bars and antitruncations are structurally unrelated phenomena, corroborating the claims derived by previous studies on the basis of relative percentages of antitruncations in barred and unbarred S0s. 

These results demonstrate that, contrary to the prevalent view, major mergers are a feasible mechanism to produce realistic antitruncated stellar discs in the case of resulting into S0-like remnants. The agreement in the photometric trends and scaling relations followed by real Type III S0s and by the antitruncated S0-like remnants that result from these major merger simulations strongly supports a major-merger scenario for the buildup of many antitruncated S0s.

%
% Do not delete the next line
\small  % Do not delete
%
%%% Comment the following line if you do not have acknowledgments.
\begin{acknowledgements}   % Do not delete if you declare acknowledgments

The authors thank to the anonymous referee for the provided input that helped to improve this publication significantly. The current study would not have been possible without the GalMer database, so the authors would like to acknowledge I.\,Chilingarian, P.\,Di Matteo, F.\,Combes, A.-L.\,Melchior, and B.\,Semelin for creating it, and specially P.\,Di Matteo for her kind support on the models when requested. We really appreciate the helpful discussions with Raffaella Marino (UCM) on the dust extinction of early-type galaxies, as well as the \texttt{KS2D} code, made publicly available by his author, Peter Yoachim (UW). We acknowledge the usage of the HyperLeda database (http://leda.univ-lyon1.fr). This research has made use of the NASA's Astrophysics Data System and NASA/IPAC Extragalactic Database (NED) which is operated by the Jet Propulsion Laboratory, California Institute of Technology, under contract with the National Aeronautics and Space Administration. 

\newline
\newline

Supported by the Spanish Ministry of Economy and Competitiveness (MINECO) under projects AYA2006-12955, AYA2009-10368, AYA2012-30717, AYA2012-31277, and AYA-67625-CO2-01 from the Spanish Ministry of Science, P3/86 of the Instituto de Astrof\'{\i}sica de Canarias and by de Madrid Regional Government through the AstroMadrid Project (CAM S2009/ESP-1496, http://laeff.cab.inta-csic.es/\-proyects/\-astromadrid/\-main/\-index.php). Funded by the Spanish MICINN under the Consolider-Ingenio 2010 Program grant CSD2006-0070: "First Science with the GTC" (http://www.iac.es/\-consolider-ingenio-gtc/), and by the Spanish programme of International Campus of Excellence Moncloa (CEI). 
\end{acknowledgements}

{\normalsize

\begin{appendix}

\section{Limitations of the models}
\label{Sec:limitations}

Tthe models present some limitations that affect to the comparision with real S0s. Here we discuss them.    

\subsection{Limitations due to the initial conditions}
\label{Sec:generallimitations}

The models sample a limited region of the initial conditions space (Sect.\,\ref{Sec:galmer}), so we would not expect the remnants to reproduce the whole distribution of data in the different diagrams. However, the scalable properties of the antitruncations exhibited by observations and by major merger simulations make our remnants reproduce the location and trends of real data in scale-free photometric planes (see Sect.\,\ref{Sec:scaling}). This ensures that the global results obtained here with these major merger simulations can be extrapolated to other mass ranges.

\emph{i. Mass ratios.}- The GalMer database fixes the mass ratio for each pair of progenitors, according to its morphological type (see Table\,\ref{tab:massratios}). Although the mass ratio range is wide enough (from 1:1 to 3:1), it is fixed to a given value for each pair, so we cannot analyse the effects of the different mass ratios on the properties of the final breaks for identical progenitors.

\emph{ii. Progenitor masses.}- The progenitor galaxies present fixed masses too, of $\sim 0.5$ -- $1.5\times10^{11}\Msun$ (see Table~\ref{tab:morph_params}. The final stellar masses of our remnants ($\sim 1$ -- $3\times10^{11}\Msun$) may be quite similar to the masses of the brightest galaxies in E08 and G11 samples. However, they are much brighter (see the bottom panel of Fig.\,\ref{fig:mubreak_Rbrk}), probably as a result of the youth of the recent stellar populations formed in the remnants compared to those in real nearby S0s and to the assumptions performed to obtain the $M/L$ ratios to convert mass into light in the simulations (Sect.\,\ref{Sec:profiles}). This inserts an offset in surface brightness in all the remnants that makes \mubreak, \mui, and \muo\ brighter by $\sim 1$\,mag\,arcsec$^{-2}$ than their observational counterparts (i.e. the brightest galaxies in the reference observational samples).  Moreover, our remnants are $\sim 2$ times larger on average than the galaxies in E08 and G11 samples. This is why, when we compare data and simulations in non-normalized photometric planes, the observations lie on the extrapolated trends of real data, but towards higher \rbreak\ and brighter magnitudes than observed, and offset from them by $\sim 1$\,mag\,arcsec$^{-2}$ in characteristic values of the surface brightness profile.

However, as seen in Sect.~\ref{Sec:results}, the observational data and the simulations overlap in all photometric planes once we consider normalized parameters. This clearly indicates that the underlying physical mechanism after the formation of the antitruncations is highly scalable, allowing the extrapolation of the results obtained in our major merger simulations to different mass ranges.

\emph{iii. Initial gas content.}- The models assume initial gas masses similar to the typical contents observed in local galaxies ($\sim10\%$ in gSa, $\sim20\%$ in gSb, $\sim30\%$ in gSd of the total stellar mass, see Table\,\ref{tab:morph_params}). Some authors indicate that the typical gas content for spirals was higher in the past \citep[up to $50\%$ at $z\sim1$, see][]{2005ApJ...631..101P, 2008ApJ...687...59G, 2008ApJ...680..246T, 2009ApJ...706.1364F, 2009ApJ...697.2057L}. If the bulk of S0 galaxies was assembled through major mergers of spirals at $z \sim 0.7 - 1$ (as many theoretical and observational studies support, see references in Sect.\,\ref{Sec:introduction}), the higher gas masses involved in the encounters may have affected not only the generation of antitruncations in the discs, but also the average properties of the stellar populations in the remnant discs.

\emph{iv. Orbits.}- The GalMer database surveys a wide set of orbital configurations, but these orbits have not been taken from cosmological simulations. In fact, many models do not merge during the simulation time. We have found that S0-like remnants can result from quite a variety of orbits, although they preferentially result from co-planar ones. It is known that the survival of a disc component in a major merger does not depend exclusively on the orbital parameters, but is a complex function of several conditions, mainly depending on the initial gas content \citep{2009ApJ...691.1168H,2009MNRAS.397..802H}. The fact that S0-like remnants tend to result from orbits with low inclination is not a problem, because cosmological models favour co-planar mergers \citep{2005ApJ...629..219Z,2010hsa5.conf..295G,2011hsa6.conf..148B}. This means that the formation of an S0 remnant in a major merger may be a process with a much higher probability than previously thought. In fact, many authors claim that the major mergers occurring at $z>0.5$ may have built up galaxies of even later Hubble types than S0 \citep{2009A&A...501..437Y, 2009A&A...496..381H,2010ApJ...725..542H,2013MNRAS.431.3543H, 2009A&A...493..899P}. 

\emph{v. Total simulation time.}- The available models are evolved for 3.5 \,Gyr at the most. Therefore, some experiments have not reached the full-merger time or a relaxed dynamical state. The relatively ''short'' total simulation times considered are also responsible for the youth of the central structures in the remnants that result from the merger-induced starbursts. They are $\sim$0.3 -- 1\,Gyr old, so they are very bright in the blue bands (see Paper I). 

\citet{2001ApJ...550..212B} found that the colour of a galaxy and the $M/L$ ratio in several photometric bands relate linearly for a given set of parameters defining the evolution experienced by the stellar populations in the galaxy. Considering the typical colour of nearby early-type galaxies \citep[$B-R \sim 1.5$, see][]{2009MNRAS.393.1467F}, these authors provide $M/L$ of $\sim 6$ and $\sim 1$ for the $B$ and $K$ bands, respectively, for different prescriptions of the SFHs used. This means that the brightest S0s within the E08 and G11 samples ($-20 < M_B < -22$ and $-24 < M_K < -25$) have masses comparable to our remnants, according to \citeauthor{2001ApJ...550..212B} relations ($M\sim 10^{11}\Msun$). However, the remnants of our simulations are $\sim 1$ -- 2\,mag brighter in these bands than the brightest S0s in the reference observational sample (see Paper I, but this is also noticeable in the bottom panel of Fig.\ref{fig:mubreak_Rbrk}).

These differences may be owing to the fact that the average $M/L$ ratios used to convert mass into light in our models are $\sim 2$ times lower than those derived from the \citet{2001ApJ...550..212B} relations for all bands, which explains the $\sim 1$ -- 2\,mag offset towards brighter magnitudes of our models with respect to the bright end of the observational samples. This is because of the bluer colours of our final remnants ($B-R\sim 0.9$) compared to nearby S0s. Considering these colours, the $M/L$ ratios derived from \citeauthor{2001ApJ...550..212B} relations are very similar to the average values obtained in our models for the $B$ and $K$ bands. 

The final remnants have undergone massive bursts of star formation (of $\sim 10$ -- 30\% of the stellar mass in the progenitor discs) in less than $\sim 1$ -- 2\,Gyr from the moment they are being analysed. We thus have very young (and blue) stellar populations in the centres of the remnants which bias the colour of the galaxy globally (and thus, the $M/L$ ratios). This agrees with the results by \citet{2001ApJ...550..212B}, who found that starbursts involving $\sim 10$\% of the total stellar mass in a galaxy can decrease its $M/L$ ratio by a factor of $\sim 3$ in the $B$ band and $\sim 2$ in $K$ (see their Section\,4.3). Moreover, they also indicated that these effects may persist from $\sim 1$\,Gyr up to $\sim 5$\,Gyr, depending on the colour of the underlying stellar populations (more time if the populations are redder). So, it is reasonable that the colours of our remnants are still affected by these recent starbursts. 

Accounting for the fact that real S0s of similar masses to our remnants may have experienced their last major merger at least $\sim 7$\,Gyr ago \citep{2010A&A...519A..55E,2013MNRAS.428..999P,2014arXiv1403.4932C}, our remnants would require to be evolved passively for at least $\sim 2$ -- 3\,Gyr more to present global colours analogous to those observed in present-day S0s with similar stellar masses. This passive evolution would then increase the $M/L$ ratios by a factor of $\sim 2$, making our remnants overlap with the bright end of the E08 and G11 samples in absolute magnitudes. This would correct both the offset in total magnitudes and the offset in surface brightness of our models compared to real analogs. Moreover, a merging scenario for the origin of S0s as the one simulated here could explain why the S0s at $z\sim 0.3$ -- 0.4 are bluer and brighter than their local counterparts with similar masses \citep{2009MNRAS.393.1467F}.

Therefore, the existence of young subcomponents in our S0-like remnants that make them bluer and brighter is is not a limitation of our simulations, because young inner structures are quite common in E-S0 galaxies at $z\sim 0.4$ -- 0.6 (and even in the local universe, although in S0s with lower masses). In fact, these blue substructures are usually interpreted as a trace of recent merging \citep{2009MNRAS.393.1467F,2009AJ....138..579K,2009MNRAS.400..687M,2010A&A...515A...3H,2010ApJ...714L.171T,2010ApJ...708..841W,2011MNRAS.411.2148K}.

\subsection{Limitations due to the assumptions adopted for the mass-to-light conversion}
\label{Sec:age_effects}

\subsubsection{IMFs and SFHs considered in the progenitors}
\label{Sec:IMF_SFH}

The assumptions adopted to model the SFH of the different types of particles in the simulations may be also affecting the $M/L$ ratios used to convert mass into light (Sect.\,\ref{Sec:S0identification}). The offsets in total magnitudes and in surface brightness of our simulations compared to real (massive) analogs in the reference observational samples may be partially due to this assumptions, besides the existence of young stellar populations (Sect.\,\ref{Sec:generallimitations}).

In particular, the IMF assumed may affect significantly. The $M/L$ ratios derived using a Chabrier IMF in \citet{2003MNRAS.344.1000B} models are a factor of $\sim 0.3$ -- 0.5 lower than those obtained from a Salpeter IMF under the same conditions \citep{2001ApJ...550..212B,2009MNRAS.394..774L}. Therefore, the offset in magnitudes between our models and real data may be completely explained by accounting for uncertainties inherent to the modelling of the SFHs, such as the details of the IMF \citep[see also][]{2012Natur.484..485C,2013MNRAS.435.2764M}. 

We have also assumed the same SFH for all old stellar particles coming from the progenitors of a given Hubble type. Although the parametrization of the SFH of each type (E, Sa, Sb, or Sd) has been set to typical observational values, this is another oversimplification of reality. Old stellar particles represent mass blocks of $\sim 10^5\Msun$, so it would be more realistic to consider that each particle has experienced a different SFH prior to the interaction, mostly depending on its location in the progenitor galaxy. Nevertheless, the selection of these SFHs would be highly subjective and this would make the analysis much more complex, without providing any important advantages. A similar reasoning can be done for the SSP models used to approximate the SFH of hybrid particles.

Summarizing, the offset in total magnitudes and in surface brightness observed between our remnants and real S0s is not a limitation to the global results presented here, as it may be corrected easily by adjusting the parameters used to model the SFHs.

\subsubsection{Age assumed for the old stellar particles}
\label{Sec:age}

Collisionless stellar particles in the GalMer simulations have an assigned age of 10\,Gyr. We have assumed an average age of 10\,Gyr for them too to convert mass into light (see Sect.\,\ref{Sec:profiles}), but considering that this is the typical age of old stellar populations in the outer discs of real S0 galaxies \citep{2012MNRAS.427..790S}. However, this assumption is an obvious oversimplification of reality. First, because this age has been considered independent of the morphological type of the progenitor which hosted the particle at the beginning of the simulation. Nevertheless, the average age of stars in galaxy discs depends on the Hubble type, in the sense that earlier types tend to have older stellar populations at intermediate radii in their discs. In addition, average stellar ages show a wide dispersion even within a given Hubble type, as well as radial gradients, and we have not considered these \citep{2001ApJ...553...90V, 2004ApJS..152..175M, 2012MNRAS.427..790S}. 

The age considered for the old stellar particles obviously affects to the $M/L_R$ ratio used to convert projected density profiles into surface brightness profiles, and thus it may be partially responsible for the offset of $\sim 1$\,mag\,arcsec$^{-2}$ found in the characteristic surface brightness between real and simulated data and for the difference in $\sim 1$ --2\,mag in total absolute magnitudes between the brightest galaxies in E08 and G11 samples and our remnants. An exhaustive study on how the assumed stellar age profiles affect the properties of the formed antitruncations is beyond the scope of this paper. However, we have tested whether this offset may be an artefact of this assumption by analysing the effects of assuming different ages for the old stellar particles in the photometric properties of the breaks. We have selected four models that differ in their initial gas content, morphology of the progenitors, and orbital configuration to test this: gSagSao5, gSagSdo41, gSbgSbo22, and gSdgSdo5. 

The luminosity-weighted average ages of Sa and Sb galaxies range from $\sim 7$ to $\sim 11$\,Gyr and from $\sim 6$ to $\sim 10$\,Gyr in Sd galaxies \citep{2004ApJS..152..175M}. So, considering that the simulated mergers take $\sim 3$\,Gyr, we have assumed ages spanning an even wider range for the old stellar particles in these four models (from 6 to 14 \,Gyr). For simplicity, we have used the same stellar age for the old stellar particles of both progenitors independently of their Hubble types. Therefore, we re-simulated the $R$-band surface brightness profiles of these four models for different ages spanning the range of 6 -- 14\,Gyr, and performed again the fits to define the properties of the break in each case. Some ages represent extreme cases of progenitor galaxies hosting very young or very old average stellar populations, but they are realistic attending to the properties of real galaxies \citep{2004ApJS..152..175M}. 

In Fig.~\ref{fig:trends_age}, we plot the characteristic photometric parameters of the antitruncations and of the inner and outer discs of each model as a function of the assumed age for the old stellar particles. Two immediate facts are evident from these plots:

\begin{itemize}
\item[i.] The characteristic surface brightness of the break and of their inner and outer discs become brighter as we decrease the assumed stellar age (panels on the left of the figure). The difference can be up to $\sim 1$\,mag\,arcsec$^{-2}$. Therefore, we can expect that a more realistic (and complex) distribution of stellar ages in the progenitors may partially correct the offset in the characteristic surface brightness values that we have obtained between real and simulated data. 

\item[ii] Interestingly, the break radius and scalelengths of the inner and outer discs do not vary significantly with the age of the old stellar particles (panels on the right of the figure). The maximum change is less than $\sim 10$\% of the value in all cases. This result provides more reliability to our results, because it means that the agreement between the scaling relations found in our simulations and in observations is robust against this assumption. 
\end{itemize}

This simple test demonstrates that, by assuming a more complex age distribution for the old stellar particles of the progenitors, the offset in the characteristic surface brightness of the breaks and of their inner and outer discs found between models and real data can be overridden, without significantly affecting their characteristic scalelengths. Therefore, the offsets in surface brightness and total magnitudes found between models and real data do not establish any limitation to the global results derived in the present study.

\begin{figure*}[th!]
\center
\includegraphics[angle=270,width = 0.49\textwidth, bb = 60 50 560 750, clip]{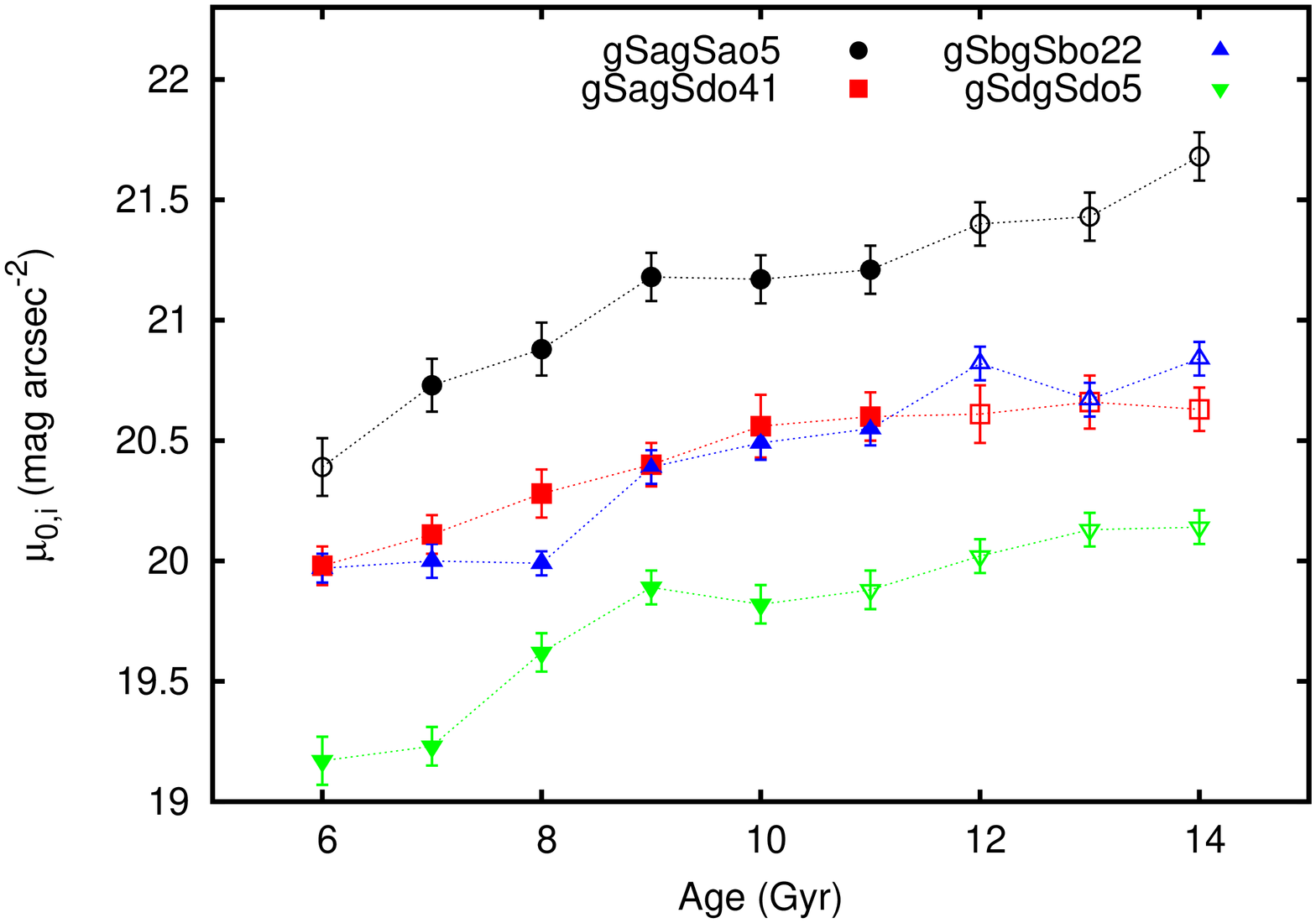}  
\includegraphics[angle=270,width = 0.49\textwidth, bb = 60 50 560 750, clip]{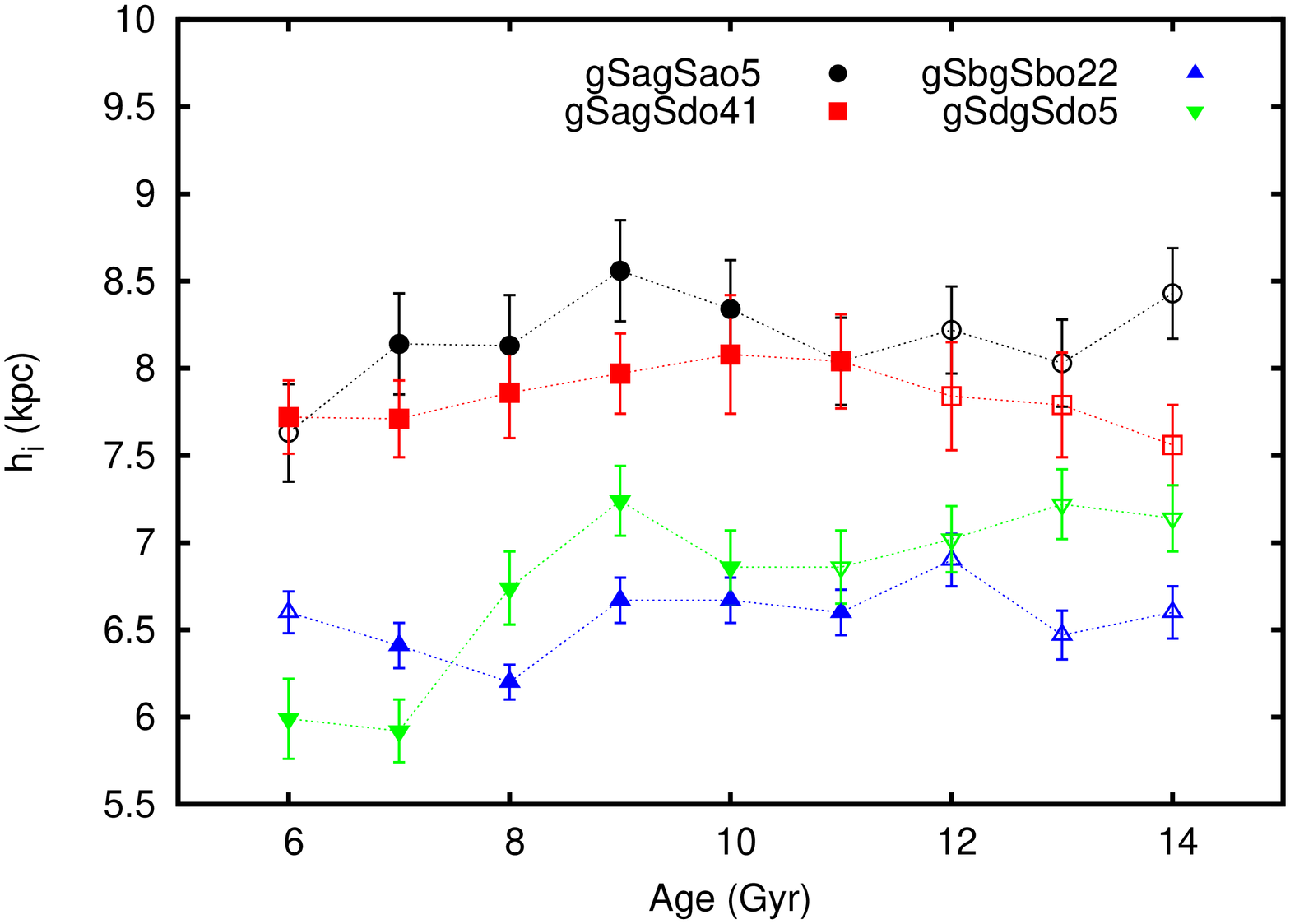}  
\includegraphics[angle=270,width = 0.49\textwidth, bb = 60 50 560 750, clip]{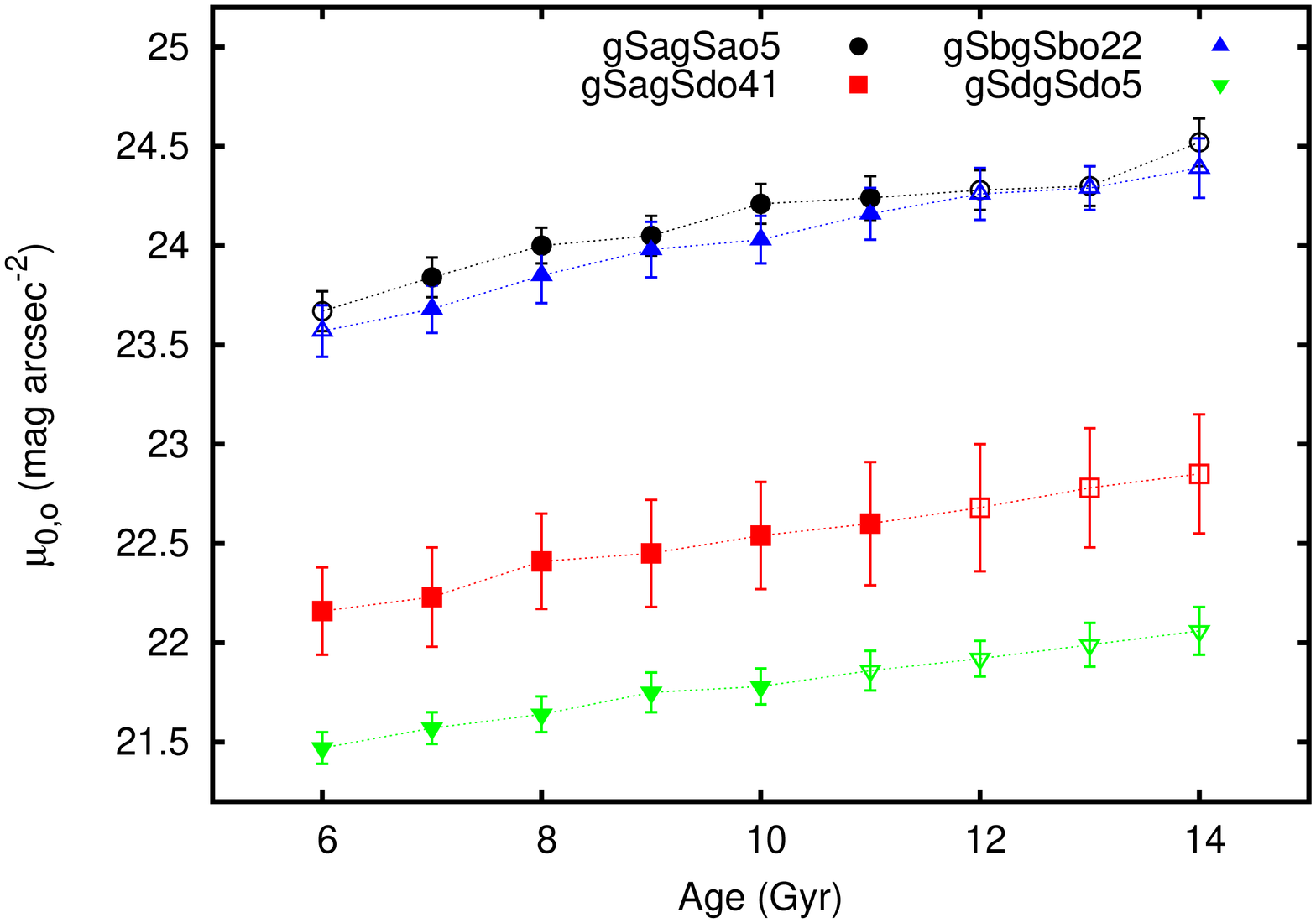}
\includegraphics[angle=270,width = 0.49\textwidth, bb = 60 50 560 750, clip]{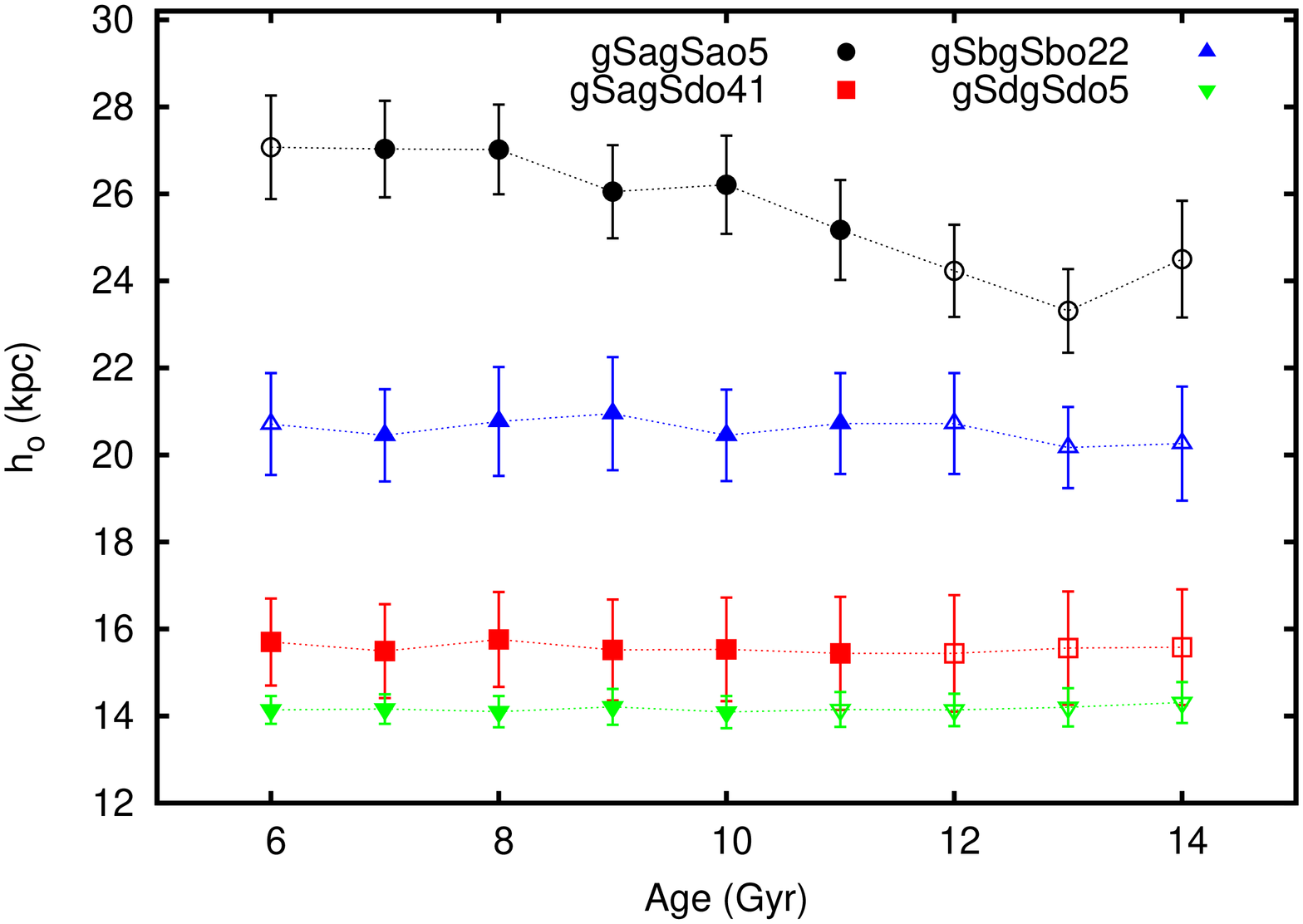}
\includegraphics[angle=270,width = 0.49\textwidth, bb = 60 50 560 750, clip]{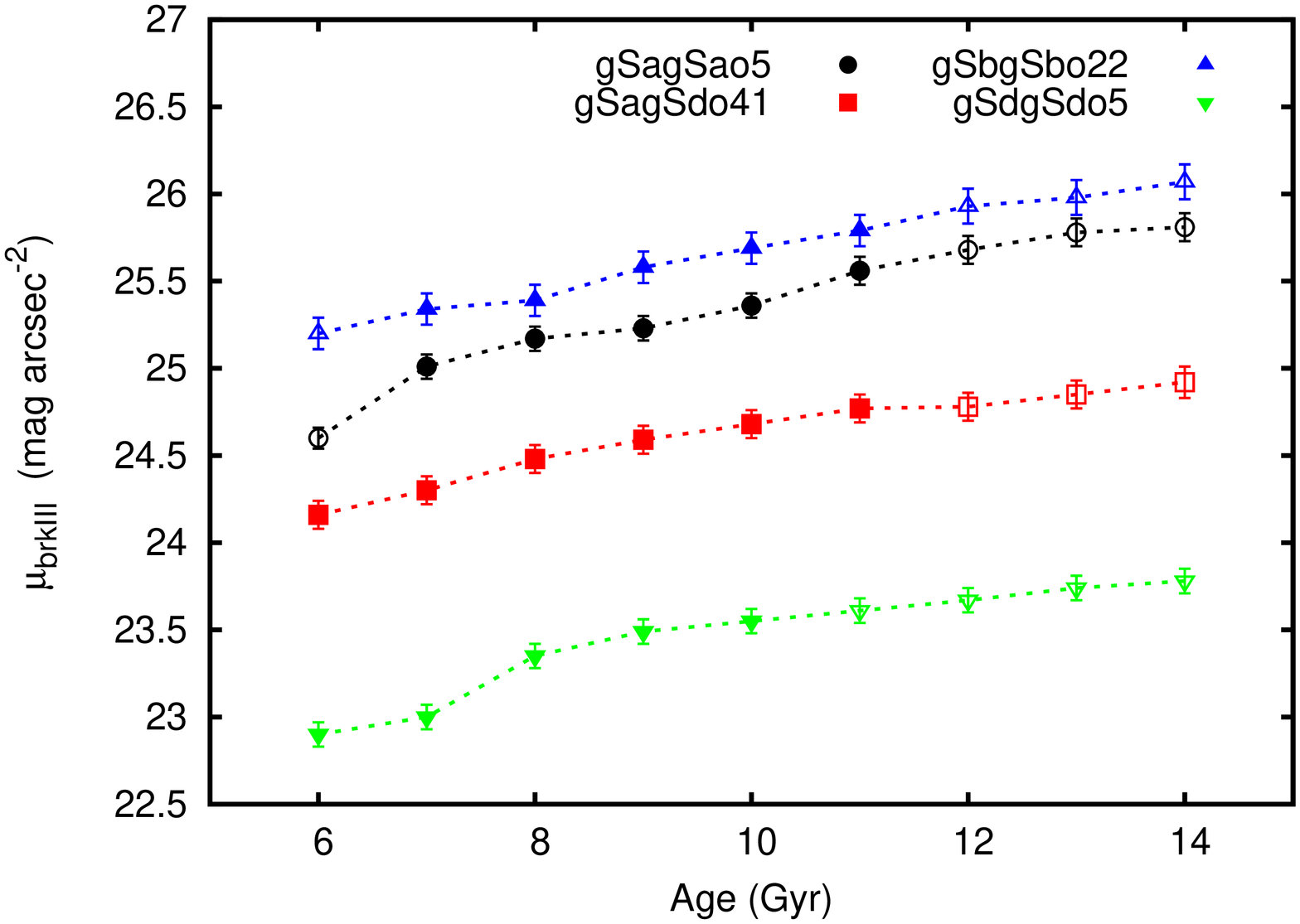}  
\includegraphics[angle=270,width = 0.49\textwidth, bb = 60 50 560 750, clip]{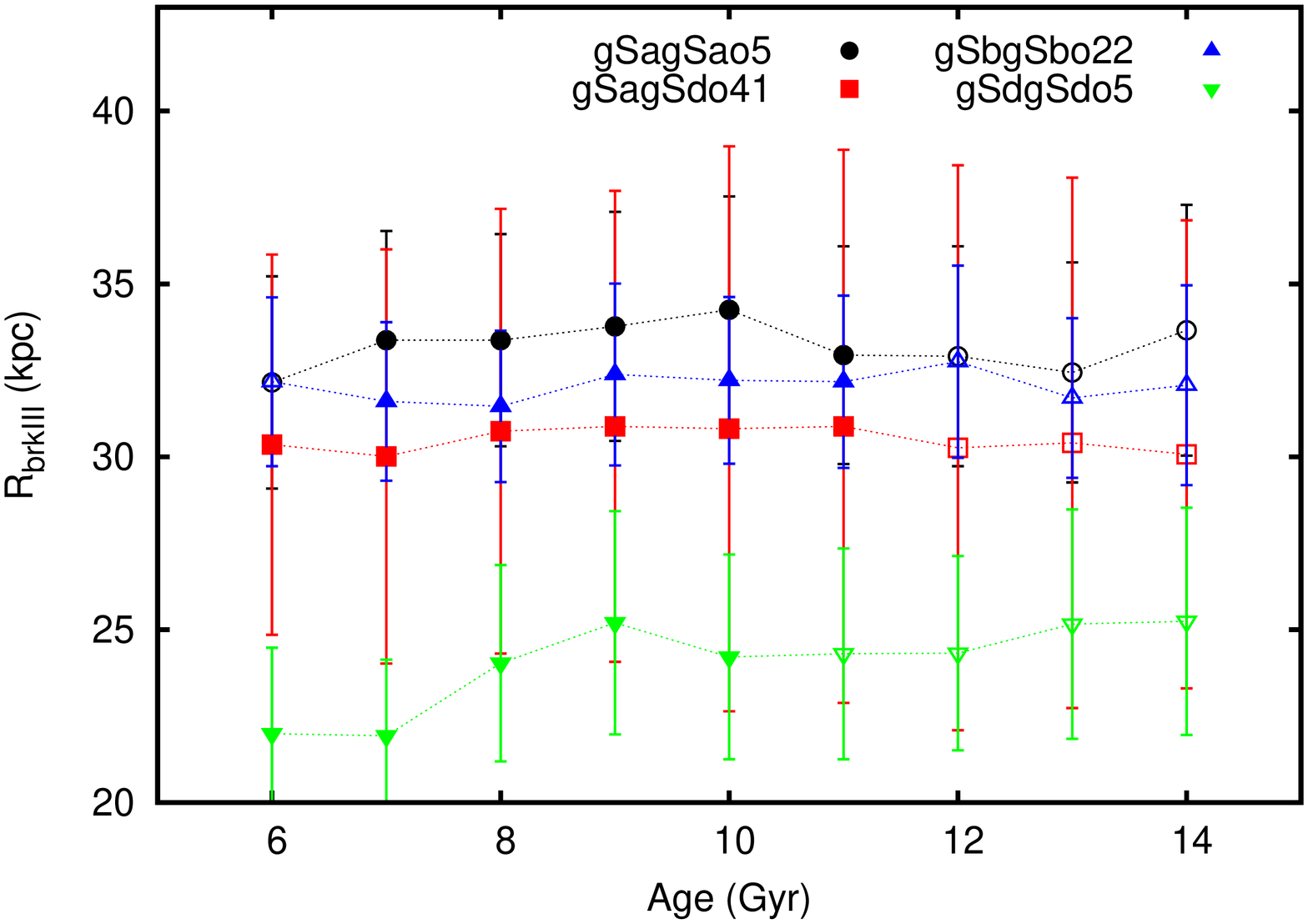}
\caption{Dependence on the average age assumed for old stellar particles of the characteristic photometric parameters of the antitruncations and of the inner and outer discs in four models (gSagSao5, gSagSdo41, gSbgSbo22, and gSdgSdo5). Filled symbols remark the most likely ages for the old stellar populations coming from a progenitor of a given Hubble type (see ~\ref{Sec:age_effects}). Consult the legend in the figure.  [\emph{A colour version of this figure is available in the online edition.}]
}
\label{fig:trends_age}
\end{figure*}

\subsubsection{Dust extinction effects}
\label{Sec:dust}

As commented in Sect.\,\ref{Sec:profiles}, no dust effects have been assumed to derive the surface brightness profiles. It would make the analysis very complex, without giving significant advantages. 

We have estimated the extinction that might be experienced by a typical E-S0 galaxy to check whether our remnant S0s might overlap with real data in the bottom panel of Fig.\,\ref{fig:mubreak_Rbrk}, just accounting for dust extinction effects.  We have combined the estimates of relative extinction derived by \citet{2010MNRAS.407.2475F} for E-S0 galaxies with ionized gas and dust lanes with the reddening values obtained by \citet{2010A&A...519A..40A} for nearby S0s. Considering that E-S0 galaxies typically have $A_B/E(B-V)\sim 4$, $A_V/E(B-V)\sim 3$, and $A_R/E(B-V)\sim 2.3$ and that $E(B-V)\sim 0.3$ on average for them, the average extinction of an E-S0 galaxy ranges between $\sim 1.2$ and $\sim 0.9$\,mag in the $B$, $V$, and $R$ bands. Although considering that the reddening of S0s can be up to 1.7\,mag, the extinction can achieve values up to $\sim 7$, 5, and 4\,mag in the $B$, $V$, and $R$ bands respectively. This means that the offset in total magnitud between our remnant models and real S0s with analogous masses could be overridden just by taking into account dust extinction effects. 

Moreover, the effects of dust extinction are expected to be quite limited in the simulated surface brightness profiles. First, because the profiles are obtained as azimuthal averages with the galaxies in face-on views (which minimizes the effects of dust extinction), and secondly, because the extinction in the $R$ band is $\sim 1$\,mag for the whole galaxy. So including dust extinction barely affect the most important results of this study.

\end{appendix}

}

{
%\twocolumn
\small  

\bibliographystyle{aa}
\bibliography{borlaff_13.bib}{}

\begin{thebibliography}{119}
\expandafter\ifx\csname natexlab\endcsname\relax\def\natexlab#1{#1}\fi

\bibitem[{{Aguerri} {et~al.}(2001){Aguerri}, {Balcells}, \&
  {Peletier}}]{2001A&A...367..428A}
{Aguerri}, J.~A.~L., {Balcells}, M., \& {Peletier}, R.~F. 2001, \aap, 367, 428

\bibitem[{{Annibali} {et~al.}(2010){Annibali}, {Bressan}, {Rampazzo},
  {Zeilinger}, {Vega}, \& {Panuzzo}}]{2010A&A...519A..40A}
{Annibali}, F., {Bressan}, A., {Rampazzo}, R., {et~al.} 2010, \aap, 519, A40

\bibitem[{{Baillard} {et~al.}(2011){Baillard}, {Bertin}, {de Lapparent},
  {Fouqu{\'e}}, {Arnouts}, {Mellier}, {Pell{\'o}}, {Leborgne}, {Prugniel},
  {Makarov}, {Makarova}, {McCracken}, {Bijaoui}, \&
  {Tasca}}]{2011A&A...532A..74B}
{Baillard}, A., {Bertin}, E., {de Lapparent}, V., {et~al.} 2011, \aap, 532, A74

\bibitem[{{Bakos} {et~al.}(2011){Bakos}, {Trujillo}, {Azzollini}, {Beckman}, \&
  {Pohlen}}]{2011MSAIS..18..113B}
{Bakos}, J., {Trujillo}, I., {Azzollini}, R., {Beckman}, J.~E., \& {Pohlen}, M.
  2011, Memorie della Societa Astronomica Italiana Supplementi, 18, 113

\bibitem[{{Barr} {et~al.}(2007){Barr}, {Bedregal}, {Arag{\'o}n-Salamanca},
  {Merrifield}, \& {Bamford}}]{2007A&A...470..173B}
{Barr}, J.~M., {Bedregal}, A.~G., {Arag{\'o}n-Salamanca}, A., {Merrifield},
  M.~R., \& {Bamford}, S.~P. 2007, \aap, 470, 173

\bibitem[{{Barway} {et~al.}(2013){Barway}, {Wadadekar}, {Vaghmare}, \&
  {Kembhavi}}]{2013MNRAS.432..430B}
{Barway}, S., {Wadadekar}, Y., {Vaghmare}, K., \& {Kembhavi}, A.~K. 2013,
  \mnras, 432, 430

\bibitem[{{Bekki} \& {Couch}(2011)}]{2011MNRAS.415.1783B}
{Bekki}, K. \& {Couch}, W.~J. 2011, \mnras, 415, 1783

\bibitem[{{Bell} \& {de Jong}(2001)}]{2001ApJ...550..212B}
{Bell}, E.~F. \& {de Jong}, R.~S. 2001, \apj, 550, 212

\bibitem[{{Benjouali} {et~al.}(2011){Benjouali}, {G{\'o}mez Flechoso},
  {Dom{\'{\i}}nguez-Tenreiro}, {Mart{\'{\i}}nez-Serrano}, \&
  {Serna}}]{2011hsa6.conf..148B}
{Benjouali}, L., {G{\'o}mez Flechoso}, M.~A., {Dom{\'{\i}}nguez-Tenreiro}, R.,
  {Mart{\'{\i}}nez-Serrano}, F., \& {Serna}, A. 2011, in Highlights of Spanish
  Astrophysics VI, ed. M.~R. {Zapatero Osorio}, J.~{Gorgas}, J.~{Ma{\'{\i}}z
  Apell{\'a}niz}, J.~R. {Pardo}, \& A.~{Gil de Paz}, 148--153

\bibitem[{{Berlind} {et~al.}(2006){Berlind}, {Frieman}, {Weinberg}, {Blanton},
  {Warren}, {Abazajian}, {Scranton}, {Hogg}, {Scoccimarro}, {Bahcall},
  {Brinkmann}, {Gott}, {Kleinman}, {Krzesinski}, {Lee}, {Miller}, {Nitta},
  {Schneider}, {Tucker}, \& {Zehavi}}]{2006ApJS..167....1B}
{Berlind}, A.~A., {Frieman}, J., {Weinberg}, D.~H., {et~al.} 2006, \apjs, 167,
  1

\bibitem[{{Bernardi} {et~al.}(2011{\natexlab{a}}){Bernardi}, {Roche},
  {Shankar}, \& {Sheth}}]{2011MNRAS.412..684B}
{Bernardi}, M., {Roche}, N., {Shankar}, F., \& {Sheth}, R.~K.
  2011{\natexlab{a}}, \mnras, 412, 684

\bibitem[{{Bernardi} {et~al.}(2011{\natexlab{b}}){Bernardi}, {Roche},
  {Shankar}, \& {Sheth}}]{2011MNRAS.412L...6B}
{Bernardi}, M., {Roche}, N., {Shankar}, F., \& {Sheth}, R.~K.
  2011{\natexlab{b}}, \mnras, 412, L6

\bibitem[{{Bertola} {et~al.}(1995){Bertola}, {Cinzano}, {Corsini}, {Rix}, \&
  {Zeilinger}}]{1995ApJ...448L..13B}
{Bertola}, F., {Cinzano}, P., {Corsini}, E.~M., {Rix}, H.-W., \& {Zeilinger},
  W.~W. 1995, \apjl, 448, L13

\bibitem[{{Bettoni} \& {Galletta}(1997)}]{1997A&AS..124...61B}
{Bettoni}, D. \& {Galletta}, G. 1997, \aaps, 124, 61

\bibitem[{{Bottema}(1989)}]{1989A&A...221..236B}
{Bottema}, R. 1989, \aap, 221, 236

\bibitem[{{Bournaud} {et~al.}(2011){Bournaud}, {Chapon}, {Teyssier}, {Powell},
  {Elmegreen}, {Elmegreen}, {Duc}, {Contini}, {Epinat}, \&
  {Shapiro}}]{2011ApJ...730....4B}
{Bournaud}, F., {Chapon}, D., {Teyssier}, R., {et~al.} 2011, \apj, 730, 4

\bibitem[{{Bournaud} {et~al.}(2005){Bournaud}, {Jog}, \&
  {Combes}}]{2005A&A...437...69B}
{Bournaud}, F., {Jog}, C.~J., \& {Combes}, F. 2005, \aap, 437, 69

\bibitem[{{Bruzual} \& {Charlot}(2003)}]{2003MNRAS.344.1000B}
{Bruzual}, G. \& {Charlot}, S. 2003, \mnras, 344, 1000

\bibitem[{{Cappellari} {et~al.}(2012){Cappellari}, {McDermid}, {Alatalo},
  {Blitz}, {Bois}, {Bournaud}, {Bureau}, {Crocker}, {Davies}, {Davis}, {de
  Zeeuw}, {Duc}, {Emsellem}, {Khochfar}, {Krajnovi{\'c}}, {Kuntschner},
  {Lablanche}, {Morganti}, {Naab}, {Oosterloo}, {Sarzi}, {Scott}, {Serra},
  {Weijmans}, \& {Young}}]{2012Natur.484..485C}
{Cappellari}, M., {McDermid}, R.~M., {Alatalo}, K., {et~al.} 2012, \nat, 484,
  485

\bibitem[{{Chilingarian} {et~al.}(2010){Chilingarian}, {Di Matteo}, {Combes},
  {Melchior}, \& {Semelin}}]{2010A&A...518A..61C}
{Chilingarian}, I.~V., {Di Matteo}, P., {Combes}, F., {Melchior}, A.-L., \&
  {Semelin}, B. 2010, \aap, 518, A61

\bibitem[{{Choi} {et~al.}(2014){Choi}, {Conroy}, {Moustakas}, {Graves},
  {Holden}, {Brodwin}, {Brown}, \& {van Dokkum}}]{2014arXiv1403.4932C}
{Choi}, J., {Conroy}, C., {Moustakas}, J., {et~al.} 2014, ArXiv e-prints

\bibitem[{{Comer{\'o}n} {et~al.}(2012){Comer{\'o}n}, {Elmegreen}, {Salo},
  {Laurikainen}, {Athanassoula}, {Bosma}, {Knapen}, {Gadotti}, {Sheth}, {Hinz},
  {Regan}, {Gil de Paz}, {Mu{\~n}oz-Mateos}, {Men{\'e}ndez-Delmestre},
  {Seibert}, {Kim}, {Mizusawa}, {Laine}, {Ho}, \&
  {Holwerda}}]{2012ApJ...759...98C}
{Comer{\'o}n}, S., {Elmegreen}, B.~G., {Salo}, H., {et~al.} 2012, \apj, 759, 98

\bibitem[{{Cortesi} {et~al.}(2013){Cortesi}, {Merrifield}, {Coccato},
  {Arnaboldi}, {Gerhard}, {Bamford}, {Napolitano}, {Romanowsky}, {Douglas},
  {Kuijken}, {Capaccioli}, {Freeman}, {Saha}, \&
  {Chies-Santos}}]{2013MNRAS.432.1010C}
{Cortesi}, A., {Merrifield}, M.~R., {Coccato}, L., {et~al.} 2013, \mnras, 432,
  1010

\bibitem[{{Cox} {et~al.}(2008){Cox}, {Jonsson}, {Somerville}, {Primack}, \&
  {Dekel}}]{2008MNRAS.384..386C}
{Cox}, T.~J., {Jonsson}, P., {Somerville}, R.~S., {Primack}, J.~R., \& {Dekel},
  A. 2008, \mnras, 384, 386

\bibitem[{{Crook} {et~al.}(2007){Crook}, {Huchra}, {Martimbeau}, {Masters},
  {Jarrett}, \& {Macri}}]{2007ApJ...655..790C}
{Crook}, A.~C., {Huchra}, J.~P., {Martimbeau}, N., {et~al.} 2007, \apj, 655,
  790

\bibitem[{{Desai} {et~al.}(2007){Desai}, {Dalcanton}, {Arag{\'o}n-Salamanca},
  {Jablonka}, {Poggianti}, {Gogarten}, {Simard}, {Milvang-Jensen}, {Rudnick},
  {Zaritsky}, {Clowe}, {Halliday}, {Pell{\'o}}, {Saglia}, \&
  {White}}]{2007ApJ...660.1151D}
{Desai}, V., {Dalcanton}, J.~J., {Arag{\'o}n-Salamanca}, A., {et~al.} 2007,
  \apj, 660, 1151

\bibitem[{{di Matteo} {et~al.}(2008){di Matteo}, {Bournaud}, {Martig},
  {Combes}, {Melchior}, \& {Semelin}}]{2008A&A...492...31D}
{di Matteo}, P., {Bournaud}, F., {Martig}, M., {et~al.} 2008, \aap, 492, 31

\bibitem[{{Di Matteo} {et~al.}(2007){Di Matteo}, {Combes}, {Melchior}, \&
  {Semelin}}]{2007A&A...468...61D}
{Di Matteo}, P., {Combes}, F., {Melchior}, A., \& {Semelin}, B. 2007, \aap,
  468, 61

\bibitem[{{Dressler}(1980)}]{1980ApJ...236..351D}
{Dressler}, A. 1980, \apj, 236, 351

\bibitem[{{Dressler} {et~al.}(1997){Dressler}, {Oemler}, {Couch}, {Smail},
  {Ellis}, {Barger}, {Butcher}, {Poggianti}, \&
  {Sharples}}]{1997ApJ...490..577D}
{Dressler}, A., {Oemler}, Jr., A., {Couch}, W.~J., {et~al.} 1997, \apj, 490,
  577

\bibitem[{{Duc} {et~al.}(2011){Duc}, {Cuillandre}, {Serra}, {Michel-Dansac},
  {Ferriere}, {Alatalo}, {Blitz}, {Bois}, {Bournaud}, {Bureau}, {Cappellari},
  {Davies}, {Davis}, {de Zeeuw}, {Emsellem}, {Khochfar}, {Krajnovi{\'c}},
  {Kuntschner}, {Lablanche}, {McDermid}, {Morganti}, {Naab}, {Oosterloo},
  {Sarzi}, {Scott}, {Weijmans}, \& {Young}}]{2011MNRAS.417..863D}
{Duc}, P.-A., {Cuillandre}, J.-C., {Serra}, P., {et~al.} 2011, \mnras, 417, 863

\bibitem[{{Eliche-Moral} {et~al.}(2006{\natexlab{a}}){Eliche-Moral},
  {Balcells}, {Aguerri}, \& {Gonz{\'a}lez-Garc{\'{\i}}a}}]{2006A&A...457...91E}
{Eliche-Moral}, M.~C., {Balcells}, M., {Aguerri}, J.~A.~L., \&
  {Gonz{\'a}lez-Garc{\'{\i}}a}, A.~C. 2006{\natexlab{a}}, \aap, 457, 91

\bibitem[{{Eliche-Moral} {et~al.}(2006{\natexlab{b}}){Eliche-Moral},
  {Balcells}, {Prieto}, {Garc{\'{\i}}a-Dab{\'o}}, {Erwin}, \&
  {Crist{\'o}bal-Hornillos}}]{2006ApJ...639..644E}
{Eliche-Moral}, M.~C., {Balcells}, M., {Prieto}, M., {et~al.}
  2006{\natexlab{b}}, \apj, 639, 644

\bibitem[{{Eliche-Moral} {et~al.}(2012){Eliche-Moral},
  {Gonz{\'a}lez-Garc{\'{\i}}a}, {Aguerri}, {Gallego}, {Zamorano}, {Balcells},
  \& {Prieto}}]{2012A&A...547A..48E}
{Eliche-Moral}, M.~C., {Gonz{\'a}lez-Garc{\'{\i}}a}, A.~C., {Aguerri},
  J.~A.~L., {et~al.} 2012, \aap, 547, A48

\bibitem[{{Eliche-Moral} {et~al.}(2013){Eliche-Moral},
  {Gonz{\'a}lez-Garc{\'{\i}}a}, {Aguerri}, {Gallego}, {Zamorano}, {Balcells},
  \& {Prieto}}]{2013A&A...552A..67E}
{Eliche-Moral}, M.~C., {Gonz{\'a}lez-Garc{\'{\i}}a}, A.~C., {Aguerri},
  J.~A.~L., {et~al.} 2013, \aap, 552, A67

\bibitem[{{Eliche-Moral} {et~al.}(2011){Eliche-Moral},
  {Gonz{\'a}lez-Garc{\'{\i}}a}, {Balcells}, {Aguerri}, {Gallego}, {Zamorano},
  \& {Prieto}}]{2011A&A...533A.104E}
{Eliche-Moral}, M.~C., {Gonz{\'a}lez-Garc{\'{\i}}a}, A.~C., {Balcells}, M.,
  {et~al.} 2011, \aap, 533, A104

\bibitem[{{Eliche-Moral} {et~al.}(2010{\natexlab{a}}){Eliche-Moral}, {Prieto},
  {Gallego}, {Barro}, {Zamorano}, {L{\'o}pez-Sanjuan}, {Balcells},
  {Guzm{\'a}n}, \& {Mu{\~n}oz-Mateos}}]{2010A&A...519A..55E}
{Eliche-Moral}, M.~C., {Prieto}, M., {Gallego}, J., {et~al.}
  2010{\natexlab{a}}, \aap, 519, A55

\bibitem[{{Eliche-Moral} {et~al.}(2010{\natexlab{b}}){Eliche-Moral}, {Prieto},
  {Gallego}, \& {Zamorano}}]{2010arXiv1003.0686E}
{Eliche-Moral}, M.~C., {Prieto}, M., {Gallego}, J., \& {Zamorano}, J.
  2010{\natexlab{b}}, ArXiv:1003.0686

\bibitem[{{Elmegreen} \& {Hunter}(2006)}]{2006ApJ...636..712E}
{Elmegreen}, B.~G. \& {Hunter}, D.~A. 2006, \apj, 636, 712

\bibitem[{{Erwin} {et~al.}(2005){Erwin}, {Beckman}, \&
  {Pohlen}}]{2005ApJ...626L..81E}
{Erwin}, P., {Beckman}, J.~E., \& {Pohlen}, M. 2005, \apjl, 626, L81

\bibitem[{{Erwin} {et~al.}(2012){Erwin}, {Guti{\'e}rrez}, \&
  {Beckman}}]{2012ApJ...744L..11E}
{Erwin}, P., {Guti{\'e}rrez}, L., \& {Beckman}, J.~E. 2012, \apjl, 744, L11

\bibitem[{{Erwin} {et~al.}(2008){Erwin}, {Pohlen}, \&
  {Beckman}}]{2008AJ....135...20E}
{Erwin}, P., {Pohlen}, M., \& {Beckman}, J.~E. 2008, \aj, 135, 20 (E08)

\bibitem[{{Finkelman} {et~al.}(2010){Finkelman}, {Brosch}, {Funes}, {Kniazev},
  \& {V{\"a}is{\"a}nen}}]{2010MNRAS.407.2475F}
{Finkelman}, I., {Brosch}, N., {Funes}, J.~G., {Kniazev}, A.~Y., \&
  {V{\"a}is{\"a}nen}, P. 2010, \mnras, 407, 2475

\bibitem[{{Fisher}(1997)}]{1997AJ....113..950F}
{Fisher}, D. 1997, \aj, 113, 950

\bibitem[{{F{\"o}rster Schreiber} {et~al.}(2009){F{\"o}rster Schreiber},
  {Genzel}, {Bouch{\'e}}, {Cresci}, {Davies}, {Buschkamp}, {Shapiro},
  {Tacconi}, {Hicks}, {Genel}, {Shapley}, {Erb}, {Steidel}, {Lutz},
  {Eisenhauer}, {Gillessen}, {Sternberg}, {Renzini}, {Cimatti}, {Daddi},
  {Kurk}, {Lilly}, {Kong}, {Lehnert}, {Nesvadba}, {Verma}, {McCracken},
  {Arimoto}, {Mignoli}, \& {Onodera}}]{2009ApJ...706.1364F}
{F{\"o}rster Schreiber}, N.~M., {Genzel}, R., {Bouch{\'e}}, N., {et~al.} 2009,
  \apj, 706, 1364

\bibitem[{{Freeman}(1970)}]{1970ApJ...160..811F}
{Freeman}, K.~C. 1970, \apj, 160, 811

\bibitem[{{Fritz} {et~al.}(2009){Fritz}, {B{\"o}hm}, \&
  {Ziegler}}]{2009MNRAS.393.1467F}
{Fritz}, A., {B{\"o}hm}, A., \& {Ziegler}, B.~L. 2009, \mnras, 393, 1467

\bibitem[{{Genzel} {et~al.}(2008){Genzel}, {Burkert}, {Bouch{\'e}}, {Cresci},
  {F{\"o}rster Schreiber}, {Shapley}, {Shapiro}, {Tacconi}, {Buschkamp},
  {Cimatti}, {Daddi}, {Davies}, {Eisenhauer}, {Erb}, {Genel}, {Gerhard},
  {Hicks}, {Lutz}, {Naab}, {Ott}, {Rabien}, {Renzini}, {Steidel}, {Sternberg},
  \& {Lilly}}]{2008ApJ...687...59G}
{Genzel}, R., {Burkert}, A., {Bouch{\'e}}, N., {et~al.} 2008, \apj, 687, 59

\bibitem[{{G{\'o}mez-Flechoso} {et~al.}(2010){G{\'o}mez-Flechoso}, {Benjouali},
  \& {Dom{\'{\i}}nguez Tenreiro}}]{2010hsa5.conf..295G}
{G{\'o}mez-Flechoso}, M.~A., {Benjouali}, L., \& {Dom{\'{\i}}nguez Tenreiro},
  R. 2010, in Highlights of Spanish Astrophysics V, ed. J.~M. {Diego}, L.~J.
  {Goicoechea}, J.~I. {Gonz{\'a}lez-Serrano}, \& J.~{Gorgas}, 295

\bibitem[{{Guti{\'e}rrez} {et~al.}(2011){Guti{\'e}rrez}, {Erwin}, {Aladro}, \&
  {Beckman}}]{2011AJ....142..145G}
{Guti{\'e}rrez}, L., {Erwin}, P., {Aladro}, R., \& {Beckman}, J.~E. 2011, \aj,
  142, 145 (G11)

\bibitem[{{Hammer} {et~al.}(2009{\natexlab{a}}){Hammer}, {Flores}, {Puech},
  {Yang}, {Athanassoula}, {Rodrigues}, \& {Delgado}}]{2009A&A...507.1313H}
{Hammer}, F., {Flores}, H., {Puech}, M., {et~al.} 2009{\natexlab{a}}, \aap,
  507, 1313

\bibitem[{{Hammer} {et~al.}(2009{\natexlab{b}}){Hammer}, {Flores}, {Yang},
  {Athanassoula}, {Puech}, {Rodrigues}, \& {Peirani}}]{2009A&A...496..381H}
{Hammer}, F., {Flores}, H., {Yang}, Y.~B., {et~al.} 2009{\natexlab{b}}, \aap,
  496, 381

\bibitem[{{Hammer} {et~al.}(2012){Hammer}, {Yang}, {Flores}, \&
  {Puech}}]{2012MPLA...2730034H}
{Hammer}, F., {Yang}, Y., {Flores}, H., \& {Puech}, M. 2012, Modern Physics
  Letters A, 27, 30034

\bibitem[{{Hammer} {et~al.}(2013){Hammer}, {Yang}, {Fouquet}, {Pawlowski},
  {Kroupa}, {Puech}, {Flores}, \& {Wang}}]{2013MNRAS.431.3543H}
{Hammer}, F., {Yang}, Y., {Fouquet}, S., {et~al.} 2013, \mnras, 431, 3543

\bibitem[{{Hammer} {et~al.}(2010){Hammer}, {Yang}, {Wang}, {Puech}, {Flores},
  \& {Fouquet}}]{2010ApJ...725..542H}
{Hammer}, F., {Yang}, Y.~B., {Wang}, J.~L., {et~al.} 2010, \apj, 725, 542

\bibitem[{{Hopkins} {et~al.}(2009{\natexlab{a}}){Hopkins}, {Cox}, {Younger}, \&
  {Hernquist}}]{2009ApJ...691.1168H}
{Hopkins}, P.~F., {Cox}, T.~J., {Younger}, J.~D., \& {Hernquist}, L.
  2009{\natexlab{a}}, \apj, 691, 1168

\bibitem[{{Hopkins} {et~al.}(2009{\natexlab{b}}){Hopkins}, {Somerville}, {Cox},
  {Hernquist}, {Jogee}, {Kere{\v s}}, {Ma}, {Robertson}, \&
  {Stewart}}]{2009MNRAS.397..802H}
{Hopkins}, P.~F., {Somerville}, R.~S., {Cox}, T.~J., {et~al.}
  2009{\natexlab{b}}, \mnras, 397, 802

\bibitem[{{Hubble}(1926)}]{1926ApJ....64..321H}
{Hubble}, E.~P. 1926, \apj, 64, 321

\bibitem[{{Huchra} \& {Geller}(1982)}]{1982ApJ...257..423H}
{Huchra}, J.~P. \& {Geller}, M.~J. 1982, \apj, 257, 423

\bibitem[{{Huertas-Company} {et~al.}(2010){Huertas-Company}, {Aguerri},
  {Tresse}, {Bolzonella}, {Koekemoer}, \& {Maier}}]{2010A&A...515A...3H}
{Huertas-Company}, M., {Aguerri}, J.~A.~L., {Tresse}, L., {et~al.} 2010, \aap,
  515, A3+

\bibitem[{{Ilyina} \& {Sil'chenko}(2012)}]{2012A&AT...27..313I}
{Ilyina}, M.~A. \& {Sil'chenko}, O.~K. 2012, Astronomical and Astrophysical
  Transactions, 27, 313

\bibitem[{{Jian} {et~al.}(2012){Jian}, {Lin}, \&
  {Chiueh}}]{2012ApJ...754...26J}
{Jian}, H.-Y., {Lin}, L., \& {Chiueh}, T. 2012, \apj, 754, 26

\bibitem[{{Jore} {et~al.}(1996){Jore}, {Broeils}, \&
  {Haynes}}]{1996AJ....112..438J}
{Jore}, K.~P., {Broeils}, A.~H., \& {Haynes}, M.~P. 1996, \aj, 112, 438

\bibitem[{{Kannappan} {et~al.}(2009){Kannappan}, {Guie}, \&
  {Baker}}]{2009AJ....138..579K}
{Kannappan}, S.~J., {Guie}, J.~M., \& {Baker}, A.~J. 2009, \aj, 138, 579

\bibitem[{{Kaviraj}(2014{\natexlab{a}})}]{2014arXiv1402.1166K}
{Kaviraj}, S. 2014{\natexlab{a}}, ArXiv e-prints

\bibitem[{{Kaviraj}(2014{\natexlab{b}})}]{2014MNRAS.437L..41K}
{Kaviraj}, S. 2014{\natexlab{b}}, \mnras, 437, L41

\bibitem[{{Kaviraj} {et~al.}(2011){Kaviraj}, {Tan}, {Ellis}, \&
  {Silk}}]{2011MNRAS.411.2148K}
{Kaviraj}, S., {Tan}, K.-M., {Ellis}, R.~S., \& {Silk}, J. 2011, \mnras, 411,
  2148

\bibitem[{{Kawata} \& {Mulchaey}(2008)}]{2008ApJ...672L.103K}
{Kawata}, D. \& {Mulchaey}, J.~S. 2008, \apjl, 672, L103

\bibitem[{{Kormendy} \& {Bender}(2012)}]{2012ApJS..198....2K}
{Kormendy}, J. \& {Bender}, R. 2012, \apjs, 198, 2

\bibitem[{{Kregel} {et~al.}(2002){Kregel}, {van der Kruit}, \& {de
  Grijs}}]{2002MNRAS.334..646K}
{Kregel}, M., {van der Kruit}, P.~C., \& {de Grijs}, R. 2002, \mnras, 334, 646

\bibitem[{{Laine} {et~al.}(2014){Laine}, {Laurikainen}, {Salo}, {Comer{\'o}n},
  {Buta}, {Zaritsky}, {Athanassoula}, {Bosma}, {Mu{\~n}oz-Mateos}, {Gadotti},
  {Hinz}, {Erroz-Ferrer}, {Gil de Paz}, {Kim}, {Men{\'e}ndez-Delmestre},
  {Mizusawa}, {Regan}, {Seibert}, \& {Sheth}}]{2014arXiv1404.0559L}
{Laine}, J., {Laurikainen}, E., {Salo}, H., {et~al.} 2014, ArXiv e-prints

\bibitem[{{Laurikainen} \& {Salo}(2001)}]{2001MNRAS.324..685L}
{Laurikainen}, E. \& {Salo}, H. 2001, \mnras, 324, 685

\bibitem[{{Laurikainen} {et~al.}(2005){Laurikainen}, {Salo}, \&
  {Buta}}]{2005MNRAS.362.1319L}
{Laurikainen}, E., {Salo}, H., \& {Buta}, R. 2005, \mnras, 362, 1319

\bibitem[{{Laurikainen} {et~al.}(2009){Laurikainen}, {Salo}, {Buta}, \&
  {Knapen}}]{2009ApJ...692L..34L}
{Laurikainen}, E., {Salo}, H., {Buta}, R., \& {Knapen}, J.~H. 2009, \apjl, 692,
  L34

\bibitem[{{Laurikainen} {et~al.}(2011){Laurikainen}, {Salo}, {Buta}, \&
  {Knapen}}]{2011MNRAS.418.1452L}
{Laurikainen}, E., {Salo}, H., {Buta}, R., \& {Knapen}, J.~H. 2011, \mnras,
  418, 1452

\bibitem[{{Laurikainen} {et~al.}(2010){Laurikainen}, {Salo}, {Buta}, {Knapen},
  \& {Comer{\'o}n}}]{2010MNRAS.405.1089L}
{Laurikainen}, E., {Salo}, H., {Buta}, R., {Knapen}, J.~H., \& {Comer{\'o}n},
  S. 2010, \mnras, 405, 1089

\bibitem[{{Law} {et~al.}(2009){Law}, {Steidel}, {Erb}, {Larkin}, {Pettini},
  {Shapley}, \& {Wright}}]{2009ApJ...697.2057L}
{Law}, D.~R., {Steidel}, C.~C., {Erb}, D.~K., {et~al.} 2009, \apj, 697, 2057

\bibitem[{{Longhetti} \& {Saracco}(2009)}]{2009MNRAS.394..774L}
{Longhetti}, M. \& {Saracco}, P. 2009, \mnras, 394, 774

\bibitem[{{Lotz} {et~al.}(2010){Lotz}, {Jonsson}, {Cox}, \&
  {Primack}}]{2010MNRAS.404..590L}
{Lotz}, J.~M., {Jonsson}, P., {Cox}, T.~J., \& {Primack}, J.~R. 2010, \mnras,
  404, 590

\bibitem[{{MacArthur} {et~al.}(2004){MacArthur}, {Courteau}, {Bell}, \&
  {Holtzman}}]{2004ApJS..152..175M}
{MacArthur}, L.~A., {Courteau}, S., {Bell}, E., \& {Holtzman}, J.~A. 2004,
  \apjs, 152, 175

\bibitem[{{Mahajan} \& {Raychaudhury}(2009)}]{2009MNRAS.400..687M}
{Mahajan}, S. \& {Raychaudhury}, S. 2009, \mnras, 400, 687

\bibitem[{{Maltby} {et~al.}(2012{\natexlab{a}}){Maltby}, {Gray},
  {Arag{\'o}n-Salamanca}, {Wolf}, {Bell}, {Jogee}, {H{\"a}u{\ss}ler},
  {Barazza}, {B{\"o}hm}, \& {Jahnke}}]{2012MNRAS.419..669M}
{Maltby}, D.~T., {Gray}, M.~E., {Arag{\'o}n-Salamanca}, A., {et~al.}
  2012{\natexlab{a}}, \mnras, 419, 669

\bibitem[{{Maltby} {et~al.}(2012{\natexlab{b}}){Maltby}, {Hoyos}, {Gray},
  {Arag{\'o}n-Salamanca}, \& {Wolf}}]{2012MNRAS.420.2475M}
{Maltby}, D.~T., {Hoyos}, C., {Gray}, M.~E., {Arag{\'o}n-Salamanca}, A., \&
  {Wolf}, C. 2012{\natexlab{b}}, \mnras, 420, 2475

\bibitem[{{Maraston} {et~al.}(2013){Maraston}, {Pforr}, {Henriques}, {Thomas},
  {Wake}, {Brownstein}, {Capozzi}, {Tinker}, {Bundy}, {Skibba}, {Beifiori},
  {Nichol}, {Edmondson}, {Schneider}, {Chen}, {Masters}, {Steele}, {Bolton},
  {York}, {Weaver}, {Higgs}, {Bizyaev}, {Brewington}, {Malanushenko},
  {Malanushenko}, {Snedden}, {Oravetz}, {Pan}, {Shelden}, \&
  {Simmons}}]{2013MNRAS.435.2764M}
{Maraston}, C., {Pforr}, J., {Henriques}, B.~M., {et~al.} 2013, \mnras, 435,
  2764

\bibitem[{{Mart{\'{\i}}nez-Delgado} {et~al.}(2010){Mart{\'{\i}}nez-Delgado},
  {Gabany}, {Crawford}, {Zibetti}, {Majewski}, {Rix}, {Fliri},
  {Carballo-Bello}, {Bardalez-Gagliuffi}, {Pe{\~n}arrubia}, {Chonis}, {Madore},
  {Trujillo}, {Schirmer}, \& {McDavid}}]{2010AJ....140..962M}
{Mart{\'{\i}}nez-Delgado}, D., {Gabany}, R.~J., {Crawford}, K., {et~al.} 2010,
  \aj, 140, 962

\bibitem[{{Mihos} \& {Hernquist}(1994)}]{1994ApJ...437..611M}
{Mihos}, J.~C. \& {Hernquist}, L. 1994, \apj, 437, 611

\bibitem[{{Miralles-Caballero} {et~al.}(2014){Miralles-Caballero},
  {D{\'{\i}}az}, {Rosales-Ortega}, {P{\'e}rez-Montero}, \&
  {S{\'a}nchez}}]{2014arXiv1403.1817M}
{Miralles-Caballero}, D., {D{\'{\i}}az}, A.~I., {Rosales-Ortega}, F.~F.,
  {P{\'e}rez-Montero}, E., \& {S{\'a}nchez}, S.~F. 2014, ArXiv e-prints

\bibitem[{{Miyamoto} \& {Nagai}(1975)}]{1975PASJ...27..533M}
{Miyamoto}, M. \& {Nagai}, R. 1975, \pasj, 27, 533

\bibitem[{{Moran} {et~al.}(2007){Moran}, {Ellis}, {Treu}, {Smith}, {Rich}, \&
  {Smail}}]{2007ApJ...671.1503M}
{Moran}, S.~M., {Ellis}, R.~S., {Treu}, T., {et~al.} 2007, \apj, 671, 1503

\bibitem[{{Naab} \& {Burkert}(2003)}]{2003ApJ...597..893N}
{Naab}, T. \& {Burkert}, A. 2003, \apj, 597, 893

\bibitem[{{Papovich} {et~al.}(2005){Papovich}, {Dickinson}, {Giavalisco},
  {Conselice}, \& {Ferguson}}]{2005ApJ...631..101P}
{Papovich}, C., {Dickinson}, M., {Giavalisco}, M., {Conselice}, C.~J., \&
  {Ferguson}, H.~C. 2005, \apj, 631, 101

\bibitem[{{Paturel} {et~al.}(2003){Paturel}, {Petit}, {Prugniel}, {Theureau},
  {Rousseau}, {Brouty}, {Dubois}, \& {Cambr{\'e}sy}}]{2003A&A...412...45P}
{Paturel}, G., {Petit}, C., {Prugniel}, P., {et~al.} 2003, \aap, 412, 45

\bibitem[{{Peirani} {et~al.}(2009){Peirani}, {Hammer}, {Flores}, {Yang}, \&
  {Athanassoula}}]{2009A&A...496...51P}
{Peirani}, S., {Hammer}, F., {Flores}, H., {Yang}, Y., \& {Athanassoula}, E.
  2009, \aap, 496, 51

\bibitem[{{Peterson}(1978)}]{1978ApJ...222...84P}
{Peterson}, C.~J. 1978, \apj, 222, 84

\bibitem[{{Poggianti} {et~al.}(2009){Poggianti}, {Arag{\'o}n-Salamanca},
  {Zaritsky}, {DeLucia}, {Milvang-Jensen}, {Desai}, {Jablonka}, {Halliday},
  {Rudnick}, {Varela}, {Bamford}, {Best}, {Clowe}, {Noll}, {Saglia},
  {Pell{\'o}}, {Simard}, {von der Linden}, \& {White}}]{2009ApJ...693..112P}
{Poggianti}, B.~M., {Arag{\'o}n-Salamanca}, A., {Zaritsky}, D., {et~al.} 2009,
  \apj, 693, 112

\bibitem[{{Poggianti} {et~al.}(2001){Poggianti}, {Bridges}, {Carter},
  {Mobasher}, {Doi}, {Iye}, {Kashikawa}, {Komiyama}, {Okamura}, {Sekiguchi},
  {Shimasaku}, {Yagi}, \& {Yasuda}}]{2001ApJ...563..118P}
{Poggianti}, B.~M., {Bridges}, T.~J., {Carter}, D., {et~al.} 2001, \apj, 563,
  118

\bibitem[{{Pohlen} \& {Trujillo}(2006)}]{2006A&A...454..759P}
{Pohlen}, M. \& {Trujillo}, I. 2006, \aap, 454, 759

\bibitem[{{Prieto} {et~al.}(2001){Prieto}, {Aguerri}, {Varela}, \&
  {Mu{\~n}oz-Tu{\~n}{\'o}n}}]{2001A&A...367..405P}
{Prieto}, M., {Aguerri}, J.~A.~L., {Varela}, A.~M., \&
  {Mu{\~n}oz-Tu{\~n}{\'o}n}, C. 2001, \aap, 367, 405

\bibitem[{{Prieto} {et~al.}(2013){Prieto}, {Eliche-Moral}, {Balcells},
  {Crist{\'o}bal-Hornillos}, {Erwin}, {Abreu}, {Dom{\'{\i}}nguez-Palmero},
  {Hempel}, {L{\'o}pez-Sanjuan}, {Guzm{\'a}n}, {P{\'e}rez-Gonz{\'a}lez},
  {Barro}, {Gallego}, \& {Zamorano}}]{2013MNRAS.428..999P}
{Prieto}, M., {Eliche-Moral}, M.~C., {Balcells}, M., {et~al.} 2013, \mnras,
  428, 999

\bibitem[{{Puech} {et~al.}(2009){Puech}, {Hammer}, {Flores}, {Neichel}, \&
  {Yang}}]{2009A&A...493..899P}
{Puech}, M., {Hammer}, F., {Flores}, H., {Neichel}, B., \& {Yang}, Y. 2009,
  \aap, 493, 899

\bibitem[{{Roche} {et~al.}(2010){Roche}, {Bernardi}, \&
  {Hyde}}]{2010MNRAS.407.1231R}
{Roche}, N., {Bernardi}, M., \& {Hyde}, J. 2010, \mnras, 407, 1231

\bibitem[{{Sil'chenko}(2013)}]{2013MSAIS..25...93S}
{Sil'chenko}, O. 2013, Memorie della Societa Astronomica Italiana Supplementi,
  25, 93

\bibitem[{{Sil'chenko}(2009)}]{2009IAUS..254..173S}
{Sil'chenko}, O.~K. 2009, in IAU Symposium, Vol. 254, IAU Symposium, ed.
  J.~{Andersen}, {Nordstr{\"o}ara}, B.~{m}, \& J.~{Bland-Hawthorn}, 173--178

\bibitem[{{Sil'Chenko} {et~al.}(2011){Sil'Chenko}, {Chilingarian}, {Sotnikova},
  \& {Afanasiev}}]{2011MNRAS.414.3645S}
{Sil'Chenko}, O.~K., {Chilingarian}, I.~V., {Sotnikova}, N.~Y., \& {Afanasiev},
  V.~L. 2011, \mnras, 414, 3645

\bibitem[{{Sil'chenko} {et~al.}(2012){Sil'chenko}, {Proshina}, {Shulga}, \&
  {Koposov}}]{2012MNRAS.427..790S}
{Sil'chenko}, O.~K., {Proshina}, I.~S., {Shulga}, A.~P., \& {Koposov}, S.~E.
  2012, \mnras, 427, 790

\bibitem[{{Simien} \& {Prugniel}(1998)}]{1998A&AS..131..287S}
{Simien}, F. \& {Prugniel}, P. 1998, \aaps, 131, 287

\bibitem[{{Simien} \& {Prugniel}(2000)}]{2000A&AS..145..263S}
{Simien}, F. \& {Prugniel}, P. 2000, \aaps, 145, 263

\bibitem[{{Simien} \& {Prugniel}(2002)}]{2002A&A...384..371S}
{Simien}, F. \& {Prugniel}, P. 2002, \aap, 384, 371

\bibitem[{{Spergel} {et~al.}(2007){Spergel}, {Bean}, {Dor{\'e}}, {Nolta},
  {Bennett}, {Dunkley}, {Hinshaw}, {Jarosik}, {Komatsu}, {Page}, {Peiris},
  {Verde}, {Halpern}, {Hill}, {Kogut}, {Limon}, {Meyer}, {Odegard}, {Tucker},
  {Weiland}, {Wollack}, \& {Wright}}]{2007ApJS..170..377S}
{Spergel}, D.~N., {Bean}, R., {Dor{\'e}}, O., {et~al.} 2007, \apjs, 170, 377

\bibitem[{{Tacconi} {et~al.}(2008){Tacconi}, {Genzel}, {Smail}, {Neri},
  {Chapman}, {Ivison}, {Blain}, {Cox}, {Omont}, {Bertoldi}, {Greve},
  {F{\"o}rster Schreiber}, {Genel}, {Lutz}, {Swinbank}, {Shapley}, {Erb},
  {Cimatti}, {Daddi}, \& {Baker}}]{2008ApJ...680..246T}
{Tacconi}, L.~J., {Genzel}, R., {Smail}, I., {et~al.} 2008, \apj, 680, 246

\bibitem[{{Tapia} {et~al.}(2014){Tapia}, {Eliche-Moral}, {Querejeta},
  {Balcells}, {C{\'e}sar Gonz{\'a}lez-Garc{\'{\i}}a}, {Prieto}, {Aguerri},
  {Gallego}, {Zamorano}, {Rodr{\'{\i}}guez-P{\'e}rez}, \&
  {Borlaff}}]{2014A&A...565A..31T}
{Tapia}, T., {Eliche-Moral}, M.~C., {Querejeta}, M., {et~al.} 2014, \aap, 565,
  A31

\bibitem[{{Thilker} {et~al.}(2010){Thilker}, {Bianchi}, {Schiminovich}, {Gil de
  Paz}, {Seibert}, {Madore}, {Wyder}, {Rich}, {Yi}, {Barlow}, {Conrow},
  {Forster}, {Friedman}, {Martin}, {Morrissey}, {Neff}, \&
  {Small}}]{2010ApJ...714L.171T}
{Thilker}, D.~A., {Bianchi}, L., {Schiminovich}, D., {et~al.} 2010, \apjl, 714,
  L171

\bibitem[{{van Dokkum} \& {Franx}(2001)}]{2001ApJ...553...90V}
{van Dokkum}, P.~G. \& {Franx}, M. 2001, \apj, 553, 90

\bibitem[{{Vijayaraghavan} \& {Ricker}(2013)}]{2013MNRAS.435.2713V}
{Vijayaraghavan}, R. \& {Ricker}, P.~M. 2013, \mnras, 435, 2713

\bibitem[{{Wei} {et~al.}(2010){Wei}, {Kannappan}, {Vogel}, \&
  {Baker}}]{2010ApJ...708..841W}
{Wei}, L.~H., {Kannappan}, S.~J., {Vogel}, S.~N., \& {Baker}, A.~J. 2010, \apj,
  708, 841

\bibitem[{{Wilman} {et~al.}(2009){Wilman}, {Oemler}, {Mulchaey}, {McGee},
  {Balogh}, \& {Bower}}]{2009ApJ...692..298W}
{Wilman}, D.~J., {Oemler}, Jr., A., {Mulchaey}, J.~S., {et~al.} 2009, \apj,
  692, 298

\bibitem[{{Yang} {et~al.}(2009){Yang}, {Hammer}, {Flores}, {Puech}, \&
  {Rodrigues}}]{2009A&A...501..437Y}
{Yang}, Y., {Hammer}, F., {Flores}, H., {Puech}, M., \& {Rodrigues}, M. 2009,
  \aap, 501, 437

\bibitem[{{Younger} {et~al.}(2007){Younger}, {Cox}, {Seth}, \&
  {Hernquist}}]{2007ApJ...670..269Y}
{Younger}, J.~D., {Cox}, T.~J., {Seth}, A.~C., \& {Hernquist}, L. 2007, \apj,
  670, 269 (Y07)

\bibitem[{{Zentner} {et~al.}(2005){Zentner}, {Kravtsov}, {Gnedin}, \&
  {Klypin}}]{2005ApJ...629..219Z}
{Zentner}, A.~R., {Kravtsov}, A.~V., {Gnedin}, O.~Y., \& {Klypin}, A.~A. 2005,
  \apj, 629, 219

\end{thebibliography}
}

\end{document}